\documentclass[aps,prx,reprint,showkeys,floatfix]{revtex4-2}

\usepackage{amsmath}
\usepackage{amsfonts}
\usepackage{amssymb}
\usepackage{array,longtable,booktabs}
\newcolumntype{L}{>{$}l<{$}}
\newcolumntype{C}{>{$}c<{$}}
\usepackage[dvipsnames,table]{xcolor}
\usepackage{hyperref}
\hypersetup{
	colorlinks,
	linkcolor={red!50!black},
	citecolor={red!70!black},
	urlcolor={red!80!black}
}
\usepackage[capitalize]{cleveref}
\usepackage{multirow}
\usepackage{physics}
\usepackage{placeins}
\usepackage{graphicx}
\usepackage[caption=false]{subfig}
\usepackage{siunitx}
\usepackage{tensor}
\usepackage{tikz}
\usepackage{mhchem}

\usetikzlibrary{arrows.meta,shapes.multipart}
\usetikzlibrary{arrows,shapes,positioning,matrix,fit,patterns,backgrounds}
\usetikzlibrary{decorations.markings}
\usetikzlibrary{decorations.text}
\tikzstyle arrowstyle=[scale=1]

\DeclareMathOperator{\Span}{span}
\DeclareMathOperator{\Lie}{Lie}
\DeclareMathOperator{\ad}{ad}

\newcommand{\Pu}{P{\mathrel{\uparrow}}}
\newcommand{\Pd}{P{\mathrel{\downarrow}}}
\newcommand{\Qu}{Q{\mathrel{\uparrow}}}
\newcommand{\Qd}{Q{\mathrel{\downarrow}}}
\newcommand{\Ru}{R{\mathrel{\uparrow}}}
\newcommand{\Rd}{R{\mathrel{\downarrow}}}
\newcommand{\Su}{S{\mathrel{\uparrow}}}
\newcommand{\Sd}{S{\mathrel{\downarrow}}}

\newcommand{\intS}{\tensor*[^{[0]}]{A}{_{PQ}^{RS}}}
\newcommand{\intT}{\tensor*[^{[1]}]{A}{_{PQ}^{RS}}}

\newcommand{\sm}{Supplemental Material}
\newcommand{\foreign}[1]{\textit{#1}}

\definecolor{setgray}{gray}{0.55} 
\newcommand{\CommuteBoundary}{%
	\arrayrulecolor{Aquamarine!50}
	\cmidrule[3pt]{1-2}
	\arrayrulecolor{black}
}
\newcommand{\AntiHBoundary}{\cmidrule(l{0.125\textwidth}r{0.125\textwidth}){1-2}}

\begin{document}

\title{Spin-Adapted Fermionic Unitaries:\linebreak
	From Lie Algebras to Compact Quantum Circuits}

\author{Ilias Magoulas}
\email{ilias.magoulas@emory.edu}
\author{Francesco A. Evangelista}
\affiliation{Department of Chemistry and Cherry Emerson Center for Scientific Computation, Emory University, Atlanta, Georgia 30322, USA}

\begin{abstract}

Conservation of symmetries is crucial for reliable quantum simulations of molecular systems, yet compact circuit implementations of fully symmetry-adapted fermionic unitaries have remained elusive beyond the simplest excitation classes.
Here we address this issue for the set of singlet spin-adapted generalized singles and doubles operators (saGSD).
Using the Wei--Norman approach, we derive exact product formulas that express spin-adapted fermionic unitaries as products of elementary spin-orbital unitaries.
To obtain closed-form parameters in the more challenging 28- and 84-dimensional dynamical Lie algebras, we develop a computational discovery-and-verification protocol combining numerical optimization, parameter-structure identification, closed-form inference, and exact validation against reduced Wei--Norman equations.
We also introduce an algorithm for constructing closed-form fermionic unitary transformations on Krylov subspaces and extend the fermionic-excitation-based circuit formalism to generators consisting of an anti-Hermitian fermionic string multiplied by arbitrary linear combinations of number-operator products.
Together, these developments yield the most compact circuits to date for exact implementation of saGSD unitaries.
Finally, for non fully spin-polarized systems, we identify a compact universal symmetry-adapted subset of saGSD that further reduces the quantum resources required for chemically relevant simulations.

\end{abstract}

\keywords{
fermionic algebra, quantum many-body theory, quantum computing, spin-adaptation
}

\maketitle
\section{Introduction}
The simulation of quantum many-body systems, such as those encountered in chemistry, condensed matter physics, and materials science, has long been identified as a killer application for quantum computers \cite{Feynman.1982.10.1007/BF02650179,Lloyd.1996.10.1126/science.273.5278.107,Georgescu.2014.10.1103/RevModPhys.86.153,Reiher.2017.10.1073/pnas.1619152114}.
In the non-relativistic regime, the correlated motion of particles comprising such systems is governed by the Schr\"{o}dinger equation.
Although the computational resources required to solve the Schr\"{o}dinger equation exactly in a finite basis grow combinatorially with system size \cite{Paldus.1974.10.1063/1.1681883}, an ideal quantum device would require, in principle, a number of qubits that scales only linearly \cite{Lloyd.1996.10.1126/science.273.5278.107,Abrams.1997.10.1103/PhysRevLett.79.2586,Abrams.1999.10.1103/PhysRevLett.83.5162,Berry.2007.10.1007/s00220-006-0150-x,Babbush.2018.10.1103/PhysRevX.8.011044}.
Beyond academic interest, pushing the boundaries of system sizes amenable to computational treatment, both in the numbers of particles and basis functions, will have a transformative effect in many fields, including, for example, catalysis, photochemistry, energy-storage materials, biology, and strongly correlated systems \cite{Cao.2019.10.1021/acs.chemrev.8b00803,Bauer.2020.10.1021/acs.chemrev.9b00829,McArdle.2020.10.1103/RevModPhys.92.015003}.
However, access to a fault-tolerant quantum computer capable of rapid, noise-free simulations of large quantum systems would be of no practical value if the results were unreliable or outright wrong.

A major barrier hindering the practical utility of quantum simulations is the proper enforcement of the symmetries characterizing the system of interest.
In the context of chemical applications, for example, the electronic Hamiltonian conserves the number of particles $N$, the projection of the total spin on the $z$ axis $S_z$, the total spin squared $S^2$, and the irreducible representation of the molecular point group.
Arguably, the most important aspect of symmetry adaptation is the disentangling of states with different symmetry properties.
Indeed, if a quantum algorithm violates even one of the inherent symmetries of the problem of interest, there is always a risk of collapsing into a state with undesired symmetry.

\begin{figure}[h!]
	\centering
	\includegraphics[width=8.5cm]{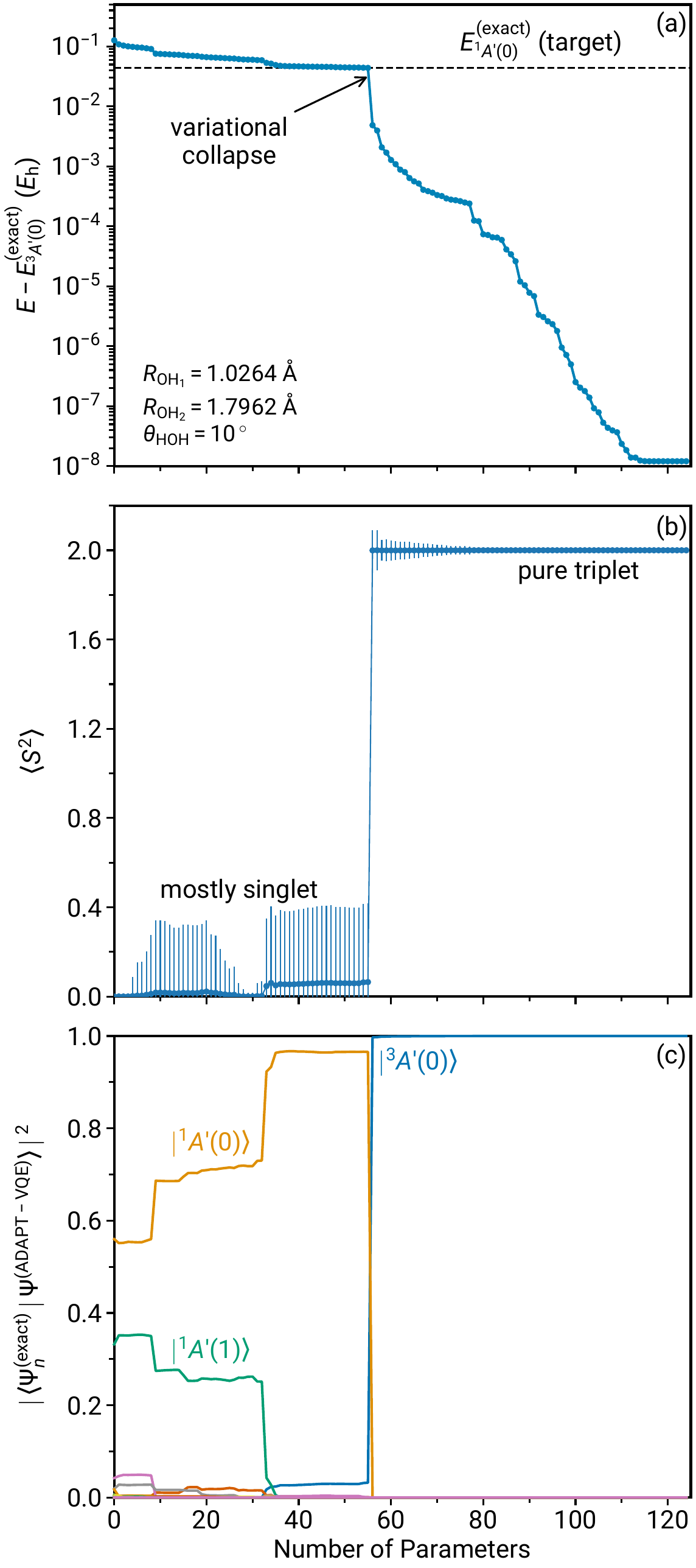}
	\caption{
		Convergence of ADAPT-VQE simulations using the $S^2$-breaking GSD operator pool for the water molecule.
		(a) Energy errors relative to the exact lowest-energy totally symmetric eigenstate in the ($N = 10$, $S_z = 0$) symmetry sector.
		(b) Expectation values of the total spin squared $S^2$ operator and their standard deviations.
		(c) Weights of the exact totally symmetric eigenstates in the ($N = 10$, $S_z = 0$) symmetry sector.
	}
	\label{fig:h2o_gsd}
\end{figure}

We illustrate such a failure with a numerical example based on the adaptive derivative-assembled pseudo-Trotter (ADAPT) \cite{Grimsley.2019.10.1038/s41467-019-10988-2} variational quantum eigensolver (VQE) \cite{Peruzzo.2014.10.1038/ncomms5213}, an adaptive algorithm that constructs the ansatz iteratively.
In \cref{fig:h2o_gsd}, we show the convergence of ADAPT-VQE simulations using the $S^2$-breaking generalized singles and doubles (GSD) \cite{Nooijen.2000.10.1103/PhysRevLett.84.2108,Nakatsuji.2000.10.1063/1.1287275} operator pool for a $C_s$-symmetric conformation of the water molecule, with one doubly occupied core spatial orbital frozen.
As shown in the {\sm}, this particular conformation was selected because the target lowest-energy $^1A^\prime$ state lies above the lowest-energy $^3A^\prime$ state.
The latter is the global energy minimum in the ($A^\prime$, $N = 10$, $S_z = 0$) symmetry sector of the many-electron Hilbert space.
The simulations are initiated from the closed-shell restricted Hartree--Fock Slater determinant, which is a pure singlet.
As shown in \cref{fig:h2o_gsd}, for the first 55 macro-iterations, the ADAPT-VQE state retains a mostly singlet character, correctly tracking the target $^1A^\prime$ state as demonstrated by the substantial increase in the weight of this eigenvector.
However, once the 56th operator is appended to the ansatz, the VQE optimization variationally collapses to an incorrect state with nearly pure triplet character.
As shown in panel (c) of \cref{fig:h2o_gsd}, the new ADAPT-VQE state is practically orthogonal to the target $^1A^\prime$ state and has weight close to 1 on the $^3A^\prime$ state.

Such dramatic failures due to variational collapse are particularly problematic because error-mitigation techniques \cite{Cai.2023.10.1103/RevModPhys.95.045005}, \foreign{e.g.}, postselection \cite{Aaronson.2005.10.1098/rspa.2005.1546}, are generally ineffective in this scenario.
To be precise, although postselection based on the targeted symmetry eigenvalue is useful in restoring symmetries artificially broken due to device noise \cite{Bonet-Monroig.2018.PhysRevA.98.062339,McArdle.2019.10.1103/PhysRevLett.122.180501}, it cannot be applied when the state produced by the noiseless quantum algorithm belongs to a wrong symmetry sector, as in the collapse to the incorrect triplet state shown in panel (c) of \cref{fig:h2o_gsd}.
These observations emphasize the need for quantum algorithms that rigorously respect all available symmetries.
Moreover, respecting symmetries in quantum simulations offers several advantages \cite{Magoulas.2025.10.1080/00268976.2025.2534672}.
To begin with, the dimension of the many-electron Hilbert space is significantly reduced when symmetry constraints are enforced, resulting in far fewer optimization parameters to obtain accurate or even numerically exact results.
Furthermore, for adaptive algorithms \cite{Grimsley.2019.10.1038/s41467-019-10988-2,Ryabinkin.2020.10.1021/acs.jctc.9b01084,Stair.2021.10.1103/PRXQuantum.2.030301}, symmetry adaptation leads to smaller operator pools and, hence, fewer measurements during the operator selection process.

In this work, we employ Lie-algebraic techniques to design compact quantum circuits that rigorously enforce all the symmetries of the electronic Hamiltonian relevant to chemical systems.
Specifically, we use the Wei--Norman decomposition \cite{Wei.1963.10.1063/1.1703993,Wei.1964.10.2307/2034065} to express  unitaries generated by symmetry-adapted double excitations exactly as products of elementary, spin-orbital unitaries.
The individual components respect the particle-number, $S_z$, and spatial symmetries by construction, while the exactness of the decomposition ensures that their product enforces $S^2$ symmetry as well.
Furthermore, we extend the fermionic-excitation-based (FEB) formalism \cite{Yordanov.2020.10.1103/PhysRevA.102.062612,Magoulas.2023.10.1021/acs.jctc.2c01016}, which leads to compact fermionic circuits, to unitaries generated by an anti-Hermitian fermionic string multiplied by an arbitrary linear combination of number operator products.
This allows us to construct compact circuit implementations of spin-adapted unitaries, assuming the Jordan--Wigner \cite{Jordan.1928.10.1007/BF01331938} mapping and all-to-all connectivity.
Finally, to further reduce the required quantum resources, we introduce a compact universal subset of the fully symmetry-adapted generalized singles and doubles operator pool \cite{Nooijen.2000.10.1103/PhysRevLett.84.2108,Nakatsuji.2000.10.1063/1.1287275}.

Beyond the resulting circuit constructions, this work introduces several methodological advances that may be useful in broader applications of Lie-algebraic quantum circuit synthesis.
First, we obtain closed-form expressions for Wei--Norman decompositions involving dynamical Lie algebras of dimensions 28 and 84, providing, to the best of our knowledge, one of the most extensive closed-form Wei--Norman parametrizations.
Second, we develop a computational discovery-and-verification strategy for otherwise intractable Wei--Norman systems.
In particular, we combine numerical optimization, identification of parameter symmetries and proportionalities, reduction to independent parameter families, closed-form inference, and final validation by exact satisfaction of the Wei--Norman equations in a reduced space.
Third, we introduce an automatic Krylov-subspace algorithm for deriving exact closed-form unitary transformations of fermionic operators involving finite linear combinations of fermionic strings, with the minimum number of nested commutators required by the adjoint action.
Together with an extension of the FEB circuit-synthesis formalism to generators consisting of anti-Hermitian fermionic strings multiplied by linear combinations of number-operator products, these tools make it possible to compress many elementary exponentials into single compact circuit blocks.	
	
While finalizing this manuscript, we became aware of concurrent work by Jain \foreign{et al.} \cite{Jain.2026.10.1063.5.0326865}, who independently derived exact factorizations of unitaries generated by singlet spin-adapted fermionic double excitations.
Although the basis elements of the pertinent Lie algebras have the same algebraic structure, they adopted a different strategy for implementing the quantum circuits.
Specifically, Jain \foreign{et al.} map the individual generators to linear combinations of mutually commuting Pauli strings \cite{Romero.2019.10.1088/2058-9565/aad3e4} within the Jordan--Wigner \cite{Jordan.1928.10.1007/BF01331938} fermionic encoding.
For example, this procedure results in 640 Pauli strings in the most complicated case of the double excitation going through an intermediate triplet.
Considering that each exponentiated Pauli string requires two controlled-NOT (CNOT) ladders, the circuits reported by Jain \foreign{et al.} are long and deep.
In contrast, our extension of the FEB formalism leads to compact quantum circuits that require only two CNOT ladders in total for each spin-adapted unitary.

\section{Symmetry enforcement in quantum simulations}

One can classify techniques to enforce symmetries into three categories: 1) a broad set of approaches that do not require substantial algorithmic or state-preparation modifications, 2) techniques to produce a symmetry-adapted initial state, and 3) symmetry-preserving quantum gates.
Within the first category, one approach is symmetry restoration via projection \cite{Whitfield.2013.10.1063/1.4812566,Izmaylov.2019.10.1021/acs.jpca.9b01103}.
Although, as already mentioned, projection after variation, \foreign{i.e.}, postselection, is not ideal, variation after projection is more robust and requires fewer optimization parameters because the optimization occurs in the symmetry-adapted Hilbert space.
This technique is general and can be used to restore any symmetries of interest, including spatial (\foreign{e.g.}, point-group) \cite{Seki.2020.10.1103/PhysRevA.101.052340,Seki.2022.10.1103/PhysRevA.105.032419}, spin \cite{Tsuchimochi.2020.10.1103/PhysRevResearch.2.043142,Tsuchimochi.2022.10.1103/PhysRevResearch.4.033100,Siwach.2021.10.1103/PhysRevA.104.062435,Seki.2022.10.1103/PhysRevA.105.032419,Guzman.2023.10.1103/PhysRevC.107.034310}, and particle-number \cite{Khamoshi.2021.10.1088/2058-9565/abc1bb,Guzman.2022.10.1103/PhysRevC.105.024324,Guzman.2023.10.1103/PhysRevC.107.034310} symmetries.
Nevertheless, symmetry projectors are typically implemented as linear combinations of unitaries (LCU) \cite{Childs.2012.10.26421/QIC12.11-12,Berry.2015.10.1103/PhysRevLett.114.090502}, which introduce ancilla registers and many controlled operations, inflating the CNOT count (see also Ref.\ \cite{Bastidas.2025.10.1103/PhysRevA.111.052433} for the use of LCU in unifying finite symmetries in the quantum simulation of many-body systems).
Furthermore, their success depends on the non-zero overlap of the initial symmetry-broken state with the target symmetry sector.

An alternative strategy is the incorporation of penalty terms in the Hamiltonian that artificially raise the energy of states with undesired symmetries \cite{McClean.2016.10.1088/1367-2630/18/2/023023,Ryabinkin.2019.10.1021/acs.jctc.8b00943,Greene-Diniz.2021.10.1002/qua.26352,Kuroiwa.2021.10.1103/PhysRevResearch.3.013197} (see, also, \cite{Selvarajan.2022.10.3390/sym14030457}).
In doing so, the quantum simulations are biased toward states with desired symmetry properties.
Although such approaches are general, they typically require a large number of measurements, depend on user-defined hyperparameters, and the optimization is formally occurring on the full, rather than symmetry-adapted, Hilbert space.

In qubit reduction techniques \cite{Bravyi.2017.1701.08213,Setia.2020.10.1021/acs.jctc.0c00113,Picozzi.2023.10.1088/2058-9565/acd86c}, one takes advantage of $\mathbb{Z}_2$ symmetries of the Hamiltonian, including particle-number parities and certain point group symmetries, to block diagonalize the Hamiltonian and eliminate redundant qubits.
The quantum simulation is subsequently performed on the Hamiltonian block with the targeted $\mathbb{Z}_2$ symmetries.
Although these schemes reduce the required quantum resources, in their current formulation they are not applicable to non-Abelian or continuous symmetries, such as $S^2$.

In the second family of methods, a few strategies have been proposed for the preparation of spin eigenfunctions on quantum devices \cite{Sugisaki.2016.10.1021/acs.jpca.6b04932,Sugisaki.2019.10.1016/j.cpletx.2018.100002,Gard.2020.10.1038/s41534-019-0240-1,Carbone.2022.10.3390/sym14030624}.
In \cite{Sugisaki.2016.10.1021/acs.jpca.6b04932,Sugisaki.2019.10.1016/j.cpletx.2018.100002,Carbone.2022.10.3390/sym14030624}, a single configuration state function is prepared on the quantum device, serving as a good initial guess for algorithms such as quantum phase estimation \cite{Kitaev.1995.quant-ph/9511026}.
However, the approximate implementation of the time-evolution operator can still introduce symmetry contaminants.
The symmetry-preserving preparation circuits of \cite{Gard.2020.10.1038/s41534-019-0240-1} construct a hyperspherical parameterization of the exact ground state that requires a combinatorial number of CNOT gates, and, thus, limits its application to small active spaces.
Alternatively, one can transform the computational basis states from encoding Slater determinants to representing configuration state functions via the quantum Schur transform \cite{Bacon.2006.10.1103/PhysRevLett.97.170502}.
The issue with this approach is that it requires a register of qudits whose count and local dimension depend on the system size.
For spin Hamiltonians, it has been demonstrated that this difficulty can be circumvented by employing a truncated spin-adapted basis \cite{Gandon.2025.10.1103/dbnd-sl4j}, albeit introducing an approximation.
Recently, the Quantum Paldus Transform has been introduced \cite{Burkat.2025.2506.09151}, which transforms the computational basis states from Slater determinants to a spin-adapted Gel'fand--Tsetlin basis associated with the unitary group approach.
The algorithm is efficient, having an asymptotic Toffoli complexity that scales with the number of spatial orbitals $d$ as $\mathcal{O}(d^3)$, although it shifts the computational complexity to the representation and implementation of chemically relevant operators in the transformed basis.

Methods in the third category employ unitaries that preserve symmetries by construction.
In general, a unitary operator can be expressed either as $\exp(-iH)$ with a Hermitian generator $H$, as in time evolution and quantum phase estimation \cite{Kitaev.1995.quant-ph/9511026}, or as $\exp(A)$ with an anti-Hermitian generator $A$, which is typical in ansatz construction.
Of these two forms, the latter has received considerable attention, because electronic-structure simulations are a key target for current noisy and near-fault-tolerant quantum hardware.
Notable examples in this family of methods are the quantum-number-preserving (QNP) gate fabric \cite{Anselmetti.2021.10.1088/1367-2630/ac2cb3}, the discretely optimized variational quantum eigensolver (DISCO-VQE) \cite{Burton.2023.10.1038/s41534-023-00744-2}, and the tiled unitary product states (tUPS) approach \cite{Burton.2024.10.1103/PhysRevResearch.6.023300}.
These schemes rely on essentially the same set of elementary, symmetry-preserving unitaries, namely, the singlet spin-adapted singles and perfect pairing doubles operator pool.
This pool typically enforces the particle-number, $S_z$, and $S^2$ symmetries, but generally requires spatial-symmetry-violating operators to attain universality \cite{Magoulas.2026.10.1063/5.0316482}.
As a result, quantum simulations based on these techniques may variationally collapse to states with undesired symmetry.

\section{Generators of symmetry-adapted fermionic unitaries}

Arguably the most natural approach to locally enforce the key symmetries relevant to molecular quantum simulations is to employ operator pools relying on fermionic excitation operators, typically based on the unitary extension \cite{Kutzelnigg.1977.10.1007/978-1-4757-0887-5_5,Kutzelnigg.1982.10.1063/1.444231,Kutzelnigg.1983.10.1063/1.446313,Kutzelnigg.1984.10.1063/1.446736,Bartlett.1989.10.1016/S0009-2614(89)87372-5,Szalay.1995.10.1063/1.469641,Taube.2006.10.1002/qua.21198,Cooper.2010.10.1063/1.3520564,Evangelista.2011.10.1063/1.3598471,Harsha.2018.10.1063/1.5011033,Filip.2020.10.1063/5.0026141,Freericks.2022.10.3390/sym14030494,Anand.2022.10.1039/d1cs00932j} of coupled-cluster theory \cite{Coester.1958.10.1016/0029-5582(58)90280-3,Coester.1960.10.1016/0029-5582(60)90140-1,Cizek.1966.10.1063/1.1727484,Cizek.1969.10.1002/9780470143599.ch2,Cizek.1971.10.1002/qua.560050402,Paldus.1972.10.1103/PhysRevA.5.50}.
The most widely used pool in this category consists of anti-Hermitian fermionic generalized single,
\begin{equation}
	\label{eq:gs}
	A_p^q = a_q^\dagger a_p - a_p^\dagger a_q,
\end{equation}
and double,
\begin{equation}
	\label{eq:gd}
	A_{pq}^{rs} = a_r^\dagger a_s^\dagger a_q a_p - a_p^\dagger a_q^\dagger a_s a_r,
\end{equation}
excitation operators \cite{Nooijen.2000.10.1103/PhysRevLett.84.2108,Nakatsuji.2000.10.1063/1.1287275}, commonly denoted as GSD.
Here, indices $p$, $q$, $r$, and $s$ denote generic spin orbitals, and $a_p$ ($a_p^\dagger$) is the annihilation (creation) operator associated with the $p$th spin orbital.
The popularity of the GSD pool can be attributed to the following advantages.
Although composed of low-rank many-body operators, it is universal if each operator can appear multiple times in the ansatz with independent optimization parameters \cite{Evangelista.2019.10.1063/1.5133059}.
The construction of efficient quantum circuit implementations of GSD unitaries is enabled through the FEB formulation \cite{Yordanov.2020.10.1103/PhysRevA.102.062612,Xia.2021.10.1088/2058-9565/abbc74}.
Finally, GSD operators satisfy particle-number symmetry by construction, while the enforcement of $S_z$ and point-group symmetries is straightforward by imposing appropriate restrictions on the excitation indices \cite{Magoulas.2025.10.1080/00268976.2025.2534672}.

A persistent limitation of the GSD operator pool is the enforcement of $S^2$ symmetry, which requires the replacement of \cref{eq:gs,eq:gd} by their singlet spin-adapted counterparts \cite{Paldus.1977.10.1002/qua.560110511,Paldus.1977.10.1063/1.434526,Adams.1979.10.1103/PhysRevA.20.1,Chiles.1981.10.1063/1.441643,Piecuch.1989.10.1002/qua.560360402,Geertsen.1991.10.1016/S0065-3276(08)60364-0,Piecuch.1992.10.1007/BF01113244,Piecuch.1994.10.1063/1.467304}.
Using the notation in which $P$, $Q$, $R$, and $S$ index generic spatial orbitals, and the $s_z = -\tfrac{1}{2}$ and $s_z = \tfrac{1}{2}$ spin functions are designated as $\downarrow$ and $\uparrow$, respectively, the singlet spin-adapted single excitations take the form
\begin{equation}
	\label{eq:sa_gs}
	\begin{split}
	A_P^Q &= \frac{1}{\sqrt{2}} \left( A_{\Pu}^{\Qu} + A_{\Pd}^{\Qd} \right)\\
	&= \frac{1}{\sqrt{2}} \left( a_{\Qu}^\dagger a_{\Pu} + a_{\Qd}^\dagger a_{\Pd} - \text{h.c.} \right),
	\end{split}
\end{equation}
with ``h.c.'' denoting the Hermitian conjugate.
The corresponding double excitations can be grouped into four distinct cases:
\begin{equation}
	\label{eq:sa_gd_ppqq}
	\begin{split}
	A_{PP}^{QQ} &= A_{\Pu\Pd}^{\Qu\Qd}\\
	&= a_{\Qu}^\dagger a_{\Qd}^\dagger a_{\Pd} a_{\Pu} - \text{h.c.},
	\end{split}
\end{equation}
\begin{equation}
	\label{eq:sa_gd_ppqr}
	A_{PP}^{QR} = \frac{1}{\sqrt{2}} \left( A_{\Pu\Pd}^{\Qu\Rd} - A_{\Pu\Pd}^{\Qd\Ru} \right),
\end{equation}
\begin{equation}
	\label{eq:sa_gd_pqrs_int0}
	\intS = \frac{1}{2} \left( A_{\Pu\Qd}^{\Ru\Sd} - A_{\Pu\Qd}^{\Rd\Su} - A_{\Pd\Qu}^{\Ru\Sd} + A_{\Pd\Qu}^{\Rd\Su} \right),
\end{equation}
and
\begin{equation}
	\label{eq:sa_gd_pqrs_int1}
	\begin{split}
	\intT ={}& \frac{1}{\sqrt{3}} \left[\vphantom{\frac{1}{2}} A_{\Pu\Qu}^{\Ru\Su} + A_{\Pd\Qd}^{\Rd\Sd}\right.\\
	&\left.+ \frac{1}{2} \left( A_{\Pu\Qd}^{\Ru\Sd} + A_{\Pu\Qd}^{\Rd\Su} + A_{\Pd\Qu}^{\Ru\Sd} + A_{\Pd\Qu}^{\Rd\Su} \right) \right],
	\end{split}
\end{equation}
with the ``[0]'' and ``[1]'' superscripts in \cref{eq:sa_gd_pqrs_int0,eq:sa_gd_pqrs_int1} denoting the common intermediate spin quantum number characterizing the $PQ$ and $RS$ orbital pairs.
The singlet spin-adapted GSD pool is comprised of the operators in \cref{eq:sa_gs,eq:sa_gd_ppqq,eq:sa_gd_ppqr,eq:sa_gd_pqrs_int0,eq:sa_gd_pqrs_int1} and is typically denoted as saGSD.

The design of efficient quantum circuits representing unitaries generated by singlet spin-adapted operators is straightforward in the case of single [\cref{eq:sa_gs}] and perfect-pairing double [\cref{eq:sa_gd_ppqq}] excitations, which together define the operator pool typically denoted as saGSpD \cite{Anselmetti.2021.10.1088/1367-2630/ac2cb3,Burton.2023.10.1038/s41534-023-00744-2,Burton.2024.10.1103/PhysRevResearch.6.023300}.
The issue with designing efficient quantum circuits for unitaries generated by the more complicated singlet spin-adapted double excitation operators [\cref{eq:sa_gd_ppqr,eq:sa_gd_pqrs_int0,eq:sa_gd_pqrs_int1}] lies in the fact that finite-order product formulas based on, for example, truncated Trotter--Suzuki or Zassenhaus decompositions, violate spin symmetry (see Ref.\ \cite{Magoulas.2025.10.1080/00268976.2025.2534672} for the case of Trotterization).
	
Nevertheless, when the relevant operators generate a finite-dimensional Lie algebra, exact product formulas may be obtainable.
Let us consider, for example, the exponential $\exp[\theta (A+B)]$ with $A$ and $B$ being noncommuting operators.
An $m$th-order Trotter--Suzuki decomposition approximates this exponential by a finite product of alternating exponentials of $A$ and $B$ \cite{Hatano.2005.10.1007/11526216_2,Barthel.2020.10.1016/j.aop.2020.168165}:
\begin{equation}\label{eq:trotter}
	\begin{split}
	e^{\theta (A+B)} &\approx e^{\alpha_1 \theta A} e^{\alpha_2 \theta B} e^{\alpha_3 \theta A} e^{\alpha_4 \theta B} \cdots e^{\alpha_k \theta B}\\
	&= e^{\theta (A+B) + \mathcal{O}(\theta^{m+1})}.
	\end{split}
\end{equation}
The coefficients $\alpha_i$ appearing in the exponentials are chosen so that all error terms up to order $\theta^m$ cancel, yielding an error of order $\theta^{m+1}$.
If the operators $A$ and $B$ give rise to a finite-dimensional Lie algebra, it may be possible to construct an exact product formula analogous to \cref{eq:trotter}, albeit with $\theta$-dependent optimization parameters $\alpha_i (\theta)$.
As shown in Section SI of the {\sm}, in the case of $\exp(\theta A_{PP}^{QR})$, we were able to find an exact product formula with six exponentials using an alternating pattern of $A_{\Pu\Pd}^{\Qu\Rd}$ and $A_{\Pu\Pd}^{\Qd\Ru}$ operators.
However, this construction requires an additional exponential compared to the exact decomposition derived in \cref{sec:wn_ppqr}.
To make matters worse, the parameters are highly oscillatory functions of $\theta$ (see Fig.\ S1 in the {\sm}), rendering the resulting quantum circuits more sensitive to device noise and complicating the computation of derivatives.
	
An alternative exact strategy is based on LCU methods.
We have recently shown that spin-adapted unitaries can be exactly represented as a finite linear combination of spin-orbital operators \cite{Magoulas.2025.10.1080/00268976.2025.2534672} (see also \cite{Kjellgren.2025.10.1063/5.0278717}), enabling their circuit implementation via the LCU approach \cite{Childs.2012.10.26421/QIC12.11-12,Berry.2015.10.1103/PhysRevLett.114.090502}.
However, as discussed in Section SII in the {\sm}, even state-of-the-art LCU implementations \cite{Berry.2014.10.1145/2591796.2591854,Berry.2015.10.1109/FOCS.2015.54,Berry.2015.10.1103/PhysRevLett.114.090502,Gilyen.2019.10.1145/3313276.3316366,Low.2017.10.1103/PhysRevLett.118.010501,Low.2019.10.22331/q-2019-07-12-163} typically require ancilla registers and multiple multi-controlled gates to implement a single spin-adapted unitary.
The situation is further complicated by the fact that products of spin-adapted unitaries quickly become unwieldy.

\section{Overview of Wei--Norman decomposition}\label{sec:wn}

To address these difficulties, we introduce exact decompositions of unitaries generated by singlet spin-adapted double excitation operators into products of spin-orbital unitaries.
This is accomplished via the Lie-algebraic technique known as the Wei--Norman decomposition \cite{Wei.1963.10.1063/1.1703993,Wei.1964.10.2307/2034065}.
In brief, this approach allows us to express $\exp(\theta A)$, a rotation generated by $A = \sum_i c_i A_i$, as an ordered product $\prod_i \exp[\alpha_i(\theta) E_i]$, where $\{E_i\}$ is a fixed ordered basis for the dynamical Lie algebra $\mathfrak{g}$ obtained from the Lie closure of $\{A_i\}$, and the $\theta$-dependent coefficients $\alpha_i (\theta)$ are determined by the Wei--Norman equations.
From a bird's-eye view, the Wei--Norman decomposition algorithm proceeds as follows \cite{Altafini.2002.20270,Altafini.2002.10.1023.A:1019825109040,Tannor.2007.IntroductionQuantumMechanics,Charzynski.2013.10.1088/1751-8113/46/26/265208}:
\begin{enumerate}
	\item Select the target unitary, $U (\theta) \equiv \exp(\theta \sum_i c_i A_i)$.
	\item Select an ordered basis $\{E_i\}$ of the dynamical Lie algebra $\mathfrak{g} \equiv \Lie\left(\{A_i\}\right)$ and express the generator $A$ of the target unitary in this basis:
	\begin{equation}
		A = \sum_i d_i E_i.
	\end{equation}
	\item Use $\theta$-dependent parameters $\{\alpha_i (\theta)\}$ to postulate the factorization
	\begin{equation}
		\label{eq:target_unitary}
		e^{\theta \sum_i d_i E_i} = \prod_i e^{\alpha_i(\theta) E_i},
	\end{equation}
	with initial conditions $\alpha_i (0) = 0$.
	\item Differentiate \cref{eq:target_unitary} with respect to $\theta$ to obtain
	\begin{equation}\label{eq:diff_target_unitary}
		\sum_i d_i E_i U(\theta) = \sum_i \alpha_i^\prime (\theta) (\bar{E}_i)_{i-1\ldots1} U(\theta),
	\end{equation}
	where $\alpha_i^\prime (\theta) = \dv{\alpha_i(\theta)}{\theta}$ and, omitting the $\theta$-dependence of the $\alpha_i$ parameters for clarity,
	\begin{equation}\label{eq:wn_gen_fst}
		\begin{split}
		(\bar{E}_i)_{i-1\ldots1} &= e^{\alpha_1 E_1} \cdots e^{\alpha_{i-1}E_{i-1}}E_i e^{-\alpha_{i-1}E_{i-1}} \cdots e^{-\alpha_1 E_1}\\
		&=\sum_j M_{ji}(\alpha_1, \ldots, \alpha_{i-1}) E_j.
		\end{split}
	\end{equation}	
	\item Right-multiply \cref{eq:diff_target_unitary} by $U^{-1}(\theta)$ and equate the coefficients of the basis elements $\{E_i\}$ on both sides of the equality to obtain the Wei–Norman system of coupled ordinary differential equations (ODEs) for $\{\alpha_i (\theta)\}$, which can be expressed in matrix form as
	\begin{equation}\label{eq:wn_system}
		\begin{split}
		\mathbf{M}(\alpha_1,\alpha_2,\ldots) (\alpha_1^\prime, \alpha_2^\prime, \ldots)^\intercal = (d_1, d_2, \ldots)^\intercal \Rightarrow\\
		(\alpha_1^\prime, \alpha_2^\prime, \ldots)^\intercal = \mathbf{M}^{-1}(\alpha_1,\alpha_2,\ldots) (d_1, d_2, \ldots)^\intercal.
		\end{split}
	\end{equation}
	\item Solve \cref{eq:wn_system} in the nonsingular $\det(\mathbf{M})\neq 0$ region of interest to obtain the numerical values of the $\theta$-dependent parameters $\alpha_i (\theta)$.
\end{enumerate}
There are two degrees of freedom that affect the Wei--Norman decomposition.
The first is the choice of basis of the dynamical Lie algebra, and the second is the ordering of the exponentials in the product formula.
By appropriately choosing these degrees of freedom, one may obtain Wei--Norman systems that are nonsingular on the region of interest, easier to solve, or even admit closed-form solutions.
In our applications of the Wei--Norman decomposition to spin-adapted unitaries, we further impose the restriction that the elementary unitary factors admit compact circuit implementations.
In addition, we choose the ordering of the exponentials to maximize gate cancellations between adjacent circuit blocks.

The practical utility of the Wei--Norman decomposition for expressing symmetry-adapted unitaries faces 
three main challenges: i) the availability of analytic expressions for the unitary transformations in \cref{eq:wn_gen_fst}, ii) the existence of compact quantum circuit implementations for the elementary unitaries, and iii) the computational complexity of symbolically constructing and analytically solving the Wei--Norman system of ODEs.
We address these issues through three complementary developments.
First, we employ  exact, closed-form expressions for fermionic unitary transformations.
When both the generator and the operator to be transformed are single anti-Hermitian fermionic strings, we employ the closed-form expressions that we recently derived \cite{Evangelista.2025.10.1103/PhysRevA.111.042825}.
Inspired by the work of Jayakumar \foreign{et al.} \cite{Jayakumar.2026.10.1021.acs.jctc.5c02089}, we introduce a novel algorithm for cases involving finite linear combinations of anti-Hermitian fermionic strings.
Given the generator and operator to be transformed, the algorithm automatically generates exact, closed-form transformations with a minimal number of nested commutators, based on the matrix representation of the adjoint action on a Krylov subspace.

Second, we extend the FEB formalism \cite{Yordanov.2020.10.1103/PhysRevA.102.062612,Magoulas.2023.10.1021/acs.jctc.2c01016} to fermionic operators multiplied by linear combinations of number operators.
This allows us to generate substantially more compact circuits than with the standard FEB framework.
Third, we develop an AI-assisted discovery-and-verification framework for obtaining closed-form solutions of the Wei--Norman ODEs.
This formalism combines numerical optimization and model reduction to identify reduced parameterizations, uses symbolic inference to suggest candidate
closed-form expressions, and verifies these expressions directly in the reduced Wei--Norman system.

\section{Spin-Adapted Unitary generated by $\boldsymbol{A_{PP}^{QR}}$}\label{sec:wn_ppqr}

Here, we focus on the simplest non-trivial singlet spin-adapted fermionic unitary, namely, the one generated by $A_{PP}^{QR}$ [\cref{eq:sa_gd_ppqr}].
In the first step, we compute the Lie closure of $\{A_{\Pu\Pd}^{\Qu\Rd}, A_{\Pu\Pd}^{\Qd\Ru}\}$, and obtain that the dynamical Lie algebra $\mathfrak{g}$ is spanned by
\begin{equation}
	\label{eq:Lie_basis}
	\begin{split}
		\mathfrak{g} &= \Span
		\begin{Bmatrix}
			A_{\Pu\Pd}^{\Qu\Rd},\\
			A_{\Pu\Pd}^{\Qu\Rd} \left( h_{\Qd\Ru} + n_{\Qd\Ru} \right),\\
			A_{\Pu\Pd}^{\Qd\Ru},\\
			A_{\Pu\Pd}^{\Qd\Ru} \left( h_{\Qu\Rd} + n_{\Qu\Rd} \right),\\
			A_{\Qu\Rd}^{\Qd\Ru} \left( h_{\Pu\Pd} - n_{\Pu\Pd} \right)
		\end{Bmatrix}\\
		&\equiv
		\Span\left\{E_1,E_2,E_3,E_4,E_5\right\},
	\end{split}
\end{equation}
where, for the sake of brevity, we denote the basis elements as $\{E_i\}_{i=1}^5$.
As far as the many-body nature of the basis elements is concerned, $E_1$ and $E_3$ are the spin orbital excitation operators defining the singlet spin-adapted generator $A_{PP}^{QR}$.
The remaining elements are four-body operators, each containing a two-body anti-Hermitian excitation component multiplied by a linear combination of two particle--hole-conjugate number operator strings.
In particular, $E_2$ gives rise to a controlled version of the rotation generated by $E_1$, conditioned on the occupancy of the $\Qd\Ru$ spin-orbital pair.
Likewise, $E_4$ is the counterpart to $E_3$, conditioned on the occupancy of the $\Qu\Rd$ spin-orbital pair.
Finally, $E_5$ is a spin-flip excitation operator whose action is restricted to configurations in which spatial orbital $P$ is not singly occupied, with the sign determined by whether $P$ is empty or doubly occupied.
This basis was chosen because the corresponding unitaries can have compact quantum circuit implementations (\foreign{vide infra}).
Regarding its classification, as shown in Section SIII of the {\sm}, an alternative basis exists in which this 5-dimensional Lie algebra becomes isomorphic to $\mathbb{R}^2\oplus\mathfrak{so}(3)\cong\mathbb{R}^2\oplus\mathfrak{su}(2)$ (see also Ref.\ \cite{Jain.2026.10.1063.5.0326865}).

In the next step, we postulate the factorization of the spin-adapted unitary as
\begin{equation}
	\label{eq:wn_decomposition}
	e^{\frac{\theta}{\sqrt{2}} \left(E_1 - E_3\right)} = \prod_{i=1}^5 e^{\alpha_i E_i},
\end{equation}
where the $\alpha_i$ are $\theta$-dependent parameters.
Here, we adopt a notation in which products with ascending indices are ordered from left to right and conversely for descending indices.
This particular ordering of generators was chosen for two main reasons.
First, commuting generators, namely, $[E_1,E_2] = 0$ and $[E_3,E_4] = 0$, are placed next to one another, simplifying the derivation of the Wei--Norman system of coupled ODEs.
Second, in this order, the cancellation of gates between adjacent circuit blocks is maximized, resulting in an even more compact circuit representation of the spin-adapted unitary (\foreign{vide infra}).

To determine the values of the parameters \cite{Charzynski.2013.10.1088/1751-8113/46/26/265208}, we differentiate \cref{eq:wn_decomposition} with respect to $\theta$ and obtain:
\begin{equation}
	\label{eq:wn_derivative}
	\begin{split}
		\frac{1}{\sqrt{2}}&\left( E_1 - E_3 \right)e^{\frac{\theta}{\sqrt{2}} \left(E_1 - E_3\right)}\\ &=\sum_{i=1}^5\alpha_i^\prime \prod_{j=1}^{i-1} e^{\alpha_j E_j}E_i \prod_{k = i}^5 e^{\alpha_k E_k}\\
		&=\sum_{i=1}^5\alpha_i^\prime \prod_{j=1}^{i-1} e^{\alpha_j E_j}E_i \prod_{k = i-1}^1 e^{-\alpha_k E_k} \prod_{l = 1}^5 e^{\alpha_l E_l}\\
		&=\sum_{i=1}^{5} \alpha_i^\prime (\bar{E}_i)_{i-1\ldots1}e^{\theta \frac{1}{\sqrt{2}} \left(E_1 - E_3\right)},
	\end{split}
\end{equation}
where in the last step we used \cref{eq:wn_decomposition}, and the similarity transformations are defined as
\begin{subequations}
	\label{eq:wn_similarity}
	\begin{align}
		(\bar{E}_2)_1 &\equiv e^{\alpha_1 E_1} E_2 e^{-\alpha_1 E_1}, \\
		(\bar{E}_3)_{21} &\equiv e^{\alpha_1 E_1} e^{\alpha_2 E_2} E_3 e^{-\alpha_2 E_2} e^{-\alpha_1 E_1}, \quad \ldots.
	\end{align}
\end{subequations}
In general, unitary transformations of arbitrary fermionic operators cannot be expressed in a closed form, as they give rise to a non-terminating Hausdorff expansion.
Nevertheless, we have recently shown that when both the generator and the operator to be transformed are single (anti-)Hermitian fermionic strings the commutator expansion can be re-summed to a closed form involving only the single and doubly nested commutators multiplied by trigonometric functions \cite{Evangelista.2025.10.1103/PhysRevA.111.042825}.
However, the unitary transformations required for the Wei--Norman decomposition of $\exp(\theta A_{PP}^{QR})$ involve generators that contain an anti-Hermitian fermionic string multiplied, at most, by a linear combination of two particle--hole-conjugate number-operator strings [see \cref{eq:Lie_basis}].
Despite this complication, in Section SIV of the {\sm} we demonstrate that these types of generators satisfy $E_j^3 = - E_j$ and $E_j [E_i,E_j]E_j = 0$.
These are the necessary and sufficient conditions for the unitary transformation to be exactly expressed as \cite{Evangelista.2025.10.1103/PhysRevA.111.042825}
\begin{equation}
	\label{eq:wn_fst}
	\begin{split}
		e^{\alpha_j E_j} E_i e^{-\alpha_j E_j} ={}& E_i - \sin\left(\alpha_j\right) [E_i,E_j]\\
		&+ \left[1 - \cos\left(\alpha_j\right)\right] [[E_i, E_j],E_j].
	\end{split}
\end{equation}

Using \cref{eq:wn_fst}, we compute all pertinent similarity transformations appearing in \cref{eq:wn_derivative}.
By equating the coefficients of the generators $E_i$ appearing on both sides of \cref{eq:wn_derivative}, we arrive at a system of coupled ODEs for the $\theta$-dependent parameters $\alpha_i$.
Due to the structure of the Lie algebra (see Section SIII in the {\sm}), two equations immediately decouple, yielding
\begin{equation}
	\alpha_1^\prime = \frac{1}{\sqrt{2}}
\end{equation}
and
\begin{equation}
	\alpha_3^\prime = -\frac{1}{\sqrt{2}}.
\end{equation}
Using the initial conditions $\alpha_i(0) = 0$, they give
\begin{equation}
	\label{eq:ppqr_1}
	\alpha_1(\theta) = \frac{\theta}{\sqrt{2}}
\end{equation}
and
\begin{equation}
	\label{eq:ppqr_3}
	\alpha_3(\theta) = -\frac{\theta}{\sqrt{2}}.
\end{equation}
Using these relations, the remaining differential equations, expressed in matrix form, read
\begin{equation}
	\mathbf{M}
	\begin{pmatrix}
		\alpha_2^\prime \\
		\alpha_4^\prime \\
		\alpha_5^\prime
	\end{pmatrix}
	=
	\begin{pmatrix}
		0\\
		\frac{\cos\left(\alpha_2+\frac{\theta}{\sqrt{2}}\right) - 1}{\sqrt{2}}\\
		\frac{\sin\left(\alpha_2+\frac{\theta}{\sqrt{2}}\right)}{\sqrt{2}}
	\end{pmatrix}
\end{equation}
with the Wei--Norman coefficient matrix $\mathbf{M}$ given by
\begin{equation}
	\resizebox{\linewidth}{!}{$
	\mathbf{M} =
	\begin{pmatrix}
		1 & 0 & \sin\left(\alpha_4 - \frac{\theta}{\sqrt{2}}\right)\\
		0 & \cos\left(\alpha_2 + \frac{\theta}{\sqrt{2}}\right) & -\sin\left(\alpha_2 + \frac{\theta}{\sqrt{2}}\right) \cos\left(\alpha_4 - \frac{\theta}{\sqrt{2}}\right) \\
		0 & \sin\left(\alpha_2 + \frac{\theta}{\sqrt{2}}\right) & \cos\left(\alpha_2 + \frac{\theta}{\sqrt{2}}\right) \cos\left(\alpha_4 - \frac{\theta}{\sqrt{2}}\right)
	\end{pmatrix}.
	$}
\end{equation}
The above system of ODEs is sufficiently small and simple that it can be solved analytically, yielding
\begin{equation}
	\label{eq:ppqr_2}
	\alpha_2(\theta) = \frac{\sqrt{2} - 1}{\sqrt{2}}\theta
	-\arctan\left[\frac{(3-2\sqrt{2})\sin(2\theta)}{1+(3-2\sqrt{2})\cos(2\theta)}\right],
\end{equation}
\begin{equation}
	\label{eq:ppqr_4}
	\alpha_4(\theta) = \frac{\theta}{\sqrt{2}}
	-\arcsin\left[\frac{\sin(\theta)}{\sqrt{2}}\right],
\end{equation}
and
\begin{equation}
	\label{eq:ppqr_5}
	\alpha_5(\theta) = \frac{\pi}{4}
	-\arctan\left[\cos(\theta)\right].
\end{equation}
Furthermore, considering that
\begin{equation}
	\begin{split}
	\det(\mathbf{M}) &= \cos(\alpha_4 - \frac{\theta}{\sqrt{2}})\\
	&=\cos\left(\arcsin\left[\frac{\sin(\theta)}{\sqrt{2}}\right]\right)\\
	&= \sqrt{1 - \frac{\sin^2(\theta)}{2}}\\
	&>0,
	\end{split}
\end{equation}
the Wei--Norman decomposition is globally valid.

\begin{figure}[h!]
	\centering
	\includegraphics[width=8.5cm]{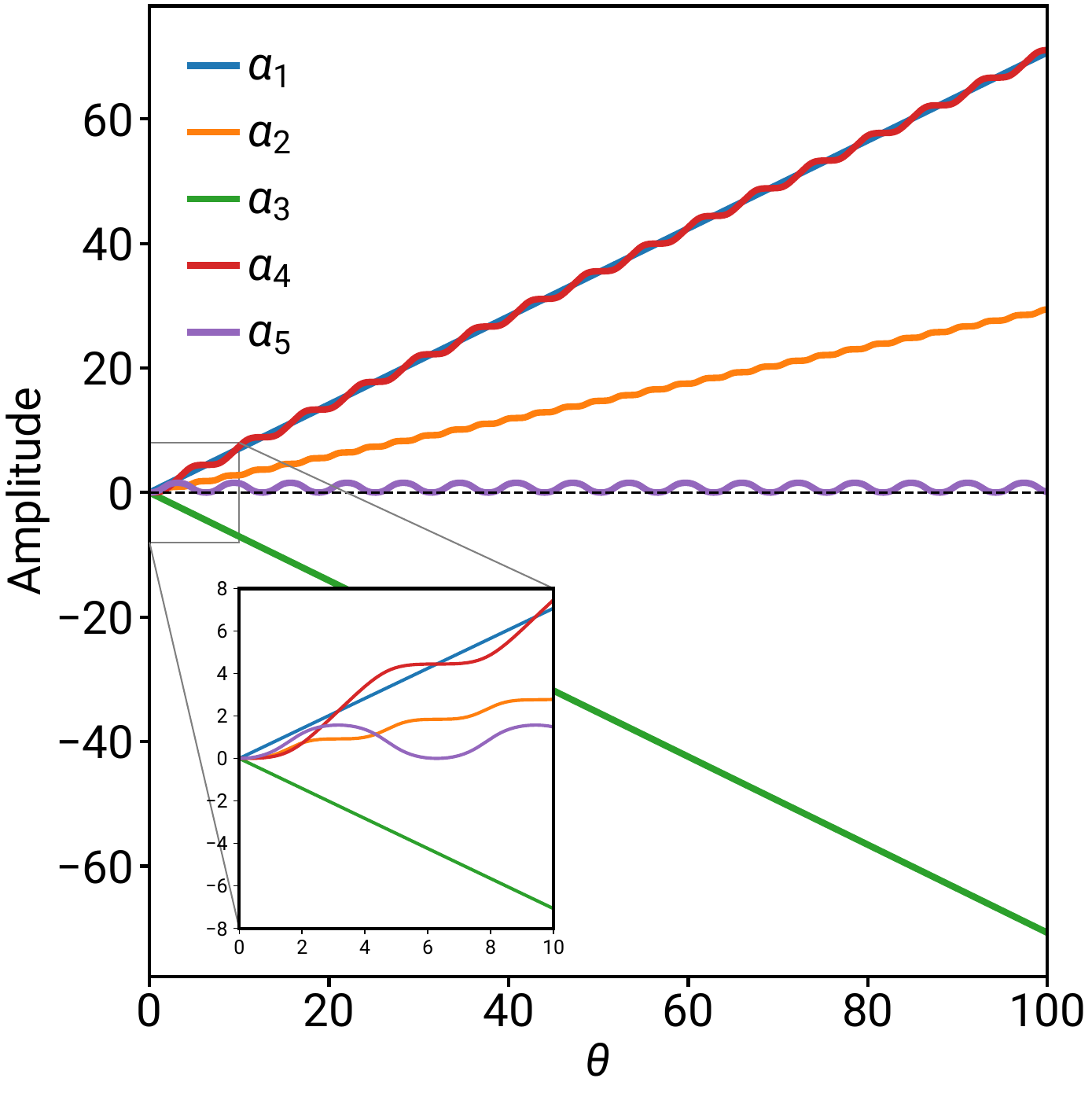}
	\caption{
		Plot of the $\theta$-dependent parameters, \cref{eq:ppqr_1,eq:ppqr_3,eq:ppqr_2,eq:ppqr_4,eq:ppqr_5}, defining the Wei--Norman decomposition of $\exp\left( \theta A_{PP}^{QR}\right)$ [\cref{eq:wn_decomposition}].
	}
	\label{fig:wn_ppqr}
\end{figure}

The closed-form solutions reveal several useful structural features.
The parameters multiplying $E_1$ and $E_3$ coincide with the corresponding coefficients in the definition of $A_{PP}^{QR}$.
The parameters multiplying the excitation-type generators $E_1$ through $E_4$ are odd functions of $\theta$, while the parameter multiplying the spin-flip generator $E_5$ is even.
Furthermore, the parameters associated with products of excitation operators with linear combinations of number-operator strings contain an oscillatory contribution superimposed on a linear term, whereas the spin-flip parameter is purely periodic.
These observations will prove crucial for uncovering the closed-form expressions for the parameters appearing in the Wei--Norman decompositions of the more complicated singlet spin-adapted unitaries discussed in the next section.
In \cref{fig:wn_ppqr}, we plot \cref{eq:ppqr_1,eq:ppqr_2,eq:ppqr_3,eq:ppqr_4,eq:ppqr_5} for $\theta$ values ranging from 0 to 100.

\begin{figure}[h!]
	\centering
	
	\subfloat[]{
		\includegraphics[width=8.5cm]{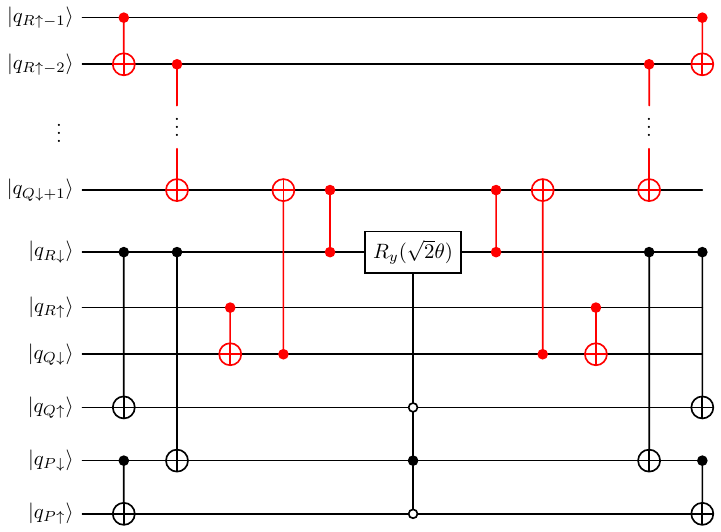}
		\label{fig:feb_ppqr}
	}\\
	\subfloat[]{
		\includegraphics[width=8.5cm]{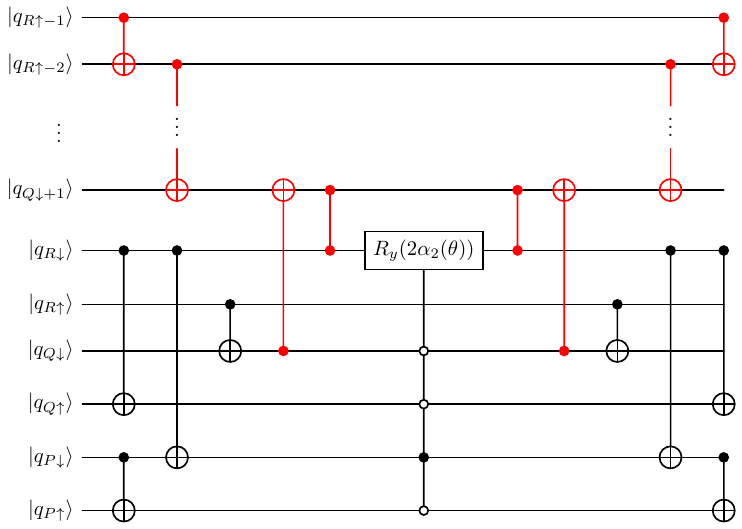}
		\label{fig:feb_ppqr_nn}
	}\\
	\subfloat[]{
		\includegraphics[width=8.5cm]{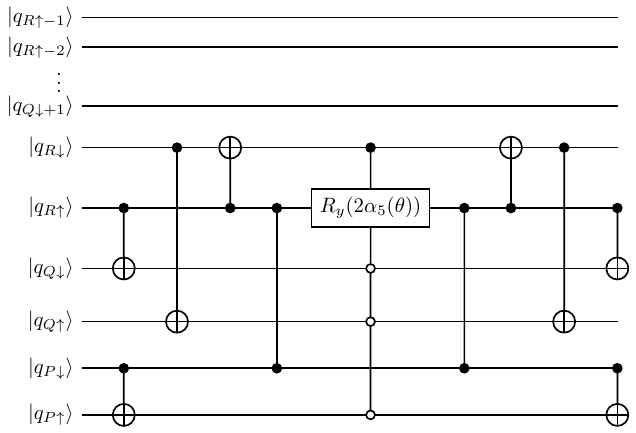}
		\label{fig:qeb_qrqr}
	}
	\caption{
		Quantum circuits implementing the (a) $\exp(\alpha_1 A_{P\uparrow P\downarrow}^{Q\uparrow R\downarrow})$, (b) $\exp[\alpha_2 A_{P\uparrow P\downarrow}^{Q\uparrow R\downarrow} (h_{Q\downarrow R\uparrow} + n_{Q\downarrow R\uparrow})]$, and (c) $\exp[\alpha_5 A_{Q\uparrow R\downarrow}^{Q\downarrow R\uparrow} (h_{P\uparrow P\downarrow} - n_{P\uparrow P\downarrow})]$ unitaries.
	}
	\label{fig:ppqr_comp_circs}
\end{figure}

\begin{figure*}[h!]
	\centering
	\includegraphics[width=17cm]{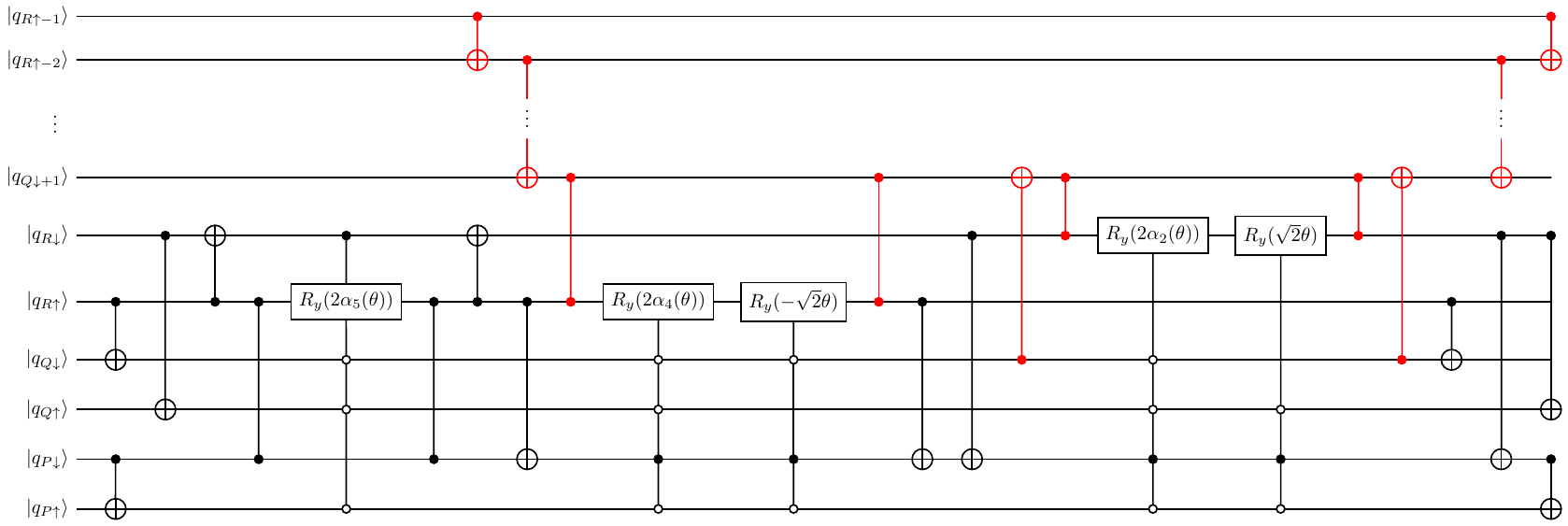}
	\caption{
		Quantum circuit implementing the spin-adapted unitary $\exp(\theta A_{PP}^{QR})$.
	}
	\label{fig:ppqr_circ}
\end{figure*}

Having obtained a globally valid Wei–Norman parameterization, we now turn to compact quantum circuit implementations of the individual exponential factors.
For the unitaries $\exp(\alpha_1 E_1)$ and $\exp(\alpha_3 E_3)$, the circuits are obtained through a straightforward application of the FEB formalism, with the circuit of the former shown in \cref{fig:ppqr_comp_circs}(a) as an example.
The situation with the remaining unitaries is more complicated.
The $\exp(\alpha_2 E_2)$ and $\exp(\alpha_4 E_4)$ unitaries can be viewed as controlled versions of $\exp(\alpha_1 E_1)$ and $\exp(\alpha_3 E_3)$, respectively.
For example, $E_2 \equiv E_1 (h_{Q\downarrow R\uparrow} + n_{Q\downarrow R\uparrow})$, so that $\exp(\alpha_2 E_2) \equiv \exp(\alpha_2 E_1)$ if the spin orbitals $\Qd$ and $\Ru$ have the same occupancy, acting as the identity otherwise.
The quantum circuit implementing $\exp(\alpha_2 E_2)$ is given in \cref{fig:ppqr_comp_circs}(b), and compared with \cref{fig:ppqr_comp_circs}(a), it contains 6 additional CNOT gates and one extra control in the multi-controlled $R_y$ gate.
The generator $E_5$ has two factors, namely, a spin-flip excitation operator ($A_{Q\uparrow R\downarrow}^{Q\downarrow R\uparrow}$) and the difference between two particle--hole-conjugate number operator strings ($h_{P\uparrow P\downarrow} - n_{P\uparrow P\downarrow}$).
The bare spin-flip operator can be implemented via the standard FEB formalism.
The number-operator component ensures that i) the unitary is applied only when spatial orbital $P$ is not singly occupied, and ii) the sign of the rotation is flipped when spatial orbital $P$ is doubly occupied.
The corresponding quantum circuit is shown in \cref{fig:ppqr_comp_circs}(c).
Compared to the FEB circuit of the bare spin-flip operator, it contains two additional CNOTs, two extra controlled-$Z$ (C$Z$) gates, and one additional control in the multi-controlled $R_y$ gate.
The complete circuit is obtained by concatenating the circuits of the five unitaries, in reverse order, and canceling gates between block pairs.
The final optimized circuit of the spin-adapted unitary $\exp(\theta A_{PP}^{QR})$ is shown in \cref{fig:ppqr_circ}.

At this point, it is worth commenting on the aforementioned two degrees of freedom affecting the Wei--Norman decomposition.
Since the dynamical Lie algebras characterizing the examined spin-adapted unitaries are not abelian, different orderings of exponentials in the product formula will lead to different systems of coupled differential equations.
To that end, we constructed and numerically integrated using SciPy \cite{Virtanen.2020.10.1038/s41592-019-0686-2} the Wei--Norman equations arising from all 120 permutations of the five basis elements of the dynamical Lie algebra [\cref{eq:Lie_basis}].
In Figs.\ S3--S7 in the {\sm}, we show the results of the numerical integration in the extended region $\theta \in [-100,100]$.
These figures reveal that $\alpha_1$--$\alpha_4$ are always odd functions of $\theta$, while $\alpha_5$ is even.
Furthermore, for more than one third of the permutations we encountered singularities, limiting the numerical integration to a small region.
As anticipated from the structure of the underlying dynamical Lie algebra, the angles of the unitaries generated by $A_{\Pu\Pd}^{\Qu\Rd}$ and $A_{\Pu\Pd}^{\Qd\Ru}$ matched their coefficients from the definition of $A_{PP}^{QR}$.
However, the remaining three parameters were quite sensitive to the ordering of the exponentials.
Nevertheless, in the nonsingular cases, the parameters varied smoothly with $\theta$, suggesting that small errors in $\theta$ do not induce large changes in the final rotation angles.

	\section{Spin-Adapted Unitaries generated by $\boldsymbol{\intS}$ and $\boldsymbol{\intT}$}\label{sec:wn_intT_intT}

\subsection{Challenges of Direct Wei--Norman Construction}

In this section, we discuss the Wei--Norman decompositions of the more challenging singlet spin-adapted unitaries, namely those generated by the $\intS$ and $\intT$ double excitations.
To facilitate the rapid derivation and verification of analytic expressions, we developed a prototype code, built on SymPy \cite{Meurer.2017.10.7717/peerj-cs.103}, for the symbolic manipulation of fermionic operators.
We found that the Lie algebras associated with $\intS$ and $\intT$ have dimensions $\dim (\mathfrak{g}_0) = 28$ and $\dim (\mathfrak{g}_1) = 84$, respectively.
The symbolic construction and analytic solution of the pertinent Wei--Norman systems is prohibitively expensive, especially in the case of $\intT$.
For example, one would need to symbolically construct an $84 \times 84$ highly non-linear coefficient matrix.
To na\"{\i}vely estimate the complexity of the problem, recall that each unitary transformation [\cref{eq:wn_similarity}] gives rise to at least three terms.
Therefore, the transformation of the 84th generator with the previous 83 would result in a number of terms on the order of $10^{39}$, before collecting like terms.

It is worth mentioning that Jain \foreign{et al.} have shown that $\mathfrak{g}_1$ is reductive \cite{Jain.2026.10.1063.5.0326865}, and when expressed in an ideal-adapted basis it is isomorphic to $\mathbb{R}^6 \oplus \mathfrak{so}(3)^{\oplus 26}$.
Although in this basis the Wei--Norman problem is dramatically simplified, the circuit implementation is highly complex as the basis elements become linear combinations of fermionic operators.
We thus opted to work with the standard bases of the $\mathfrak{g}_0$ and $\mathfrak{g}_1$ Lie algebras, where each basis element is a single spin orbital anti-Hermitian excitation operator, at most multiplied by a linear combination of two particle--hole-conjugate number-operator strings.
Subsequently, the basis elements were grouped into commuting sets and ordered similarly to those in the $A_{PP}^{QR}$ case.

To arrive at closed-form expressions for the parameters, we designed the workflow summarized in \cref{fig:wn_scheme}. 
The procedure consists of four main stages, namely,  numerical discovery of the parameter structure, reduction to a small set of independent parameter families, inference of closed-form expressions, and symbolic verification through the corresponding Wei--Norman equations.
Finally, the quantum circuits were constructed and optimized.
\begin{figure}[h!]
	\vspace{-12pt}
		\begin{tikzpicture}[every text node part/.style={align=center}]
			\tikzstyle{block} = [rectangle, draw, top color=white, bottom color=blue!20, 
			text width=2.25in, text centered, rounded corners, inner sep=1ex]
			\tikzstyle{line} = [draw, thick, -latex']
			\matrix (matrix) [matrix of nodes,row sep=18pt,column sep=0.5cm]
			{
				\node [scale=0.8, block] (lie) {\textbf{Lie algebra construction}}; \\
				\node [scale=0.8, block] (numerical) {\textbf{Numerical optimization of} \\ \textbf{Wei--Norman Parameters}}; \\
				\node [scale=0.8, block] (structure) {\textbf{Identification of}\\\textbf{parameter structure}}; \\
				\node [scale=0.8, block] (reduction) {\textbf{Reduction to families of}\\ \textbf{independent parameters}}; \\
				\node [scale=0.8, block] (fourier) {\textbf{Fourier analysis /}\\ \textbf{closed-form inference}}; 
				\\
				\node [scale=0.8, block] (symbolic) {\textbf{Symbolic verification of}\\ \textbf{Wei--Norman ODEs}}; 
				\\
				\node [scale=0.8, block] (circuit) {\textbf{Circuit construction}\\ \textbf{and optimization}}; \\
			};
			\path [line] (lie) -- (numerical);
			\path [line] (numerical) -- (structure);
			\path [line] (structure) -- (reduction);
			\path [line] (reduction) -- (fourier);
			\path [line] (fourier) -- (symbolic);
			\path [line] (symbolic) -- (circuit);
		\end{tikzpicture}
	\caption{\label{fig:wn_scheme}
		Flowchart outlining the main steps for the Wei--Norman decompositions of $\exp(\theta\intS)$ and $\exp(\theta\intT)$.}
\end{figure}

\subsection{Numerical Discovery of Parameter Structure}

In the first step, we constructed the matrix representations of the target unitaries and their factorized forms in the full Fock space generated by the eight active spin orbitals.
The optimized values of the parameters were obtained by minimizing the square of the Frobenius norm of the difference between the target unitary and its Wei--Norman decomposition for $\theta$ values ranging from 0 to 10 with increments of 0.01.
The minimization was performed using the trust-region reflective least-squares optimizer as implemented in SciPy.
We employed three-point numerical derivatives while the tolerances for the cost function, optimized parameters, and gradient norm were set to $10^{-12}$, and the maximum number of cost-function evaluations was set to $10^5$.
The solution at each $\theta$ was used as the initial guess for the next value of $\theta$.

\begin{figure*}[h!]
	\centering
	
	\subfloat[]{
		\includegraphics[width=7.8cm]{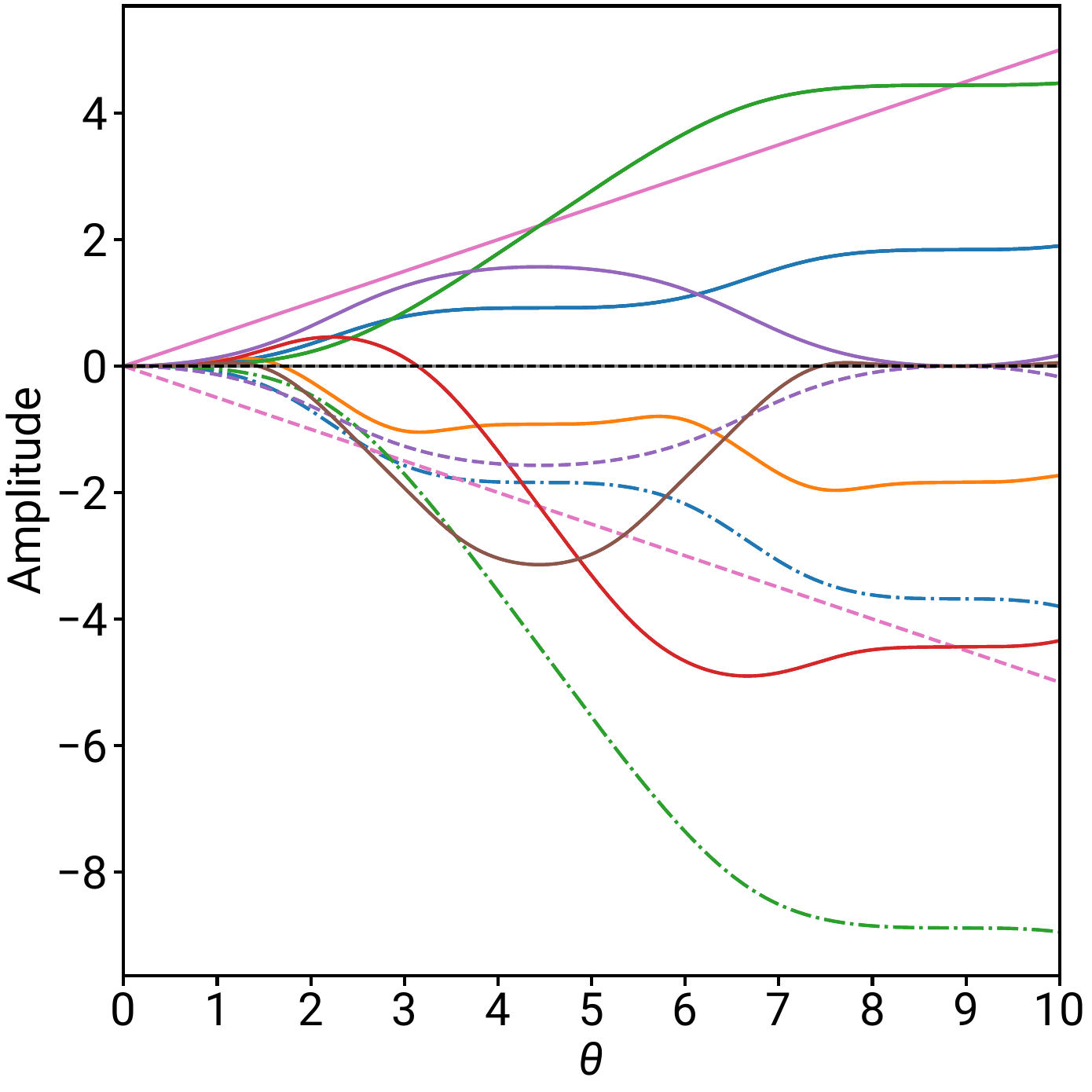}
		\label{fig:int0_params-a}
	}
	\hfill
	\subfloat[]{
		\includegraphics[width=7.8cm]{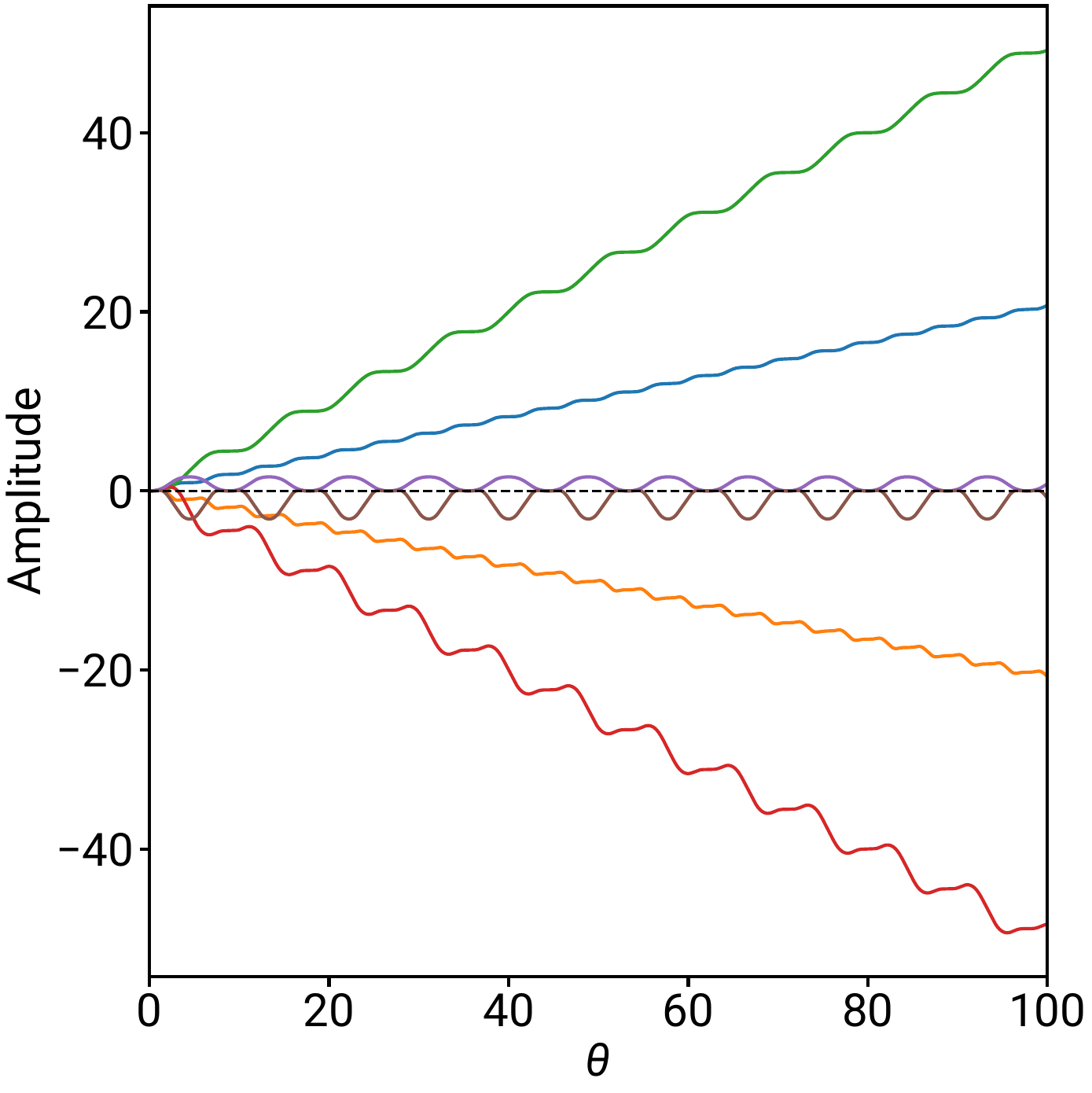}
		\label{fig:int0_params-b}
	}\\
	
	\vspace{0.3cm}
	
	\subfloat[]{
		\includegraphics[width=7.8cm]{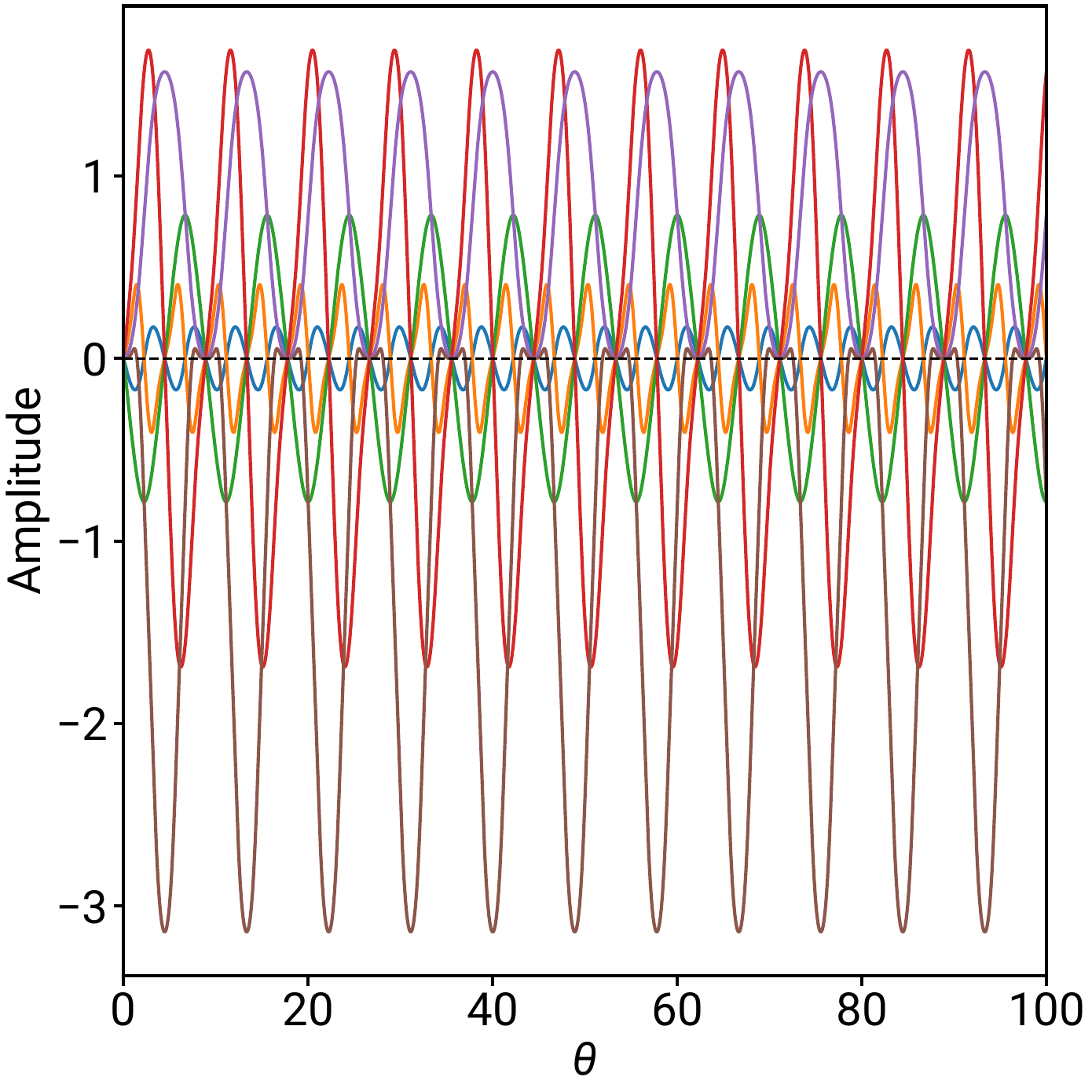}
		\label{fig:int0_params-c}
	}
	\hfill
	\subfloat[]{
		\includegraphics[width=7.8cm]{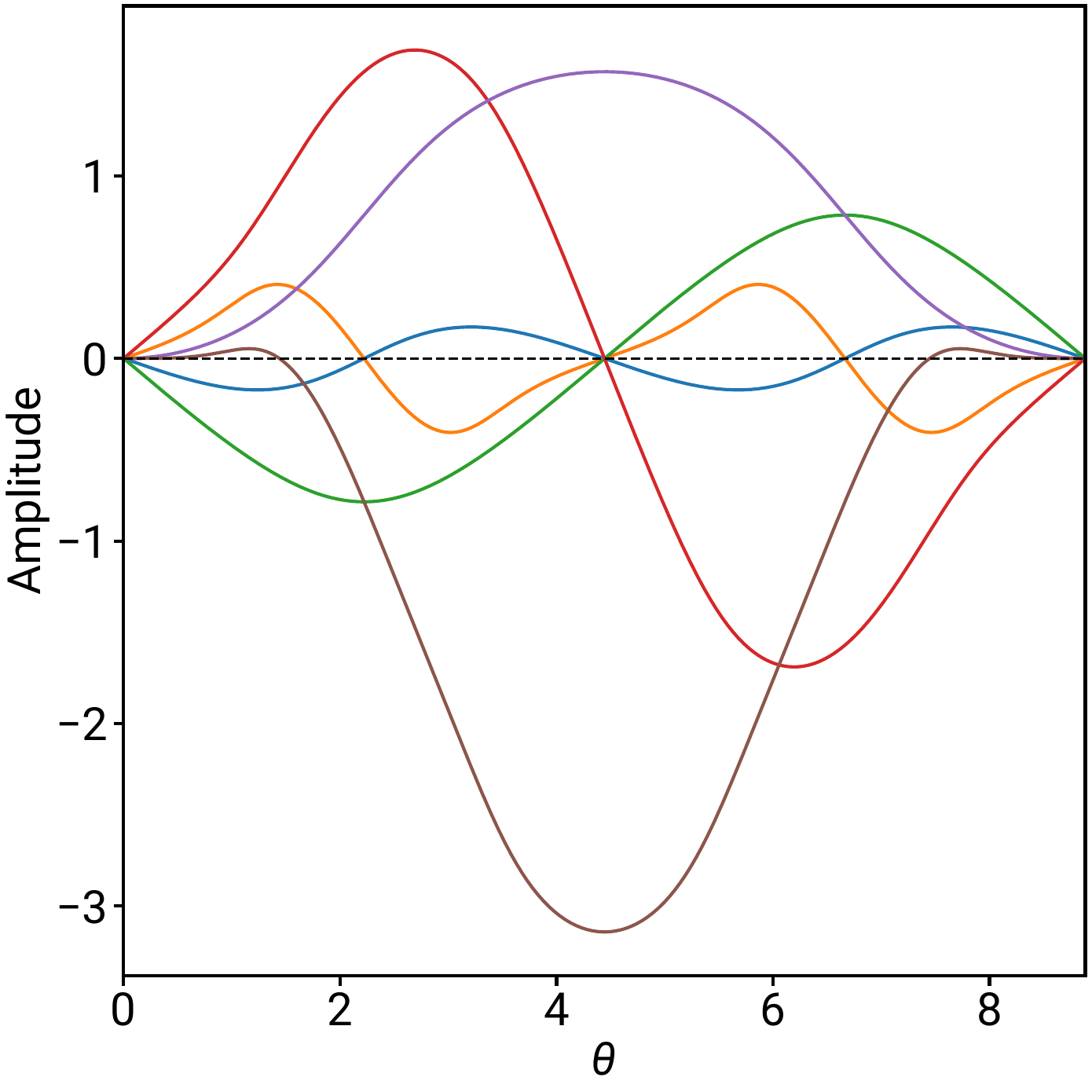}
		\label{fig:int0_params-d}
	}
	
	\caption{Numerical optimization results for the $\theta$-dependent parameters defining the Wei--Norman decomposition of $\exp(\theta\intS)$. (a) 28 parameters in the $\theta \in [0,10]$ region. Dashed/dashdotted lines denote parameters that are $\times -1$/$-2$ the corresponding solid line with the same color. (b) 6 independent parameters in the $\theta \in [0,100]$ region. (c) Shifted parameters (linear contributions eliminated) in the $\theta \in [0,100]$ region. (d) Shifted parameters in the $\theta \in [0, 2\sqrt{2}\pi]$ region.}
	\label{fig:int0_params}
\end{figure*}

\begin{figure*}[h!]
	\centering
	
	\subfloat[]{
		\includegraphics[width=7.8cm]{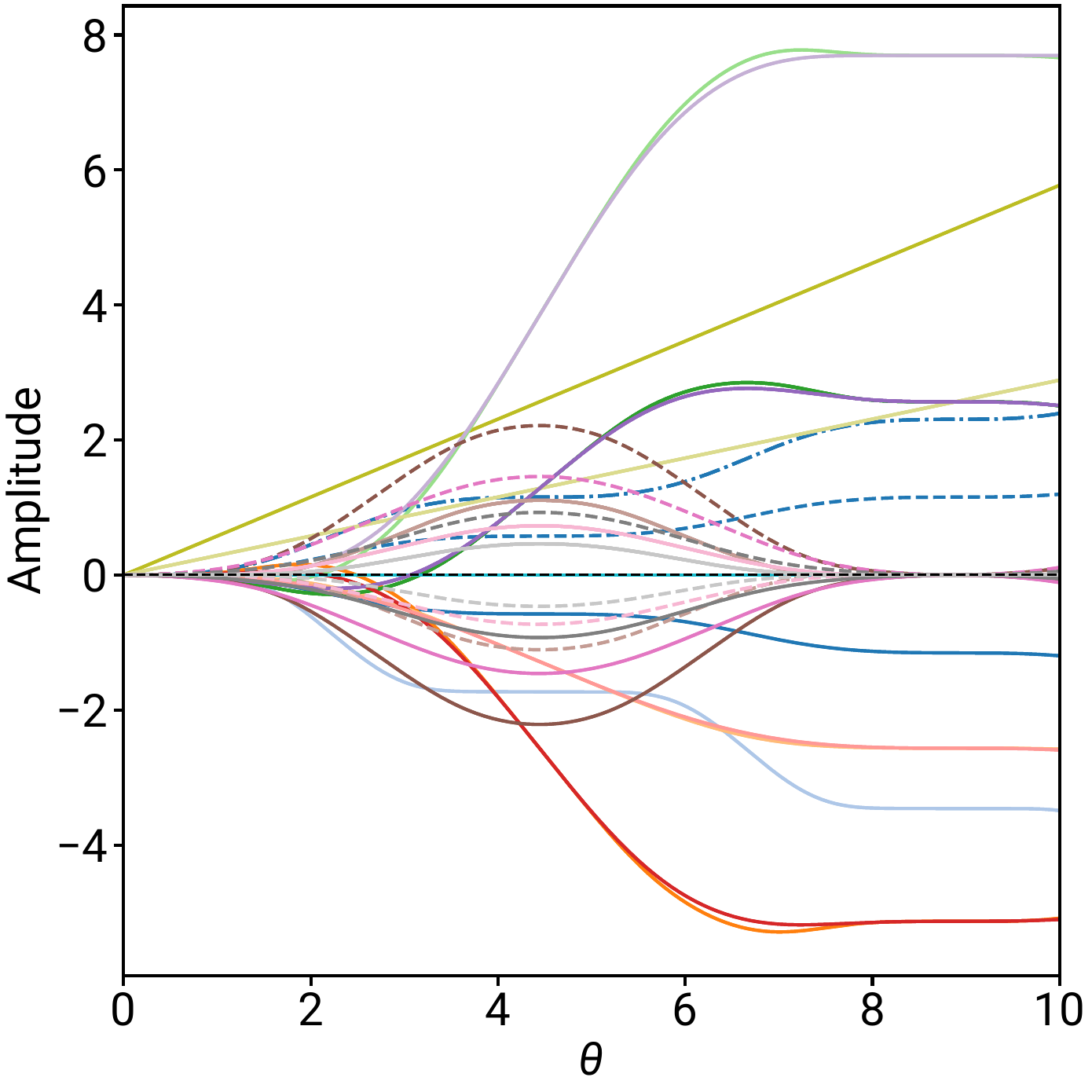}
		\label{fig:int1_params-a}
	}
	\hfill
	\subfloat[]{
		\includegraphics[width=7.8cm]{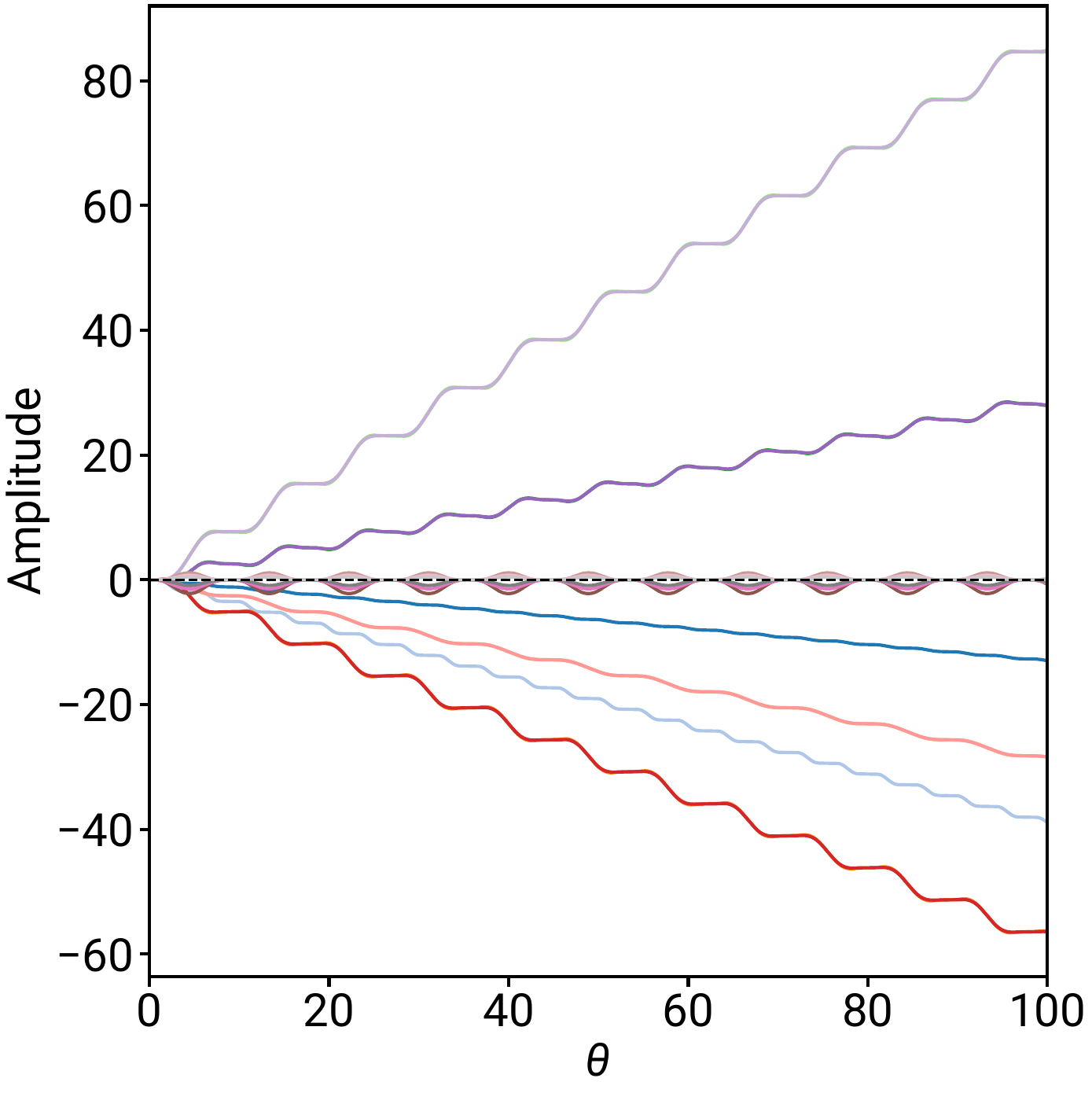}
		\label{fig:int1_params-b}
	}
	
	\vspace{0.3cm}
	
	\subfloat[]{
		\includegraphics[width=7.8cm]{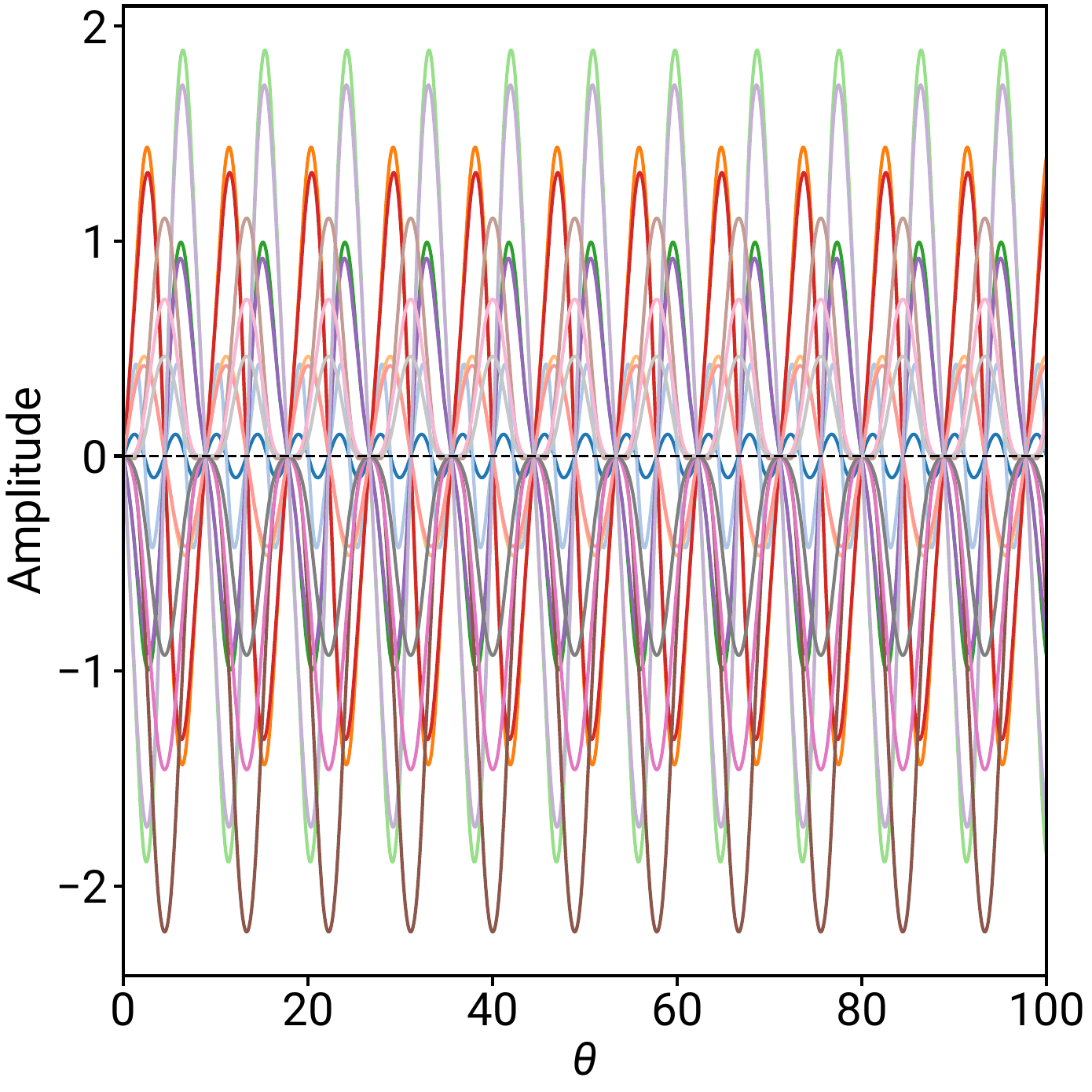}
		\label{fig:int1_params-c}
	}
	\hfill
	\subfloat[]{
		\includegraphics[width=7.8cm]{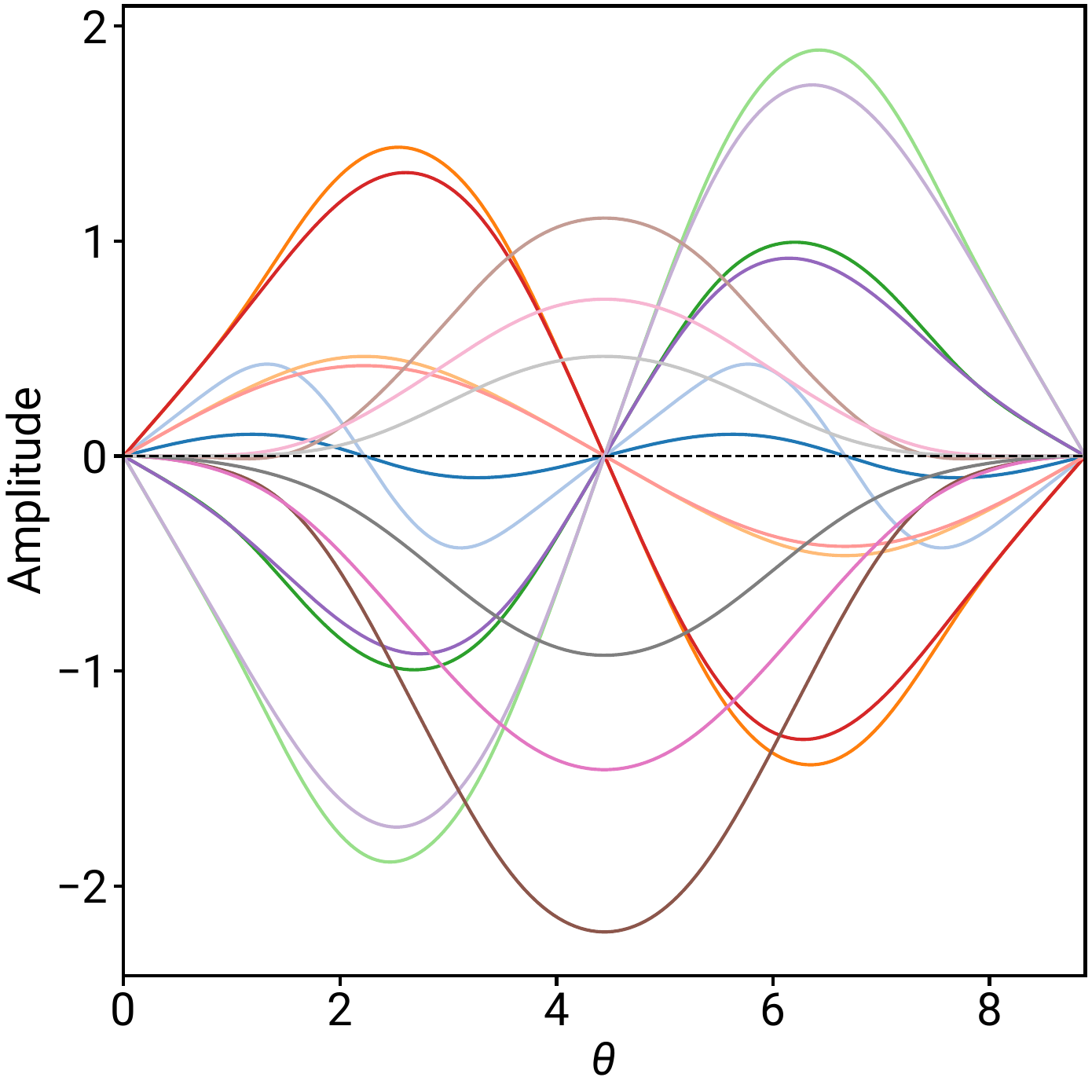}
		\label{fig:int1_params-d}
	}
	
	\caption{Numerical optimization results for the $\theta$-dependent parameters defining the Wei--Norman decomposition of $\exp(\theta\intT)$. (a) 84 parameters in the $\theta \in [0,10]$ region. Dashed/dashdotted lines denote parameters that are $\times -1$/$-2$ the corresponding solid line with the same color. (b) 16 independent parameters in the $\theta \in [0,100]$ region. (c) Shifted parameters (linear contributions eliminated) in the $\theta \in [0,100]$ region. (d) Shifted parameters in the $\theta \in [0, 2\sqrt{2}\pi]$ region.}
	\label{fig:int1_params}
\end{figure*}

After inspecting the numerical solutions shown in \cref{fig:int0_params-a,fig:int1_params-a}, we made the following observations.
Out of the 28 parameters appearing in the factorization of $\exp(\theta \intS)$, two were numerically zero ($<1.5 \times 10^{-13}$).
Similarly, 12 of the 84 parameters were numerical zeros for $\exp(\theta \intT)$ ($<3\times 10^{-14}$).
Furthermore, as was the case with $A_{PP}^{QR}$, the parameters multiplying the generators defining $\intS$ and $\intT$ had the same numerical values as in these definitions (maximum deviations on the order of $10^{-15}$).
Finally, by examining the ratios between the parameter vectors, we found that many of them were scalar multiples ($\pm1$, $\pm2$, $\pm\frac{1}{2}$) of others (maximum deviations on the order of $10^{-13}$).
As a result of this numerical endeavor, we were able to reduce the numbers of independent parameters from 28 and 84 to 6 and 16, respectively.

Having dramatically reduced the numbers of independent parameters, we were able to optimize the parameters in the extended region $\theta \in [0,100]$, using the same optimization protocol as before.
In doing so, we found that the parameters for the Wei--Norman decompositions of the unitaries $\exp(\theta \intS)$ and $\exp(\theta \intT)$ share many features with their $A_{PP}^{QR}$ counterpart.
The most important observation was that parameters multiplying spin-flip operators are even periodic functions, while those multiplying conditional excitations are odd functions containing an oscillatory component superimposed on a linear contribution (see \cref{fig:int0_params-b,fig:int1_params-b}).
To isolate the oscillations, we extracted the linear components via linear regression and subtracted them from the corresponding parameters.
As shown in \cref{fig:int0_params-c,fig:int0_params-d,fig:int1_params-c,fig:int1_params-d}, in doing so, all parameters became periodic with periods either $T=\sqrt{2}\pi$ or $2\sqrt{2}\pi$.

\subsection{Closed-Form Inference and Verification}

In the next step, we performed additional optimizations in the regions $\theta \in [0, \frac{T}{2}]$, using 1,025 uniformly spaced samples and the same optimization protocol as before.
Subsequently, we employed the trapezoid rule as implemented in NumPy \cite{Harris.2020.10.1038.s41586-020-2649-2} to determine the Fourier series representing the periodic components of the parameters.
Owing to the symmetry of the periodic functions, the integration was performed over one half-period and the resulting coefficients were multiplied by two.
Finally, candidate analytic forms for the Fourier-series expressions were explored with the assistance of artificial-intelligence (AI) tools from OpenAI and Anthropic.
In some challenging cases, none of the employed AI models were able to suggest closed-form expressions when provided only with the Fourier series.
After we identified symmetries and structures in the Fourier expansions and supplied this information to the models, we were able to guide them toward plausible closed-form expressions. 
As a first consistency check, we confirmed that the closed-form expressions reproduced the raw numerical data within the precision of the minimization procedure ($\sim 10^{-12}$).
At this point, it is worth mentioning that the Fourier series revealed additional linear dependencies among the $\exp(\theta\intT)$ parameters, namely, parameters that could be expressed as a linear combination of two others (deviations on the order of $10^{-11}$).
Although they were not supplied to the models, these linear dependencies were also satisfied by the candidate closed-form expressions.
The final numbers of linearly independent parameters for the Wei--Norman decompositions of $\exp(\theta\intS)$ and $\exp(\theta\intT)$ were 6 and 12, respectively.

Having inferred closed-form expressions for the parameters from the numerical data, we next verify that they satisfy the corresponding Wei--Norman systems of ODEs.
Although the straightforward symbolic construction of the Wei--Norman equations is computationally prohibitive, we can take advantage of the various symmetries between the parameters to reduce the complexity.
Indeed, we have numerically identified that many generators in the same commuting set share the same parameter.
Since the elementary exponentials in the Wei--Norman decompositions are grouped into commuting sets, products within each set can be combined into exponentials of aggregated generators.
For example, if $F_n$ and $G_n$ denote the generators of the Lie algebras associated with $\intS$ and $\intT$, respectively, the exponentials from the first commuting set in each Wei--Norman decomposition become
\begin{equation}
	\resizebox{\linewidth}{!}{$
	\begin{split}
e^{\frac{\theta}{2}F_1} e^{b_1 F_2} e^{b_1 F_3} e^{- 2b_1 F_4} e^{b_2 F_5} e^{\frac{\theta}{2}F_6} e^{b_1 F_7} e^{b_1 F_8} e^{- 2b_1 F_9} e^{b_2 F_{10}} \cdots =\\
e^{\frac{\theta}{2} (F_1 + F_6)} e^{b_1 (F_2 + F_3 - 2F_4 + F_7 + F_8 - 2F_9)} e^{b_2 (F_5 + F_{10})} \cdots =\\
e^{\frac{\theta}{2} \tilde{F}_1} e^{b_1 \tilde{F}_2} e^{b_2 \tilde{F}_3} \cdots
	\end{split}
	$}
\end{equation}
and
\begin{equation}
	\begin{aligned}
		& e^{\frac{\theta}{\sqrt{3}}G_1}
		e^{c_1G_2}
		e^{-c_1G_3}
		e^{-2c_1G_6}
		e^{-2c_1G_7}
		e^{c_2G_8}
		\\[-2pt]
		&\quad {}\times
		e^{\frac{\theta}{\sqrt{3}}G_9}
		e^{c_1G_{10}}
		e^{-c_1G_{11}}
		e^{-2c_1G_{14}}
		e^{-2c_1G_{15}}
		e^{c_2G_{16}}
		\cdots
		\\[3pt]
		&=
		e^{\frac{\theta}{\sqrt{3}}(G_1+G_9)}
		\\[-2pt]
		&\quad {}\times
		e^{c_1(
			G_2-G_3-2G_6-2G_7
			+G_{10}-G_{11}-2G_{14}-2G_{15}
			)}
		\\[-2pt]
		&\quad {}\times
		e^{c_2(G_8+G_{16})}
		\cdots
		\\[3pt]
		&=
		e^{\frac{\theta}{\sqrt{3}}\widetilde{G}_1}
		e^{c_1\widetilde{G}_2}
		e^{c_2\widetilde{G}_3}
		\cdots .
	\end{aligned}
\end{equation}
(see Tables SI and SII for the definitions of the generators).
This allows us to reduce the number of factors from 28 to 8 in the case of the $\intS$ generator,
\begin{equation}
	e^{\frac{\theta}{2} \left( \tilde{F}_1 - \tilde{F}_4 \right)} = e^{\frac{\theta}{2}\tilde{F}_{1}} e^{b_{1}\tilde{F}_{2}} e^{b_{2}\tilde{F}_{3}} e^{-\frac{\theta}{2}\tilde{F}_{4}} e^{b_{3}\tilde{F}_{5}} e^{b_{4}\tilde{F}_{6}} e^{b_{5}\tilde{F}_{7}} e^{b_{6}\tilde{F}_{8}},
\end{equation}
and from 84 to 19 for the $\intT$ generator,
\begin{equation}
	\begin{aligned}
		e^{\frac{\theta}{\sqrt{3}}
			\left[
			\widetilde{G}_1
			+\frac{1}{2}\left(\widetilde{G}_4+\widetilde{G}_9\right)
			\right]}
		&=
		e^{\frac{\theta}{\sqrt{3}}\widetilde{G}_1}
		e^{c_1\widetilde{G}_2}
		e^{c_2\widetilde{G}_3}
		\\[-2pt]
		&\quad {}\times
		e^{\frac{\theta}{2\sqrt{3}}\widetilde{G}_4}
		e^{c_3\widetilde{G}_5}
		e^{c_4\widetilde{G}_6}
		e^{c_5\widetilde{G}_7}
		\\[-2pt]
		&\quad {}\times
		e^{c_6\widetilde{G}_8}
		e^{\frac{\theta}{2\sqrt{3}}\widetilde{G}_9}
		e^{c_7\widetilde{G}_{10}}
		e^{c_8\widetilde{G}_{11}}
		\\[-2pt]
		&\quad {}\times
		e^{c_9\widetilde{G}_{12}}
		e^{c_{10}\widetilde{G}_{13}}
		e^{c_{11}\widetilde{G}_{14}}
		e^{c_{12}\widetilde{G}_{15}}
		\\[-2pt]
		&\quad {}\times
		e^{c_{13}\widetilde{G}_{16}}
		e^{c_{14}\widetilde{G}_{17}}
		e^{c_{15}\widetilde{G}_{18}}
		e^{c_{16}\widetilde{G}_{19}} .
	\end{aligned}
\end{equation}
At this point, it is worth mentioning that the spans of $\{\tilde{F}_i\}_{i=1}^8$ and $\{\tilde{G}_i\}_{i=1}^{19}$ are not Lie algebras, as they are not closed under the commutator.
That is to say that despite the reduction in the number of exponentials, the final Wei--Norman equations must still be expressed in terms of the original basis.

To be able to derive as compact symbolic expressions as possible, we need to have access to exact closed-form expressions for similarity transformations involving $\tilde{F}_i$ and $\tilde{G}_i$.
Inspired by the work of Jayakumar \foreign{et al.} \cite{Jayakumar.2026.10.1021.acs.jctc.5c02089}, we developed a novel approach that, given two finite linear combinations of fermionic strings $A$ and $B$, returns the closed-form expression for their similarity transformation, $\exp(\alpha A) B \exp(-\alpha A)$, with the minimum number of nested commutators.
This was accomplished by first finding the closure of the commutator,
\begin{equation}
	\ad_A^n(B) = -\sum_{i=0}^{n-1} x_i \ad_A^i(B),
\end{equation}
where $\ad_A^k(B)$ denotes the $k$-fold nested commutator of $A$ and $B$,
\begin{equation}
	\ad_A^k(B) = \underbrace{[A,[A, \cdots[A,}_k B ]\cdots]].
\end{equation}
Subsequently, we construct the matrix representation of the adjoint action of $A$ in the Krylov subspace $\mathcal{K}_n(\ad_A, B) = \Span\left\{B, \ad_A(B), \ldots, \ad_A^{n-1}(B) \right\}$ to obtain the companion matrix
\begin{equation}
	\mathbf{C} =
	\begin{pmatrix}
		0 & 0 & \ldots & 0 & -x_0\\
		1 & 0 & \ldots & 0 & -x_1\\
		0 & 1 & \ldots & 0 & -x_2\\
		\vdots & \vdots & \ddots & \vdots & \vdots\\
		0 & 0 & \ldots & 1 & -x_{n-1}	
	\end{pmatrix}.
\end{equation}
In the next step, the coefficient vector is computed by exponentiating the companion matrix and applying it to the coordinate vector representing $B$,
\begin{equation}
	e^{\alpha \mathbf C}
	\begin{pmatrix}
		1\\0\\\vdots\\0
	\end{pmatrix}
	=
	\begin{pmatrix}
		f_0(\alpha)\\f_1(\alpha)\\\vdots\\f_{n-1}(\alpha)
	\end{pmatrix}.
\end{equation}
The final transformation is exactly computed as
\begin{equation}
	e^{\alpha \ad_A}B = \sum_{i=0}^{n-1} f_i(\alpha)\ad_A^i(B).
\end{equation}

All of the above simplifications enabled us to construct the systems of 28 and 84 coupled ODEs for the Wei--Norman decompositions of the unitaries generated by the singlet spin-adapted double excitation operators $\intS$ and $\intT$.
An inspection of the ODEs revealed the following.
The parameters multiplying the generators defining $\intS$ and $\intT$ had the same numerical values as in these definitions, as was anticipated from the structure of the corresponding Lie algebras.
The parameters that were numerically equal to zero were identical zeros.
Due to the relations among the parameters, many ODEs were redundant.
After eliminating linearly dependent equations, we obtained 6 and 12 coupled ODEs associated with $\exp(\theta \intS)$ and $\exp(\theta \intT)$, respectively.
Although the analytical solution of the systems of ODEs was still impractical, we were able to verify symbolically using Mathematica \cite{Mathematica} that they were satisfied by the AI-inferred closed-form expressions.
Furthermore, the determinants of the reduced $6\times6$ and $12\times12$ Wei--Norman coefficient matrices were found to be strictly positive rendering both decompositions globally valid:
\begin{equation}
	\begin{split}
		\det(\mathbf{M}_1)
		&= \frac{1}{4}\sqrt{3 + \cos\left(\sqrt{2} \theta\right)} \sqrt{3 + \cos\left( 2\sqrt{2} \theta\right)}\\
		&>0
	\end{split}
\end{equation}
in the case of $\exp(\theta \intS)$
and
\begin{equation}
	\begin{aligned}
		\det(\mathbf{M}_2)
		&=
		\frac{1}{216}
		\Biggl(
		\left[\cos\left(\sqrt{2}\theta\right)+5\right]
		\\[-2pt]
		&\quad {}\times
		\left[
		\cos\left(\sqrt{2}\theta\right)
		+8\cos\left(\frac{\theta}{\sqrt{2}}\right)
		+27
		\right]
		\\[-2pt]
		&\quad {}\times
		\left[\cos\left(2\sqrt{2}\theta\right)+5\right]
		\\[-2pt]
		&\quad {}\times
		\left[
		8\cos\left(\sqrt{2}\theta\right)
		+\cos\left(2\sqrt{2}\theta\right)
		+27
		\right]
		\Biggr)^{1/2}
		\\[3pt]
		&>0 .
	\end{aligned}
\end{equation}
for $\exp(\theta \intT)$.
The final Wei--Norman parameters for $\exp(\theta \intS)$ and $\exp(\theta \intT)$ are provided in \cref{table:int0_closed,table:int1_closed}.
All inverse trigonometric functions appearing in the closed-form expressions are understood as continuously unwrapped functions, with branches chosen to satisfy the initial conditions at $\theta=0$ and to preserve continuity over the full parameter domain.
\begin{table}
	\caption{\label{table:int0_closed}
		Closed-form parameters for the Wei--Norman decomposition of
		$\exp(\theta \intS)$.}
	\renewcommand{\arraystretch}{1.5}
	\begin{ruledtabular}
		\begin{tabular}{c c}
			Parameter\footnote{The mapping of the parameters to the 28 generators is provided in Table SI in the \sm.} & Closed Form\\
			\hline
			$b_1(\theta)$ & $(\frac{1}{\sqrt{2}} - \frac{1}{2}) \theta - \arctan\left[ \frac{(3-2\sqrt{2})\sin(\sqrt{2} \theta)}{1+(3-2\sqrt{2})\cos(\sqrt{2}\theta)} \right]$\\
			$b_2(\theta)$ & $\frac{1}{2}b_1(2\theta) - 2b_1(\theta)$\\
			$b_3(\theta)$ & $\frac{\theta}{2} - \arcsin\left[\frac{\sin\left(\frac{\theta}{\sqrt{2}}\right)}{\sqrt{2}}\right]$\\
			$b_4(\theta)$ & $\frac{1}{2}b_3(2\theta) - 2b_3(\theta)$\\
			$b_5(\theta)$ & $\frac{\pi}{4}-\arctan\left[\cos\left(\frac{\theta}{\sqrt{2}}\right)\right]$\\
			$b_6(\theta)$ & $\frac{1}{2}b_5(2\theta) - 2b_5(\theta)$\\
		\end{tabular}
	\end{ruledtabular}
\end{table}

\begin{table*}
	\caption{\label{table:int1_closed}
		Closed-form parameters for the Wei--Norman decomposition of
		$\exp(\theta \intT)$.}
	\renewcommand{\arraystretch}{2}
	\begin{ruledtabular}
		\begin{tabular}{c c}
			Parameter\footnote{The mapping of the parameters to the 84 generators is provided in Table SII in the \sm.} & Closed Form\\
			\hline
			$c_1(\theta)$  & $-(\frac{1}{\sqrt{2}} - \frac{1}{\sqrt{3}}) \theta + \arctan\left[ \frac{(5-2\sqrt{6})\sin(\sqrt{2} \theta)}{1+(5-2\sqrt{6})\cos(\sqrt{2}\theta)} \right]$\\
			$c_2(\theta)$  & $4c_1(\theta) - \frac{1}{2} c_1(2\theta)$\\
			$c_3(\theta)$  & $-\frac{\theta}{\sqrt{3}} + 3 \arcsin\left[\frac{\sin\left(\frac{\theta}{\sqrt{2}}\right)}{\sqrt{6 - \sin^2\left(\frac{\theta}{\sqrt{2}}\right)}}\right] - \frac{1}{2} \arcsin\left[\frac{\sin\left(\sqrt{2}\theta\right)}{\sqrt{6 - \sin^2\left(\sqrt{2}\theta\right)}}\right]$\\
			$c_4(\theta)$  & $-\frac{\theta}{2\sqrt{3}} + \arcsin\left[\frac{\sin\left(\frac{\theta}{\sqrt{2}}\right)}{\sqrt{6 - \sin^2\left(\frac{\theta}{\sqrt{2}}\right)}}\right]$\\
			$c_5(\theta)$  & $c_4(\theta) - c_3(\theta)$\\
			$c_6(\theta)$  & $-c_3(\theta) - c_4(\theta)$\\
			$c_7(\theta)$  & $-\frac{\theta}{\sqrt{3}} + 3\arcsin\left[ \frac{\sin\left(\frac{\theta}{\sqrt{2}}\right)}{\sqrt{6}} \right] -\frac{1}{2} \arcsin\left[\frac{\sin\left(\sqrt{2}\theta\right)}{\sqrt{6}}\right]$\\
			$c_8(\theta)$  & $-\frac{\theta}{2\sqrt{3}} + \arcsin\left[ \frac{\sin\left(\frac{\theta}{\sqrt{2}}\right)}{\sqrt{6}} \right]$\\
			$c_9(\theta)$  & $c_8(\theta) - c_7(\theta)$\\
			$c_{10}(\theta)$ & $-c_7(\theta) - c_8(\theta)$\\
			$c_{11}(\theta)$ & $2 \arctan\left(\frac{2 \sqrt{3} \left[\cos \left(\frac{\theta}{\sqrt{2}}\right)-1\right]}{\left[\cos \left(\frac{\theta}{\sqrt{2}}\right)+2\right] \sqrt{\cos \left(\sqrt{2} \theta\right)+11}}\right)-\frac{1}{4} \arctan\left(\frac{2 \sqrt{3} \left[\cos \left(\sqrt{2} \theta\right)-1\right]}{\left[\cos \left(\sqrt{2} \theta\right)+2\right] \sqrt{\cos \left(2 \sqrt{2} \theta\right)+11}}\right)$\\
			$c_{12}(\theta)$ & $-\arctan\left(\frac{2 \sqrt{3} \left[\cos \left(\frac{\theta}{\sqrt{2}}\right)-1\right]}{\left[\cos \left(\frac{\theta}{\sqrt{2}}\right)+2\right] \sqrt{\cos \left(\sqrt{2} \theta\right)+11}}\right)+\frac{1}{4} \arctan\left(\frac{2 \sqrt{3} \left[\cos \left(\sqrt{2} \theta\right)-1\right]}{\left[\cos \left(\sqrt{2} \theta\right)+2\right] \sqrt{\cos \left(2 \sqrt{2} \theta\right)+11}}\right)$\\
			$c_{13}(\theta)$ & $2 \arctan\left(\frac{\sqrt{2} \left[\cos\left(\frac{\theta}{\sqrt{2}}\right) - 1\right]}{\sqrt{\cos^2 \left(\frac{\theta}{\sqrt{2}}\right)+4 \cos \left(\frac{\theta}{\sqrt{2}}\right)+13}}\right) - \frac{1}{4} \arctan\left(\frac{\sqrt{2} \left[\cos\left(\sqrt{2}\theta\right) - 1\right]}{\sqrt{\cos^2 \left(\sqrt{2}\theta\right)+4 \cos \left(\sqrt{2}\theta\right)+13}}\right)$\\
			$c_{14}(\theta)$ & $-\arctan\left(\frac{\sqrt{2} \left[\cos\left(\frac{\theta}{\sqrt{2}}\right) - 1\right]}{\sqrt{\cos^2 \left(\frac{\theta}{\sqrt{2}}\right)+4 \cos \left(\frac{\theta}{\sqrt{2}}\right)+13}}\right) + \frac{1}{4} \arctan\left(\frac{\sqrt{2} \left[\cos\left(\sqrt{2}\theta\right) - 1\right]}{\sqrt{\cos^2 \left(\sqrt{2}\theta\right)+4 \cos \left(\sqrt{2}\theta\right)+13}}\right)$\\
			$c_{15}(\theta)$ & $2 \arctan\left[\frac{\cos \left(\frac{\theta}{\sqrt{2}}\right) - 1}{\cos \left(\frac{\theta}{\sqrt{2}}\right)+5}\right] - \frac{1}{4} \arctan\left[\frac{\cos\left(\sqrt{2}\theta\right) - 1}{\cos \left(\sqrt{2} \theta\right)+5}\right]$\\
			$c_{16}(\theta)$ & $-\arctan\left[\frac{\cos \left(\frac{\theta}{\sqrt{2}}\right) - 1}{\cos \left(\frac{\theta}{\sqrt{2}}\right)+5}\right] + \frac{1}{4} \arctan\left[\frac{\cos\left(\sqrt{2}\theta\right) - 1}{\cos \left(\sqrt{2} \theta\right)+5}\right]$\\
		\end{tabular}
	\end{ruledtabular}
\end{table*}

\subsection{Circuit Compression}

\begin{figure*}[h!]
	\centering
	\includegraphics[width=17cm]{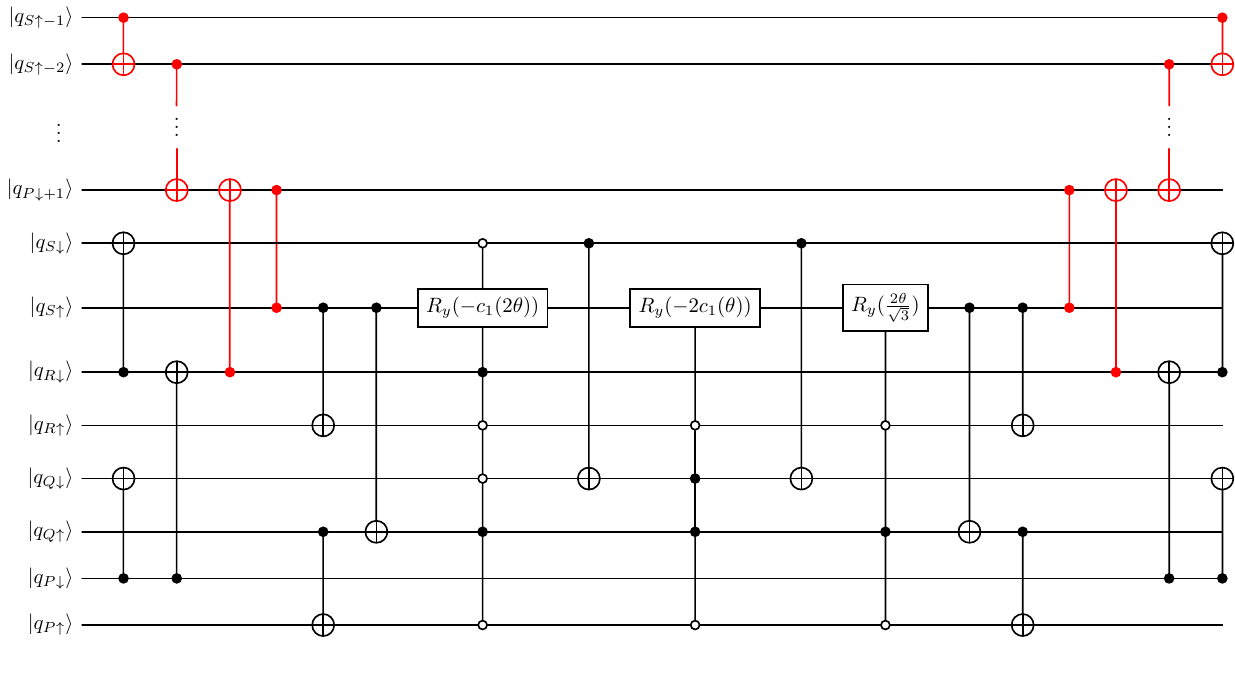}
	\caption{
		Quantum circuit implementing the 6 exponentials in \cref{eq:int1_1st_c1_block}.
	}
	\label{fig:int1_circuit_1st_c1_block}
\end{figure*}

Having found the Wei--Norman decompositions of the spin-adapted unitaries generated by $\intS$ and $\intT$, we now turn our attention to their compact circuit implementations.
As the basis elements of the Lie algebras $\mathfrak{g}_0$ and $\mathfrak{g}_1$ are anti-Hermitian operators multiplied, at most, by linear combinations of two particle--hole-conjugate number-operator strings, the elementary exponentials can be implemented using similar quantum circuits as those discussed in the previous section for the case of $\exp(\theta A_{PP}^{QR})$.
However, for $\exp(\theta \intS)$ and $\exp(\theta \intT)$ an additional level of optimization is possible that leads to substantially more compact quantum circuits.
For example, the first six exponentials in the Wei--Norman decomposition of $\exp(\theta \intT)$ read:
\begin{equation}
	\label{eq:int1_1st_c1_block}
	\begin{aligned}
		& e^{\frac{\theta}{\sqrt{3}}A_{\Pu\Qu}^{\Ru\Su}}
		e^{c_1(\theta)A_{\Pu\Qu}^{\Ru\Su}
			\left(n_{\Sd}h_{\Qd}+n_{\Qd}h_{\Sd}\right)}
		\\[-2pt]
		&\quad {}\times
		e^{-c_1(\theta)A_{\Pu\Qu}^{\Ru\Su}
			\left(n_{\Rd}h_{\Pd}+n_{\Pd}h_{\Rd}\right)}
		\\[-2pt]
		&\quad {}\times
		e^{-2c_1(\theta)A_{\Pu\Qu}^{\Ru\Su}
			\left(n_{\Rd\Sd}h_{\Qd}+n_{\Qd}h_{\Rd\Sd}\right)}
		\\[-2pt]
		&\quad {}\times
		e^{-2c_1(\theta)A_{\Pu\Qu}^{\Ru\Su}
			\left(n_{\Pd\Qd}h_{\Sd}+n_{\Sd}h_{\Pd\Qd}\right)}
		\\[-2pt]
		&\quad {}\times
		e^{\left(4c_1(\theta)-\frac{1}{2}c_1(2\theta)\right)
			A_{\Pu\Qu}^{\Ru\Su}
			\left(
			n_{\Rd\Sd}h_{\Pd\Qd}
			+n_{\Pd\Qd}h_{\Rd\Sd}
			\right)}
		\cdots .
	\end{aligned}
\end{equation}
By rearranging terms, we arrive at
\begin{equation}
	\resizebox{\linewidth}{!}{$
	e^{\frac{\theta}{\sqrt{3}} A_{\Pu\Qu}^{\Ru\Su}} e^{c_1(\theta) f_1 A_{\Pu\Qu}^{\Ru\Su}} e^{-\frac{1}{2}c_1(2\theta) A_{\Pu\Qu}^{\Ru\Su} (n_{\Rd\Sd}h_{\Pd\Qd} + n_{\Pd\Qd}h_{\Rd\Sd})},
	$}
\end{equation}
where
\begin{equation}
	\label{eq:f1_def}
	\begin{aligned}
		f_1
		={}&
		n_{\Sd}h_{\Qd}
		+n_{\Qd}h_{\Sd}
		\\[-2pt]
		&-
		\left(
		n_{\Rd}h_{\Pd}
		+n_{\Pd}h_{\Rd}
		\right)
		\\[-2pt]
		&-
		2\left(
		n_{\Rd\Sd}h_{\Qd}
		+n_{\Qd}h_{\Rd\Sd}
		\right)
		\\[-2pt]
		&-
		2\left(
		n_{\Pd\Qd}h_{\Sd}
		+n_{\Sd}h_{\Pd\Qd}
		\right)
		\\[-2pt]
		&+
		4\left(
		n_{\Rd\Sd}h_{\Pd\Qd}
		+n_{\Pd\Qd}h_{\Rd\Sd}
		\right).
	\end{aligned}
\end{equation}
Although \cref{eq:f1_def} appears to be quite involved, after evaluating it for all possible occupancy patterns of the four spin orbitals, we realize that it takes the form
\begin{equation}
	f_1 = - (n_{\Pd} \oplus n_{\Qd} \oplus n_{\Rd} \oplus n_{\Sd}),
\end{equation}
with the symbol ``$\oplus$'' denoting addition modulo two.
As a result, when compared to the bare unitary obtained by setting $f_1=1$, the circuit implementation of the middle exponential requires 6 additional CNOTs to compute and uncompute the parity of the four spin orbitals, an extra control in the $R_y$ gate, and flipping the sign of the angle.
The compact quantum circuit implementing \cref{eq:int1_1st_c1_block} is provided in \cref{fig:int1_circuit_1st_c1_block}.

\begin{figure*}[h!]
	\centering
	\includegraphics[width=17cm]{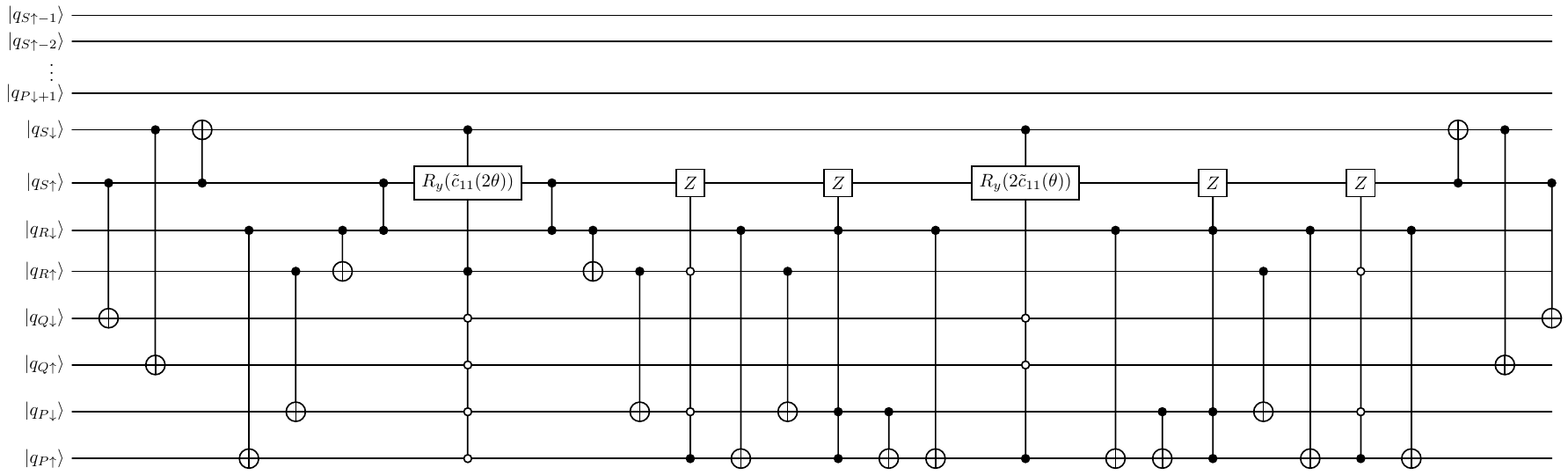}
	\caption{
		Quantum circuit implementing the 6 exponentials in \cref{eq:int1_1st_c11_block}.
	}
	\label{fig:int1_circuit_1st_c11_block}
\end{figure*}

Another sequence of operators appearing in the Wei--Norman decomposition of $\exp(\theta \intT)$ worth discussing is
\begin{equation}
	\label{eq:int1_1st_c11_block}
	\begin{aligned}
		&\cdots\,
		e^{c_{11}(\theta)A_{\Qu\Sd}^{\Qd\Su}
			\left(
			n_{\Ru}h_{\Pu}
			-n_{\Pu}h_{\Ru}
			\right)}
		\\[-2pt]
		&\quad {}\times
		e^{-c_{11}(\theta)A_{\Qu\Sd}^{\Qd\Su}
			\left(
			n_{\Rd}h_{\Pd}
			-n_{\Pd}h_{\Rd}
			\right)}
		\\[-2pt]
		&\quad {}\times
		e^{c_{12}(\theta)A_{\Qu\Sd}^{\Qd\Su}
			\left(
			n_{\Ru}h_{\Pu\Rd}
			-n_{\Pu\Rd}h_{\Ru}
			\right)}
		\\[-2pt]
		&\quad {}\times
		e^{c_{12}(\theta)A_{\Qu\Sd}^{\Qd\Su}
			\left(
			n_{\Pd\Ru}h_{\Pu}
			-n_{\Pu}h_{\Pd\Ru}
			\right)}
		\\[-2pt]
		&\quad {}\times
		e^{c_{12}(\theta)A_{\Qu\Sd}^{\Qd\Su}
			\left(
			n_{\Pd}h_{\Pu\Rd}
			-n_{\Pu\Rd}h_{\Pd}
			\right)}
		\\[-2pt]
		&\quad {}\times
		e^{-c_{12}(\theta)A_{\Qu\Sd}^{\Qd\Su}
			\left(
			n_{\Rd}h_{\Pd\Ru}
			-n_{\Pd\Ru}h_{\Rd}
			\right)}
		\cdots .
	\end{aligned}
\end{equation}
From the definitions of the $c_{11}$ and $c_{12}$ coefficients shown in \cref{table:int1_closed} and the cosine half-angle formula, it is straightforward to show that
\begin{equation}
	c_{11}(\theta) = 2\tilde{c}_{11}(\theta) - \frac{1}{4}\tilde{c}_{11}(2\theta)
\end{equation}
and
\begin{equation}
	c_{12}(\theta) = -\tilde{c}_{11}(\theta) + \frac{1}{4}\tilde{c}_{11}(2\theta),
\end{equation}
with
\begin{equation}
	\tilde{c}_{11}(\theta) = \arctan\left(\frac{\sqrt{6} \left[\cos\left(\frac{\theta}{\sqrt{2}}\right) - 1\right]}{\left[\cos\left(\frac{\theta}{\sqrt{2}}\right) + 2\right]\sqrt{\cos^2\left(\frac{\theta}{\sqrt{2}}\right) + 5}}\right).
\end{equation}
This, in conjunction with the fact that the generators of the 6 exponentials commute, enables us to rewrite \cref{eq:int1_1st_c11_block} as
\begin{equation}
	\label{eq:int1_1st_c11_block_reduced}
	\cdots e^{\tilde{c}_{11}(\theta) f_2 A_{\Qu\Sd}^{\Qd\Su}} e^{\tilde{c}_{11}(2\theta) f_3 A_{\Qu\Sd}^{\Qd\Su}} \cdots,
\end{equation}
where
\begin{equation}
	\label{eq:f2_def}
	\begin{aligned}
		f_2
		={}&
		2\left[
		\left(
		n_{\Ru}h_{\Pu}
		-n_{\Pu}h_{\Ru}
		\right)
		-
		\left(
		n_{\Rd}h_{\Pd}
		-n_{\Pd}h_{\Rd}
		\right)
		\right]
		\\[-2pt]
		&-
		\left(
		n_{\Ru}h_{\Pu\Rd}
		-n_{\Pu\Rd}h_{\Ru}
		\right)
		\\[-2pt]
		&-
		\left(
		n_{\Pd\Ru}h_{\Pu}
		-n_{\Pu}h_{\Pd\Ru}
		\right)
		\\[-2pt]
		&-
		\left(
		n_{\Pd}h_{\Pu\Rd}
		-n_{\Pu\Rd}h_{\Pd}
		\right)
		\\[-2pt]
		&+
		\left(
		n_{\Rd}h_{\Pd\Ru}
		-n_{\Pd\Ru}h_{\Rd}
		\right)
		\\[3pt]
		={}&
		\left(
		-n_{\Pu}
		+n_{\Pd}
		+n_{\Ru}
		-n_{\Rd}
		\right)
		\\[-2pt]
		&\quad {}\times
		\left(
		n_{\Pu}
		\oplus n_{\Pd}
		\oplus n_{\Ru}
		\oplus n_{\Rd}
		\right).
	\end{aligned}
\end{equation}
and
\begin{equation}
	\label{eq:f3_def}
	\begin{aligned}
		f_3
		={}&
		\frac{1}{4}
		\Biggl[
		-
		\left(
		n_{\Ru}h_{\Pu}
		-n_{\Pu}h_{\Ru}
		\right)
		\\[-2pt]
		&\quad
		+
		\left(
		n_{\Rd}h_{\Pd}
		-n_{\Pd}h_{\Rd}
		\right)
		\\[-2pt]
		&\quad
		+
		\left(
		n_{\Ru}h_{\Pu\Rd}
		-n_{\Pu\Rd}h_{\Ru}
		\right)
		\\[-2pt]
		&\quad
		+
		\left(
		n_{\Pd\Ru}h_{\Pu}
		-n_{\Pu}h_{\Pd\Ru}
		\right)
		\\[-2pt]
		&\quad
		+
		\left(
		n_{\Pd}h_{\Pu\Rd}
		-n_{\Pu\Rd}h_{\Pd}
		\right)
		\\[-2pt]
		&\quad
		-
		\left(
		n_{\Rd}h_{\Pd\Ru}
		-n_{\Pd\Ru}h_{\Rd}
		\right)
		\Biggr]
		\\[3pt]
		={}&
		\frac{1}{2}
		\left(
		n_{\Pd\Ru}h_{\Pu\Rd}
		-n_{\Pu\Rd}h_{\Pd\Ru}
		\right).
	\end{aligned}
\end{equation}
The restrictions imposed by \cref{eq:f2_def,eq:f3_def} enable us to construct the pertinent quantum circuits.
For example, according to \cref{eq:f2_def}, the first exponential in \cref{eq:int1_1st_c11_block_reduced} is acting nontrivially on the odd-parity occupation sector of the $\Pu$, $\Pd$, $\Ru$, and $\Rd$ spin orbitals.
Compared to the bare exponential, this additional check can be readily enforced by 6 extra CNOTs and an additional control on the corresponding $R_y$ gate.
However, the sign of the rotation gate depends on the occupancy of the spin orbitals as $-n_{\Pu} + n_{\Pd} + n_{\Ru} - n_{\Rd}$, which equals $\pm 1$ in the odd-parity sector.
The appropriate sign can be enforced via four three-qubit-controlled $Z$ gates and 8 CNOTs.
Finally, according to \cref{eq:f3_def}, the second exponential in \cref{eq:int1_1st_c11_block_reduced} is acting nontrivially when $n_{\Pu} = n_{\Rd} \neq n_{\Pd} = n_{\Ru}$.
When these conditions are met, the sign of the rotation angle depends on the occupancy of spin orbital $\Pu$.
Compared to the bare exponential, these additional conditions are enforced via 6 extra CNOTs, an additional C$Z$ gate, and one extra control on the $R_y$ gate.
The compact quantum circuit implementing \cref{eq:int1_1st_c11_block} is provided in \cref{fig:int1_circuit_1st_c11_block}.
The quantum circuit representations for the exponentials containing $c_{13}$ through $c_{16}$ in the Wei--Norman decomposition of $\exp(\theta \intT)$ are obtained in a similar manner, using the fact that
\begin{equation}
	c_{13}(\theta) = 2\tilde{c}_{13}(\theta) - \frac{1}{4}\tilde{c}_{13}(2\theta),
\end{equation}
\begin{equation}
	c_{14}(\theta) = -\tilde{c}_{13}(\theta) + \frac{1}{4}\tilde{c}_{13}(2\theta),
\end{equation}
\begin{equation}
	c_{15}(\theta) = 2\tilde{c}_{15}(\theta) - \frac{1}{4}\tilde{c}_{15}(2\theta),
\end{equation}
and
\begin{equation}
	c_{16}(\theta) = -\tilde{c}_{15}(\theta) + \frac{1}{4}\tilde{c}_{15}(2\theta),
\end{equation}
with
\begin{equation}
	\tilde{c}_{13}(\theta) = \arctan\left(\frac{\sqrt{2} \left[\cos\left(\frac{\theta}{\sqrt{2}}\right) - 1\right]}{\sqrt{\cos^2 \left(\frac{\theta}{\sqrt{2}}\right)+4 \cos \left(\frac{\theta}{\sqrt{2}}\right)+13}}\right)
\end{equation}
and
\begin{equation}
	\tilde{c}_{15}(\theta) = \arctan\left[\frac{\cos \left(\frac{\theta}{\sqrt{2}}\right) - 1}{\cos \left(\frac{\theta}{\sqrt{2}}\right)+5}\right].
\end{equation}

\begin{figure*}[h!]
	\centering
	\includegraphics[width=17cm]{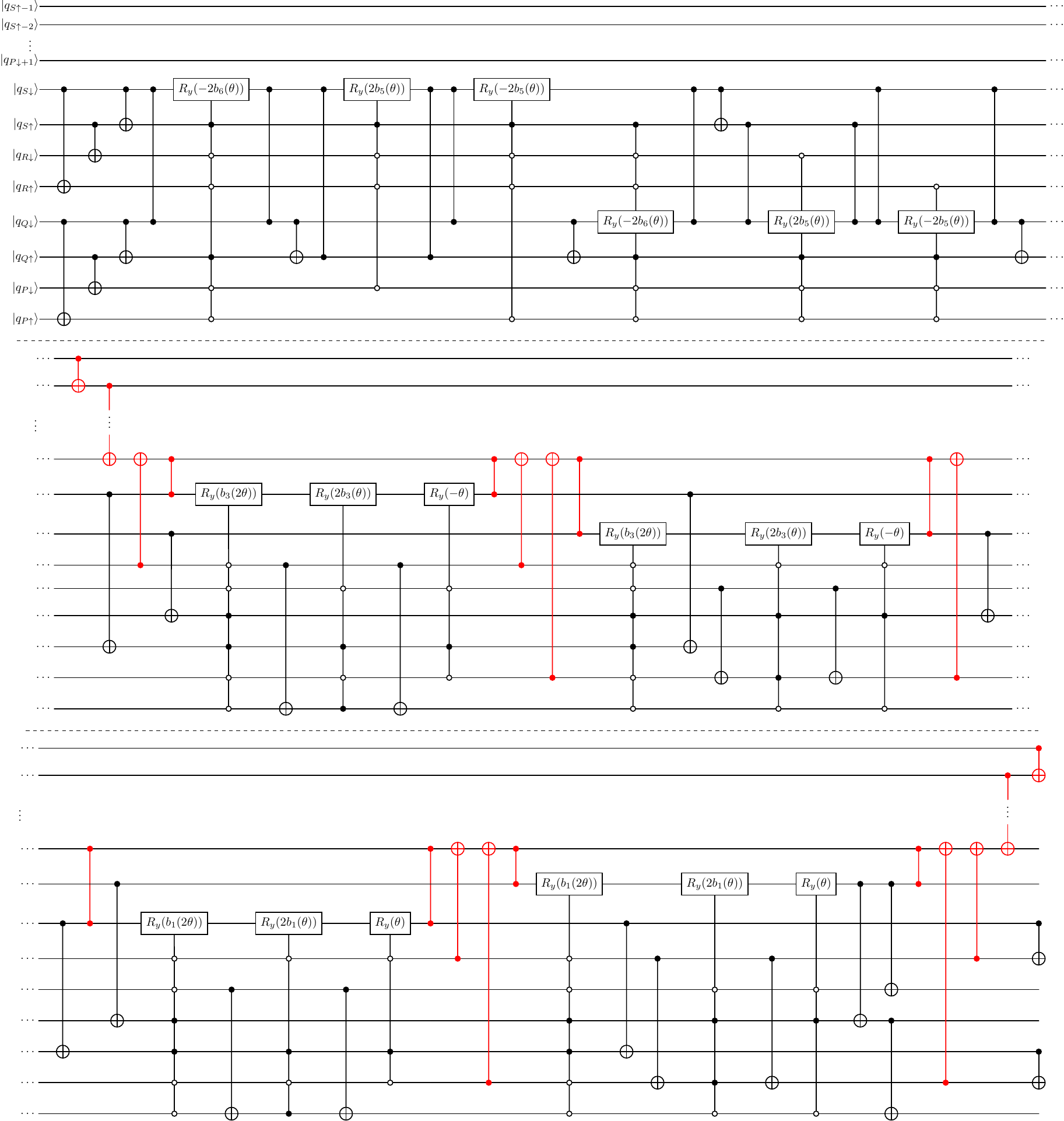}
	\caption{
		Quantum circuit implementing the spin-adapted unitary $\exp(\theta \intS)$.
	}
	\label{fig:int0_circ}
\end{figure*}

The complete circuits are obtained by concatenating the circuit blocks, in reverse order, and canceling gates between block pairs.
The final optimized circuit of the spin-adapted unitary $\exp(\theta \intS)$ is shown in \cref{fig:int0_circ}.
The corresponding circuit for $\exp(\theta \intT)$ is too long to be reproduced in the main text, but can be found in Section SVII of the {\sm}.
The validity of all spin-adapted quantum circuits generated in this study was verified by comparing their matrix representations against matrices obtained by direct exponentiation of the fermionic generators $A_{PP}^{QR}$, $\intS$, and $\intT$ for representative values of $\theta$ and spin orbital indices.

In \cref{table:gate_counts}, we provide the gate counts for the circuit implementations of the spin-adapted unitaries.
A key advantage of our compact quantum circuits is that they only require 2 CNOT ladders, independent of the singlet spin-adapted generator.
For the actual hardware implementation, the multi-controlled gates need to be decomposed into single- and two-qubit ones.
Assuming a standard Gray code implementation, the two-qubit and $R_y$ gate counts for the $\exp(\theta A_{PP}^{QR})$, $\exp(\theta \intS)$, and $\exp(\theta \intT)$ are 78 and 64, 584 and 544, and 1,498 and 1,008, respectively.
Nevertheless, depending on the type of hardware, one can employ more suitable implementations.
For noisy intermediate-scale devices, minimizing the number of two-qubit gates is crucial.
To that end, by using the deep decomposition of multi-controlled unitary gates \cite{Arsoski.2025.10.1007.s11227-025-07684-y}, the number of two-qubit gates needed to implement the spin-adapted unitaries $\exp(\theta A_{PP}^{QR})$, $\exp(\theta \intS)$, and $\exp(\theta \intT)$ is reduced to 72, 352, and 1,042.
In the case of fault-tolerant devices, Clifford gates such as CNOT and C$Z$ are essentially error-free and the main source of noise comes from the decomposition of $R_y$ into T gates.
In this case, it is possible to employ Toffoli gates, for which various optimized implementations exist \cite{Maslov.2016.10.1103/PhysRevA.93.022311}, to precompute the conditions for each multi-controlled $R_y$ gate on an ancilla qubit.
This allows us to replace all $\text{C}^nR_y$ gates by $\text{C}R_y$ gates, whose decomposition introduces only two $R_y$ gates.
Using this approach, the final numbers of $R_y$ gates are 10, 36, and 60 for the unitaries $\exp(\theta A_{PP}^{QR})$, $\exp(\theta \intS)$, and $\exp(\theta \intT)$, respectively.
\begin{table}
	\caption{\label{table:gate_counts}
		Gate counts for the quantum circuit implementations of the spin-adapted unitaries $\exp(\theta A_{PP}^{QR})$, $\exp(\theta \intS)$, and $\exp(\theta \intT)$.
		Counts are reported after circuit optimization under a Jordan--Wigner mapping with all-to-all qubit connectivity.
		CNOT and C$Z$ gates marked in red in \cref{fig:ppqr_circ,fig:int0_circ} and Figs.\ S8--S11 are part of the fermionic-sign ladders and are not included in the fixed CNOT/C$Z$ counts, since their number depends on the specific spin-orbital indices.}
	\begin{ruledtabular}
		\begin{tabular}{l c c c}
			Gate\footnote{The symbol $\text{C}^n$ is used to denote $n$-fold controlled gates, independent of whether they are controlled on the $\ket{0}$ or $\ket{1}$ states.} & $\exp(\theta A_{PP}^{QR})$ & $\exp(\theta \intS)$ & $\exp(\theta \intT)$\\
			\hline
			Ferm.-sign ladder & 2 & 2 & 2\\
			CNOT & 12 & 30 & 142\\
			C$Z$ & 2 & 10 & 12\\
			$\text{C}^3Z$ & 0 & 0 & 24\\
			$\text{C}^3R_y$ & 2 & 4 & 6\\
			$\text{C}^4R_y$ & 3 & 8 & 12\\
			$\text{C}^6R_y$ & 0 & 6 & 12
		\end{tabular}
	\end{ruledtabular}
\end{table}

The combination of exact global decompositions, closed-form expressions for the Wei--Norman parameters, and compact circuit representations enables the application of spin-adapted unitaries in quantum algorithms.
One such example is ADAPT-VQE \cite{Grimsley.2019.10.1038/s41467-019-10988-2}, an adaptive algorithm that constructs the ansatz iteratively.
Energy gradients are required in both the macro- and micro-iteration phases of the approach.
In the ansatz expansion step, the operator with the largest absolute energy gradient is selected from a predefined operator pool and is appended to the ansatz.
For this case, the evaluation of the gradients is straightforward, given by the expectation value of the commutator of the operator $A_\mu$ with the Hamiltonian $H$,
\begin{equation}
	\label{eq:selection_g}
	\begin{split}
		g_\mu =& \eval{\dv{\theta} \ev{e^{-\theta A_\mu} H e^{\theta A_\mu}}{\Psi}}_{\theta = 0}\\
		=& \ev{[H,A_\mu]}{\Psi}.
	\end{split}
\end{equation}
Note that \cref{eq:selection_g} is valid independent of whether $A_\mu$ is spin-adapted or not.
In the micro-iteration phase, the parameters of all unitaries are variationally optimized by minimizing the energy.
For unitaries generated by a single, anti-Hermitian, fermionic string, the energy gradient is computed exactly by an extension \cite{Kottmann.2021.10.1039/d0sc06627c} of the parameter shift rule \cite{Mitarai.2018.10.1103/PhysRevA.98.032309,Schuld.2019.10.1103/PhysRevA.99.032331}.
Generalizations of the parameter shift rule to generic single-parameter unitaries exist \cite{Kyriienko.2021.10.1103.PhysRevA.104.052417,Izmaylov.2021.10.1103/PhysRevA.104.062443,Wierichs.2022.10.22331.q-2022-03-30-677}, and can be naturally combined with the Wei--Norman decompositions of spin-adapted unitaries of this work,
\begin{equation}
	e^{(\theta + s_j) A_\mu} = \prod_{k=1}^{d} e^{\alpha_k(\theta + s_j) E_k},
\end{equation}
where $A_\mu$ is a spin-adapted operator, $E_k$ is an element of the $d$-dimensional dynamical Lie algebra, $\alpha_k$ is a closed-form Wei--Norman coefficient, and $s_j$ is the $j$th parameter shift.
Note that the number of shifts typically depends on the number of distinct nonzero eigenvalue differences of $A_\mu$.

\section{Cost Comparison with Alternative Quantum Algorithms Enforcing Spin Symmetry}

Here, we compare our compact quantum circuits for realizing saGSD operators with other hardware implementations of spin adaptation.
We begin with the two other known alternative exact implementations of saGSD unitaries.
The LCU approach described in the {\sm} enables, after postselection, the exact application of a spin-adapted unitary to a given state.
However, this scheme remains impractical even in a fault-tolerant setting.
For example, as shown in the {\sm}, the LCU decomposition of $\exp(\theta \intT)$ involves a linear combination of 2,985 unique Pauli strings.
Excluding the cost of the PREPARE gate, this already implies that the corresponding circuit requires 12 ancilla qubits and 2,985 12-qubit-controlled Pauli strings.
To make matters worse, this cost concerns the implementation of a single spin-adapted unitary, while typical ans\"{a}tze are expressed as a product of many such operators.

The second approach for the exact implementation of saGSD unitaries is the Pauli-string-based method of Jain \foreign{et al.} \cite{Jain.2026.10.1063.5.0326865}.
Specifically, in this scheme, each generator is mapped under the Jordan--Wigner transformation to linear combinations of commuting Pauli strings.
Subsequently, the quantum circuits are constructed via the standard sequence involving basis change, CNOT ladder, $R_z$ rotation, inverse CNOT ladder, and inverse basis change.
In this approach, the number of CNOT ladders depends on the number of Pauli strings, while our circuits require only 2 CNOT ladders for all generators.
For example, the Pauli-string-based implementation of the $\exp(\theta \intT)$ unitary requires 1,280 CNOT ladders.

An alternative strategy is to approximate the $S^2$-preserving saGSD unitaries via finite-order Trotter--Suzuki decompositions.
However, we have shown that even fourth-order Trotterization suffers from spin contamination \cite{Magoulas.2025.10.1080/00268976.2025.2534672}, and therefore remains, in principle, susceptible to variational collapse.
At the same time, it is also useful to compare the resource requirements of such high-order product formulas with those of our exact constructions.
For example, the generators $A_{PP}^{QR}$ and $\intS$ can each be partitioned into two internally commuting operator groups [see \cref{eq:sa_gd_ppqr,eq:sa_gd_pqrs_int0}].
Their fourth-order Trotter-Suzuki decomposition follows the form
\begin{equation}\label{eq:trotAB4}
	e^{\theta (X+Y)} \approx e^{\frac{s\theta}{2} X} e^{s\theta Y} e^{\frac{(1-s)\theta}{2} X} e^{(1-2s)\theta Y} e^{\frac{(1-s)\theta}{2} X} e^{s\theta Y} e^{\frac{s\theta}{2} X},
\end{equation}
where $s = \frac{1}{2-\sqrt[3]{2}}$ \cite{Hatano.2005.10.1007/11526216_2}.
The resulting individual unitaries are generated by generalized double excitations.
Their most efficient circuit implementation is based on the FEB formalism and requires, among other quantum resources, 2 CNOT ladders, 13 CNOT, 2 C$Z$, and 8 $R_y$ gates \cite{Yordanov.2020.10.1103/PhysRevA.102.062612}.
Thus, we readily obtain that the fourth-order Trotterizations of $\exp(\theta A_{PP}^{QR})$ and $\exp(\theta \intS)$ require 14 CNOT ladders and 91 CNOTs, and 28 CNOT ladders and 182 CNOTs, respectively.
Similarly, the generator $\intT$ can be partitioned into three internally commuting operator groups [see \cref{eq:sa_gd_pqrs_int1}], and its fourth-order Trotterization takes the form
\begin{equation}
	\begin{aligned}
		e^{\theta(X+Y+Z)}
		\approx{}&
		e^{\frac{s\theta}{2}X}
		e^{\frac{s\theta}{2}Y}
		e^{s\theta Z}
		e^{\frac{s\theta}{2}Y}
		\\[-2pt]
		&\quad {}\times
		e^{\frac{(1-s)\theta}{2}X}
		e^{\frac{(1-2s)\theta}{2}Y}
		e^{(1-2s)\theta Z}
		\\[-2pt]
		&\quad {}\times
		e^{\frac{(1-2s)\theta}{2}Y}
		e^{\frac{(1-s)\theta}{2}X}
		\\[-2pt]
		&\quad {}\times
		e^{\frac{s\theta}{2}Y}
		e^{s\theta Z}
		e^{\frac{s\theta}{2}Y}
		e^{\frac{s\theta}{2}X}.
	\end{aligned}
\end{equation}
with the same value of $s$ \cite{Barthel.2020.10.1016/j.aop.2020.168165}.
This approximate implementation of $\exp(\theta \intT)$ requires 52 CNOT ladders and 338 CNOTs.

These estimates should be compared with the resource requirements of our exact saGSD implementations, which rigorously preserve spin symmetry.
In particular, our exact circuits require only 2 CNOT ladders, with fixed CNOT counts of 72, 352, and 1,042 for $\exp(\theta A_{PP}^{QR})$, $\exp(\theta \intS)$, and $\exp(\theta \intT)$, respectively.
Thus, our exact implementation of $\exp(\theta A_{PP}^{QR})$ is always more economical than the corresponding fourth-order Trotter-Suzuki approximation.
For the more involved generators $\intS$ and $\intT$, our rigorously spin-adapted circuits become more efficient than their fourth-order Trotterization counterparts when the CNOT ladders involve at least 7 and 15 qubits, respectively.
Finally, even if aggressive gate cancellations reduce the CNOT ladder counts of the Trotterized approximations to two, the practical overhead of our circuits remains modest.
For example, a frozen-core simulation of benzene (\ce{C6H6}), a small aromatic molecule, using the correlation-consistent basis of quadruple $\zeta$ quality (cc-pVQZ) \cite{Dunning.1989.10.1063/1.456153} would involve 1,008 active spin orbitals, and hence 1,008 qubits under the Jordan--Wigner mapping.
For CNOT ladders of this length, our exact spin-adapted implementation of $\exp(\theta \intS)$ uses only 8\% more CNOTs than the corresponding approximate fourth-order Trotter--Suzuki implementation, while preserving spin symmetry exactly.
For the sake of comparison, the leading costs of the various exact and approximate implementations of saGSD unitaries are shown in \cref{table:saGSD_cost_comp}.
\begin{table*}
	\caption{\label{table:saGSD_cost_comp}
		Leading cost proxies for exact and approximate implementations of the
		$\exp(\theta \intT)$ saGSD unitary. Here $L$ denotes the CNOT-ladder length.}
	\renewcommand{\arraystretch}{1.5}
	\begin{ruledtabular}
		\begin{tabular}{l c c l r}
			Method & Exact $S^2$? & Exact unitary? & Main cost proxy & Leading cost \\
			\hline
			4th-order Trot.
			& No  & No  & CNOT ladders + fixed CNOTs & $52L + 338$ \\
			LCU
			& Yes & Yes & LCU terms & $2,985$ \\
			Ref \cite{Jain.2026.10.1063.5.0326865}
			& Yes & Yes & Pauli-string CNOT ladders & $1,280L$ \\
			This work
			& Yes & Yes & CNOT ladders + fixed CNOTs & $2L + 1,042$ \\
		\end{tabular}
	\end{ruledtabular}
\end{table*}

The above approaches relied on spin-adapted generators to enforce spin symmetry either exactly or approximately.
Here, we discuss methods whose generators do not enforce spin symmetry.
For the sake of consistency and comparison, we focus on approaches based on GSD operators, which can enforce all symmetries relevant to chemistry except $S^2$.
As mentioned above, the implementation of a generalized double excitation operator requires, among other resources, 2 CNOT ladders, 13 CNOT, 2 C$Z$, and 8 $R_y$ gates \cite{Yordanov.2020.10.1103/PhysRevA.102.062612}.
At a first glance, GSD circuits appear to be substantially more economical than their spin-adapted variants presented in this work.
However, in realistic molecular applications involving polyatomic systems and large one-electron basis sets, both circuits share the same leading cost, namely, the 2 CNOT ladders.
Using the same example of frozen-core \ce{C6H6}/cc-pVQZ and assuming a CNOT ladder involving 1,008 qubits, the CNOT overhead of our circuits remains a modest 17\% while eliminating the risk of variational collapse.
Even in a fault-tolerant setting, both GSD and saGSD circuits have $R_y$ gate counts that are within the same order of magnitude.
In addition, GSD-based approaches require more operators in the ansatz to reach the same level of accuracy as their spin-adapted counterparts, because they explore a much larger Hilbert space \cite{Magoulas.2025.10.1080/00268976.2025.2534672,Magoulas.2026.10.1063/5.0316482}.
Consequently, although the GSD circuits are significantly more efficient than the saGSD ones for small, proof-of-principle, molecular problems, we anticipate the disparity to decrease when considering realistic applications.

Furthermore, the violation of $S^2$ symmetry renders GSD-based quantum algorithms susceptible to spin contamination and even complete variational collapse to states with undesired spin multiplicities \cite{Tsuchimochi.2020.10.1103/PhysRevResearch.2.043142,Magoulas.2025.10.1080/00268976.2025.2534672,Magoulas.2026.10.1063/5.0316482} (see, for example, the collapse to an incorrect triplet state in \cref{fig:h2o_gsd}).
Two common approaches for mitigating or enforcing spin symmetry are spin projection \cite{Whitfield.2013.10.1063/1.4812566,Izmaylov.2019.10.1021/acs.jpca.9b01103} and spin penalty \cite{McClean.2016.10.1088/1367-2630/18/2/023023,Ryabinkin.2019.10.1021/acs.jctc.8b00943,Greene-Diniz.2021.10.1002/qua.26352,Kuroiwa.2021.10.1103/PhysRevResearch.3.013197}.
Although projection after variation, \foreign{i.e.}, post-selection, would be ineffective in the case of variational collapse, variation after projection is expected to be more robust.
To the best of our knowledge, Tsuchimochi \foreign{et al.} have designed the most efficient algorithm of this type \cite{Tsuchimochi.2020.10.1103/PhysRevResearch.2.043142}, implementing spin projection through a finite quadrature representation of the exact spin-projection integral.
This scheme requires one ancilla qubit and an additional 4$n$ CNOT gates, with a measurement cost scaling as $\mathcal{O}(N_g n^4)$, where $N_g$ is the number of quadrature grid points and $n$ is the number of spin orbitals.
Therefore, for realistic chemistry applications, augmenting GSD-based approaches with spin projection can lead to CNOT counts comparable to those of our saGSD circuits.
In addition, even this projection technique requires nonzero overlap with the target spin sector and can become numerically unstable when this overlap is small.
Finally, despite its usefulness in a VQE setting, this measurement-based approach does not directly yield a state-preparation circuit suitable for fault-tolerant algorithms.
This limitation should be contrasted with our spin-adapted circuits, which prepare symmetry-preserving states on quantum devices by construction.

In spin-penalty approaches, the bare Hamiltonian $H$ is replaced by the cost function $H^\prime$ given by
\begin{equation}
	H^\prime = H + \mu \left[S^2 -s (s + 1)\right]^2,
\end{equation}
where $\mu$ is a hyperparameter and $s(s+1)$ is the eigenvalue of $S^2$ in the target state.
Aside from its dependence on a user-dependent hyperparameter, the penalty approach has two disadvantages.
First, it introduces four-body terms in the cost function, increasing the Pauli measurement cost \cite{Tsuchimochi.2020.10.1103/PhysRevResearch.2.043142}.
Second, the quantum algorithm is still formally exploring a spin non-adapted many-electron Hilbert space, requiring more operators to reach the same level of accuracy than saGSD.
This, in turn, further inflates the number of CNOT gates and measurements.

Before we conclude this section, we comment on a couple of approaches that do not rely on either the GSD or saGSD operator sets.
The symmetry-preserving preparation circuits of Ref.\ \cite{Gard.2020.10.1038/s41534-019-0240-1} enforce particle-number, $S_z$, and $S^2$ symmetries.
However, these circuits fully span the symmetry-adapted Hilbert space of interest, requiring a combinatorial number of CNOT gates.
As such, they are only applicable to small active spaces.
In contrast, our saGSD quantum circuits provide an alternative route to symmetry-adapted state preparation that avoids the steep combinatorial scaling on quantum resources.
The Quantum Paldus Transform \cite{Burkat.2025.2506.09151} enforces particle-number, $S_z$, and $S^2$ symmetries by performing a change of basis from Slater determinants to the Gel'fand--Tsetlin basis of spin-adapted configuration state functions.
The Toffoli cost of this basis change scales with the number of spatial orbitals $d$ as $\mathcal{O}(d^3)$.
This scheme is primarily designed to enable Hamiltonian simulation in a symmetry-adapted basis, rather than to provide compact direct circuits for individual spin-adapted excitation unitaries.
To the best of our knowledge, none of the symmetry-preserving preparation circuits and Quantum Paldus Transform impose point-group symmetry.

From the above considerations, we conclude that the circuits presented in this work provide, to the best of our knowledge, the most compact exact implementation of saGSD operators to date.
Although they are not as efficient as other approximate techniques for enforcing spin symmetry in small, proof-of-principle, problems, they become competitive when realistic chemistry applications are considered.
To the best of our knowledge, for molecules that have trivial spatial symmetry, \foreign{i.e.}, they belong to the $C_1$ point group, saGSpD, which only contains spin-adapted singles and perfect-pairing doubles, remains the most efficient symmetry-adapted operator set.

	\section{Compact Universal Subset of the Symmetry-Adapted \lowercase{sa}GSD Operator Pool}

By designing compact quantum circuits for unitaries generated by the saGSD pool, we have enabled the hardware implementation of symmetry-preserving quantum simulations for many-fermion systems.
An open question is how to further reduce the required computational resources.
As shown in \cref{table:gate_counts}, the dominant contributor to the computational cost is the unitary generated by $\intT$, whose quantum circuit implementation requires about three times as many CNOT gates  as its $\intS$ counterpart.
Although it seems unlikely that more aggressive circuit optimizations would lower the CNOT count by a few hundred gates, one can question whether this type of generator is required to attain universality for systems that are not fully spin-polarized.

Indeed, for molecules without spatial symmetry, \foreign{i.e.}, those in the $C_1$ point group, and not fully spin-polarized, it has been shown that the saGSpD pool, which is comprised of singlet spin-adapted singles and perfect pairing doubles, is universal \cite{Burton.2023.10.1038/s41534-023-00744-2}.
However, we have recently shown that, for systems with higher spatial symmetry, fully symmetry-adapted saGSpD is not universal, as certain $\intS$ and $\intT$ classes are not part of its Lie algebra \cite{Magoulas.2026.10.1063/5.0316482}, \foreign{e.g.}, when $P$, $Q$, $R$, and $S$ belong to distinct irreducible representations of the point group but the total excitation is totally symmetric.
The lack of universality is caused by the spatial-symmetry restriction imposed on spin-adapted singles, $A_{P}^{Q}$, requiring both spatial orbitals to belong to the same irreducible representation.
To retain expressivity, non-totally symmetric spin-adapted single excitations are typically included in the pool \cite{Anselmetti.2021.10.1088/1367-2630/ac2cb3,Burton.2023.10.1038/s41534-023-00744-2,Burton.2024.10.1103/PhysRevResearch.6.023300}, albeit introducing spatial symmetry contaminants \cite{Magoulas.2026.10.1063/5.0316482}.

An alternative route to universality, while still enforcing all symmetries, is to augment the saGSpD pool with additional types of singlet spin-adapted double excitation operators.
Considering that the $A_{PP}^{QR}$ operators are implicitly included in the saGSpD pool via the commutator identity $A_{PP}^{QR} = [A_{QQ}^{PP}, A_Q^R]$, it is worth examining whether the inclusion of $\intS$ operators is sufficient to attain universality.
The two pools in this category are defined as $\text{saGSpDint0} \equiv \text{saGSpD} \cup \{\intS\}$ and $\text{saGSD0} \equiv \text{saGSpDint0} \cup \{A_{PP}^{QR}\}$.
Note that the saGSD0 pool contains operators reminiscent of those employed in the classical CCD0 and CCSD0 approaches \cite{Bulik.2015.10.1021/acs.jctc.5b00422}.

Before we prove the universality of saGSpDint0 and saGSD0, it is important to note that bare operators of the class $\intT$ cannot belong to their respective Lie algebras.
This is due to the fact that all elements of the saGSpDint0 and saGSD0 pools annihilate states in which all electrons have the same spin projection, while operators of the form $\intT$ do not.
Therefore, to establish the universality of saGSpDint0 and saGSD0 for non fully spin-polarized systems, it is sufficient to show that the projected action of the missing $\intT$ operators on the target symmetry sector is contained in their Lie algebras.
Considering that there is no point-group symmetry restriction between spatial orbitals $P$ and $Q$ in $A_{PP}^{QQ}$, one way to generate the spin-polarized terms appearing in the definition of $\intT$ [\cref{eq:sa_gd_pqrs_int1}] is via the commutator
\begin{equation}
	\resizebox{\linewidth}{!}{$
\begin{split}	
	[A_{PP}^{QQ}, \tensor*[^{[0]}]{A}{_{PR}^{QS}}] =& \frac{1}{2}\left[A_{\Pu\Su}^{\Qu\Ru} (n_{\Pd} - n_{\Qd}) + A_{\Pd\Sd}^{\Qd\Rd} (n_{\Pu} - n_{\Qu})\right.\\
	&\left.+ A_{\Pu\Sd}^{\Qd\Ru} (n_{\Pd} - n_{\Qu}) + A_{\Pd\Su}^{\Qu\Rd} (n_{\Pu} - n_{\Qd})\right].
\end{split}
	$}
\end{equation}
In particular, in the seniority-zero sector of the $P$ and $Q$ spatial orbitals, one readily obtains that
\begin{equation}
	[A_{PP}^{QQ}, \tensor*[^{[0]}]{A}{_{PR}^{QS}}] \propto \sqrt{3} \tensor*[^{[1]}]{A}{_{PS}^{QR}} - \tensor*[^{[0]}]{A}{_{PS}^{QR}}.
\end{equation}
Since all elements of the form $\intS$ are members of the saGSpDint0 and saGSD0 pools, the only independent elements introduced by such commutators are of the $\intT$ type.
Furthermore, saGSpDint0 and saGSD0 contain all bare singlet spin-adapted double excitation operators through an intermediate singlet.
Consequently, within any non fully spin-polarized symmetry sector, the intermediate-singlet manifold is fully explored by the pertinent Lie algebras.
In particular, sectors of the symmetry-adapted Hilbert space  in which any spatial orbital pair $PQ$ has seniority zero can be populated.
Furthermore, no additional spatial-symmetry restrictions are associated with such sectors, as spatial orbitals $P$ and $Q$ are either empty or doubly occupied and, thus, totally symmetric when the usual Abelian point groups are employed.
Thus, although bare $\intT$ operators do not belong to the saGSpDint0 and saGSD0 Lie algebras, their action on the relevant non fully spin-polarized target sector is generated indirectly through the fully explored intermediate-singlet space, proving the universality of these subsets of saGSD.
Furthermore, since spin-adapted singles belong to the Lie algebra generated by operators of the form $\intS$, we can construct the compact universal pool $\text{pDint0} = \{A_{PP}^{QQ}\} \cup \{\intS\}$.
This is the smallest saGSD subset considered here that retains universality while rigorously enforcing particle-number, point-group, $S_z$, and $S^2$ symmetry.
Although these three operator sets have the same leading CNOT cost, set by the $\intS$ operators, pDint0 contains fewer elements, resulting in a reduced number of measurements required for the expansion step of adaptive algorithms, such as ADAPT-VQE.
Finally, the fact that these reduced pools are not applicable to fully spin-polarized systems is not a major limitation.
Indeed, high-spin determinants are trivially spin-adapted, allowing the application of the standard GSD operator pool without the risk of spin contamination.

\begin{figure}[h!]
	\centering
	\includegraphics[width=8.5cm]{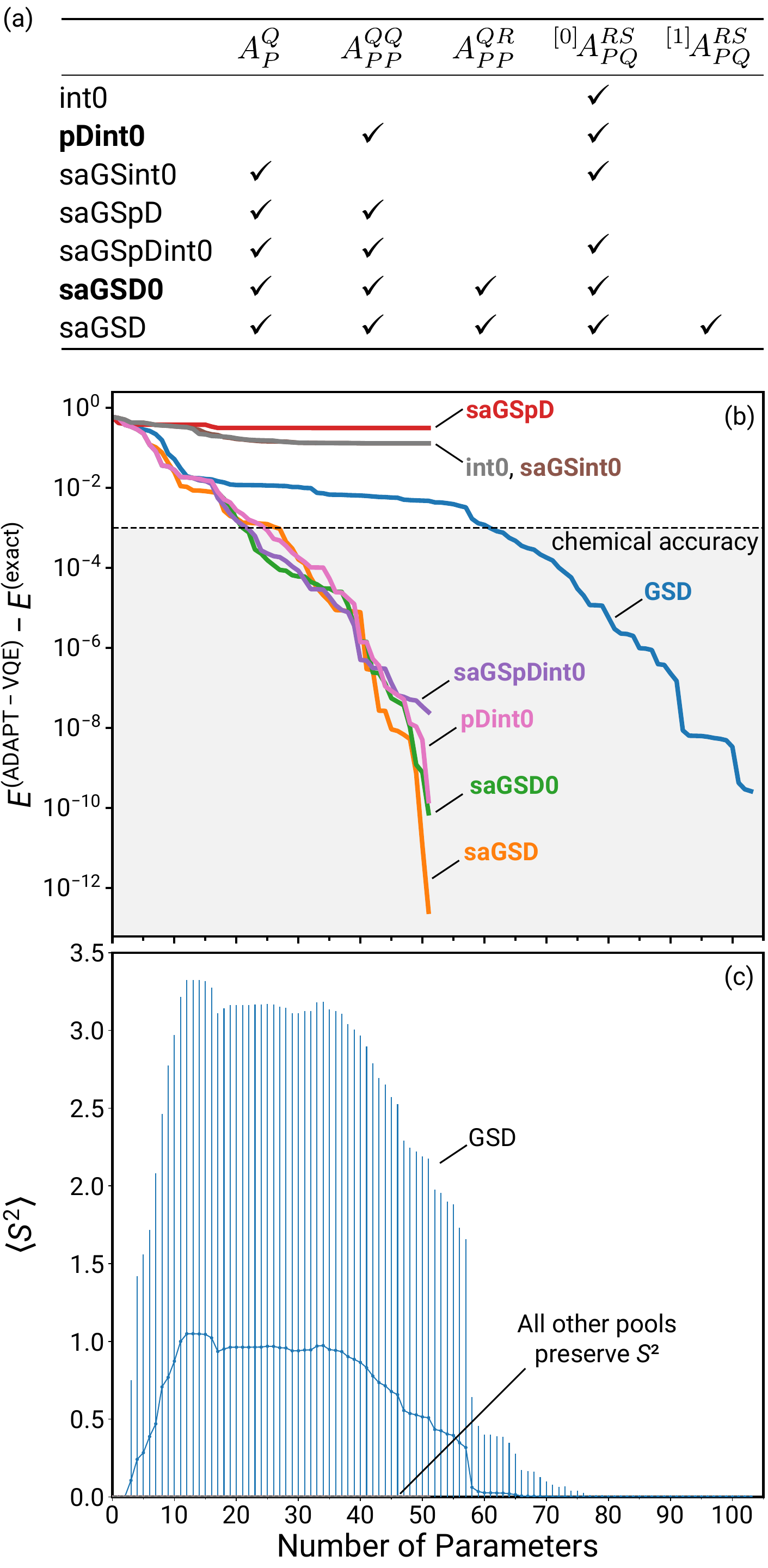}
	\caption{\label{fig:adapt}
		ADAPT-VQE simulations using the GSD and several spin-adapted operator pools for the $D_{2h}$-symmetric distorted hexagonal $\text{H}_6$/STO-6G system. (a) Definitions of the spin-adapted operator pools, (b) errors relative to the exact energy of the ground $^1 A_g$ electronic state, and (c) expectation values of the total spin squared $S^2$ operator and their standard deviations.}
\end{figure}

Here, we present a proof-of-principle numerical demonstration for the ground, $^1A_g$, electronic state of a $D_{2h}$-distorted hexagonal arrangement of six hydrogen atoms in the STO-6G minimum basis \cite{Hehre.1969.10.1063/1.1672392} (see Table SIII in the {\sm} for the nuclear coordinates).
Starting from a restricted Hartree--Fock (RHF) reference, we employed ADAPT-VQE with various operator pools, including GSD, which conserves all symmetries except $S^2$, as well as saGSD and several of its subsets defined in panel (a) of \cref{fig:adapt}, all of which respect all symmetries.

As shown in \cref{fig:adapt}(b), although ADAPT-VQE-GSD eventually converges to the exact $^1A_g$ ground electronic state, which is the global energy minimum, it requires an unnecessarily large number of parameters, as it is exploring the much larger ($A_g$, $N = 6$, $S_z = 0$) subspace of the many-electron Hilbert space, which contains singlet, triplet, quintet, and septet states.
Indeed, as shown in panel (c) of \cref{fig:adapt}, the ADAPT-VQE-GSD simulation is plagued by severe spin-symmetry breaking, which is eventually restored as the wavefunction converges to the lowest-energy $^1A_g$ state.
When employing the saGSD pool, the ADAPT-VQE ansatz rigorously preserves $S^2$ symmetry and provides chemically accurate results (within \SI{1}{\milli\textit{E}_h} of the exact energy) with only about one third of the parameters compared to its GSD counterpart, and produces numerically exact energies with about half the number of parameters.
However, the saGSD pool contains all classes of singlet spin-adapted double excitations, including the quantum-resource-demanding $\intT$ ones.

Although all subsets of the saGSD pool explored in this study rigorously enforce all symmetries, they may not be as expressive as the parent pool.
A characteristic example is the CNOT-frugal saGSpD pool, which is not universal when point-group symmetry is enforced and barely improves upon the underlying RHF energy.
As anticipated from our formal analysis, the three subsets of saGSD that produce numerically exact energies are saGSD0, saGSpDint0, and pDint0.

Before concluding, we comment on the sizes of the aforementioned operator pools for this numerical demonstration.
The largest pool is GSD, containing 196 operators.
By enforcing spin symmetry, saGSD has only 75 operators, a more than 50\% reduction compared to GSD.
The saGSD0, saGSpDint0, and pDint0 universal subsets of saGSD contain 52, 40, and 38 elements, respectively.
With about half as many operators as saGSD, pDint0 is the most compact subset considered here that rigorously enforces all relevant symmetries while retaining universality.
Finally, although the saGSpD pool is substantially smaller than pDint0, containing only 17 elements, it is not universal.

For additional numerical examples illustrating the potential of the pDint0 operator pool in ADAPT-VQE simulations of small molecular systems, see Ref.\ \cite{Magoulas.2026.10.1063/5.0316482}.

\section{Discussion}

In this work, we introduced explicit gate-level constructions for symmetry-adapted fermionic unitaries, which have traditionally been difficult to implement without sacrificing either exact symmetry preservation or circuit compactness.
The exact global Wei--Norman decompositions derived here eliminate the symmetry-breaking errors introduced by finite-order Trotter--Suzuki and Zassenhaus formulas, while extending exact circuit constructions beyond reduced operator pools based only on spin-adapted singles and perfect-pairing doubles.
Furthermore, our extension of the fermionic-excitation-based framework avoids both the ancilla overhead of LCU implementations and the proliferation of Pauli-string exponentials.
In doing so, we have closed the gap between the algebraic formulation of symmetry-adapted many-electron ans\"{a}tze and their practical implementation as compact quantum circuits, an important requirement for both near-term and fault-tolerant quantum simulations for chemistry.

A second important contribution is methodological.
For the 28- and 84-dimensional Lie algebras examined here, the direct symbolic construction and integration of the Wei--Norman systems of coupled ordinary differential equations proved to be computationally prohibitive.
We, thus, designed a discovery-and-verification strategy in which numerical optimization identifies the parameter structure, symmetry and proportionality relations reduce the problem to a small set of independent parameter families, closed-form expressions are inferred from the resulting functional patterns, and the final formulas are validated by direct substitution into the reduced Wei--Norman equations.
To further enable the symbolic construction of compact algebraic expressions, we also developed a Krylov-subspace-based algorithm for the automatic construction of exact closed-form expressions with the minimum number of nested commutators for unitary transformations involving finite linear combinations of fermionic operators.
This methodology is not limited to fermionic quantum simulations and provides a practical route for applying the Wei--Norman decomposition to Lie algebras that are too large for conventional symbolic treatment.

Finally, for the construction of compact quantum circuits, we extended the fermionic-excitation-based framework to generators in which a single anti-Hermitian fermionic string is multiplied by an arbitrary linear combination of number-operator strings.
This allowed us to combine multiple conditional exponentials into a single one, resulting in a substantially more compact circuit.
The final, CNOT-optimized, circuit implementations of $\exp(\theta A_{PP}^{QR})$, $\exp(\theta \intS)$, and $\exp(\theta \intT)$ contained 72, 352, and 1,042 CNOTs, respectively, together with two CNOT ladders that account for the fermionic sign.

To further reduce the computational cost, we provided a Lie-algebraic proof that, with the exception of fully spin-polarized systems, explicit operators of the form $\intT$ are not required for universality.
These formal considerations were further illustrated in proof-of-principle ADAPT-VQE numerical simulations, showing that the operator pools termed pDint0, saGSpDint0, and saGSD0 provided essentially exact results for the ground electronic state of a $D_{2h}$-distorted hexagonal arrangement of six hydrogen atoms.
Taken together, these results suggest that the pDint0 pool offers a good balance between computational efficiency and rigorous enforcement of all symmetries relevant to chemistry applications.

\section*{Methods}

\subsection*{Construction of dynamical Lie algebra}
All symbolic manipulations were performed with SymPy.
For each spin-adapted unitary $\exp(\theta \sum_i A_i)$, we first computed the Lie closure of the set of spin-orbital operators $\{A_i\}$.
This was accomplished by iteratively closing the set under commutation.
Specifically, for every pair $(X,Y)$ drawn from the current list of basis elements, we computed the commutator $[X,Y]$, simplified the resulting expression, and added it to the basis if it was nonzero and linearly independent from the existing elements.
The newly added operators were then commuted with both the original generators and with the other new elements until no further linearly independent operators emerged.
To enable the efficient circuit implementation of the corresponding unitaries, we took linear combinations of the basis elements so that they could take the form of an anti-Hermitian excitation operator $A_i$ multiplied, at most, by a linear combination of two particle--hole-conjugate number-operator strings $N_i$.
The final ordered basis $\{E_i\}$ of the dynamical Lie algebra was then constructed by grouping the basis elements into commuting sets.

The details of the construction and solution of the Wei--Norman systems of ordinary differential equations can be found in \cref{sec:wn_ppqr,sec:wn_intT_intT}.

To examine the effect of permutations in the decomposition of the simplest unitary $\exp(\theta A_{PP}^{QR})$, we symbolically built the Wei--Norman system of coupled ordinary differential equations in matrix form,
\begin{equation}
	\mathbf{A}(\theta, \boldsymbol{\alpha}) \boldsymbol{\alpha}^\prime(\theta) = \mathbf{d}(\theta),
\end{equation}
where $\mathbf{A}$ is the Wei--Norman coefficient matrix, $\boldsymbol{\alpha}$ is the vector of $\theta$-dependent parameters, $\boldsymbol{\alpha}^\prime = \dv{\boldsymbol{\alpha}}{\theta}$, and $\mathbf{d}$ is the vector containing the coefficients multiplying the basis elements of the dynamical Lie algebra in the target spin-adapted unitary.
The symbolic expressions for $\mathbf{A}$ and $\mathbf{d}$ were converted to numerical callables.
The numerical integration of the Wei--Norman system of equations was performed using the initial-value-problem solver from SciPy.
The interval of integration was set to [-100,100] and the solution was sampled on 20,001 uniformly spaced points.
We used the DOP853 integration method, with relative and absolute tolerances set to $\text{rtol} = 10^{-9}$ and $\text{atol} = 10^{-12}$, respectively.

\subsection*{Alternative exact product formulas}
The computational protocol we adopted for the construction of an exact, Trotter--Suzuki-like product formula of $\exp(\theta A_{PP}^{QR})$ was as follows.
We constructed the matrix representation of both the target unitary and the product ansatz in a minimal Fock space of 6 spin orbitals.
For each $\theta \in [0,10]$, sampled on 2,001 uniformly spaced points, we minimized the Frobenius norm of the difference between these two matrices using the least-squares optimizer of SciPy.
We employed tight convergence criteria ($\text{xtol}=\text{ftol}=\text{gtol}=10^{-12}$) and set the maximum number of function evaluations to $\text{max\_nfev}=3,000$.
To increase confidence that the global minimum was obtained, for every value of $\theta$ we performed 20 restarts with random initial parameter values.
The maximum residual Frobenius error over the grid was $\sim 10^{-16}$.

\subsection*{ADAPT-VQE numerical simulations}
In each ADAPT-VQE macro-iteration, we selected the operator with the largest energy gradient from the corresponding pool and appended it to the ansatz.
The simulations were terminated when the ansatz contained as many parameters as the dimension of the corresponding many-electron Hilbert space minus one (to account for normalization).
During the VQE steps, all parameters were optimized variationally using the Broyden--Fletcher--Goldfarb--Shanno (BFGS) optimizer \cite{Broyden.1970.10.1093/imamat/6.3.222,Fletcher.1970.10.1093/comjnl/13.3.317,Goldfarb.1970.10.1090/S0025-5718-1970-0258249-6,Shanno.1970.10.1090/S0025-5718-1970-0274029-X} as implemented in SciPy.
The convergence criterion for the micro-iterations was set to \SI{e-6}{\textit{E}_h} for the gradient norm.
All numerical simulations were performed using a developmental version of QForte \cite{Stair.2022.10.1021/acs.jctc.1c01155}, based on sparse matrices and employing closed-form expressions of spin-adapted unitaries \cite{Magoulas.2025.10.1080/00268976.2025.2534672,Kjellgren.2025.10.1063/5.0278717}.
These closed-form expressions enable more efficient simulations than the sequential application of each gate in the corresponding quantum circuits, without affecting the results of the numerical demonstration.
The one- and two-electron integrals were obtained with RHF as implemented in Psi4 \cite{Smith.2020.10.1063/5.0006002}.

\section*{Data availability}

Data generated and analyzed during the current study are available from the corresponding author upon reasonable request.

\section*{Code availability}

The code used in the current study is available from the corresponding author upon reasonable request.

\section*{Acknowledgments}

The authors would like to thank Muhan Zhang for insightful discussions regarding the quantum circuits presented in this study.
The authors would also like to acknowledge the use of AI tools (ChatGPT, Claude) for the inference of candidate closed-form expressions based on Fourier series.
All candidate formulas were subsequently verified analytically by showing that the underlying Wei--Norman systems were satisfied.
AI tools (ChatGPT) were also used for a light polish of the grammar, clarity, and readability of the text.
All AI-assisted text in the article was reviewed and approved by the authors.

\section*{Funding}

This work was supported by the U.S.\ Department of Energy under Award No.\ DE-SC0019374.

\section*{Author contributions}

I.M.\ and F.A.E.\ contributed equally to the conceptualization of the project.
I.M.\ wrote the code for symbolic manipulations of fermionic operators and the sparse-matrix ADAPT-VQE code.
I.M.\ designed the quantum circuits presented in this work.
I.M.\ and F.A.E.\ contributed equally to the development of the project, the discussion of the results, and the writing of the manuscript.

\section*{Competing interests}

The authors declare no competing interests.

\section*{Supplemental Material}
The supplemental document contains the details on alternative exact product formulas, the linear-combination-of-unitaries implementation of spin-adapted unitaries, the classification of the dynamical Lie algebra generated by $A_{PP}^{QR}$, the proof of the closed-form unitary transformations used in the Wei--Norman decomposition of $A_{PP}^{QR}$, the effect of operator permutations on the decomposition of $A_{PP}^{QR}$, the dynamical Lie algebras associated with $\intS$ and $\intT$, the full compact circuit for $\exp(\theta \intT)$, the molecular geometry of \ce{H6} used in the numerical simulations, and the exact potential-energy curves for the $C_s$-symmetric bending motion of \ce{H2O}/STO-6G.

\clearpage

\renewcommand{\theequation}{S\arabic{equation}}
\setcounter{equation}{0}

\renewcommand{\thetable}{S\arabic{table}}
\setcounter{table}{0}

\renewcommand{\thefigure}{S\arabic{figure}}
\setcounter{figure}{0}

\renewcommand{\thesection}{S\arabic{section}}
\setcounter{section}{0}

\renewcommand{\thepage}{S\arabic{page}}
\setcounter{page}{1}

\onecolumngrid
\fontsize{12}{22}\selectfont
\begin{center}
	\textbf{\large Supplemental Material:\\
		Spin-Adapted Fermionic Unitaries:\linebreak From Lie Algebras to Compact Quantum Circuits
	}\\[.2cm]
	Ilias Magoulas$^{*}$ and Francesco A.\ Evangelista\\[.1cm]
	{\itshape Department of Chemistry and Cherry Emerson Center for Scientific Computation,\\ 
		Emory University, Atlanta, Georgia 30322, USA\\}
	${}^*$Corresponding author; e-mail: ilias.magoulas@emory.edu.
\end{center}

\newpage

	The Supplemental Material document is organized as follows.
In Section SI, we describe the construction of exact, Trotter-like, finite product formulas.
In Section SII, we discuss the design of spin-adapted quantum circuits based on the linear-combination-of-unitaries approach.
The classification of the Lie algebra associated with the $A_{PP}^{QR}$ generator is discussed in Section SIII.
The proof of the closed-form unitary transformations used in the derivation of the Wei--Norman equations for the $\exp(\theta A_{PP}^{QR})$ unitary is presented in Section SIV.
In Section SV, we show in a graphical form the effect of permutations on the Wei--Norman decompositions of the unitary $\exp(\theta A_{PP}^{QR})$.
The generators of the dynamical Lie algebras associated with $\intS$ and $\intT$ are shown in Section SVI.
The compact quantum circuit implementing the spin-adapted unitary $\exp(\theta \intT)$ is given in Section SVII.
The nuclear coordinates of the $\text{H}_6$ system used in our numerical study are given in Section SVIII.
The exact potential-energy curves for the $C_s$-symmetric bending motion of $\text{H}_2\text{O}$/STO-6G are provided in Section SIX.
Reference numbers in the Supplemental Material document correspond to those in the main text.

\pagebreak 

\section{Alternative Exact Product Formulas}

\begin{figure}[h!]
	\centering
	\includegraphics[width=3.375in]{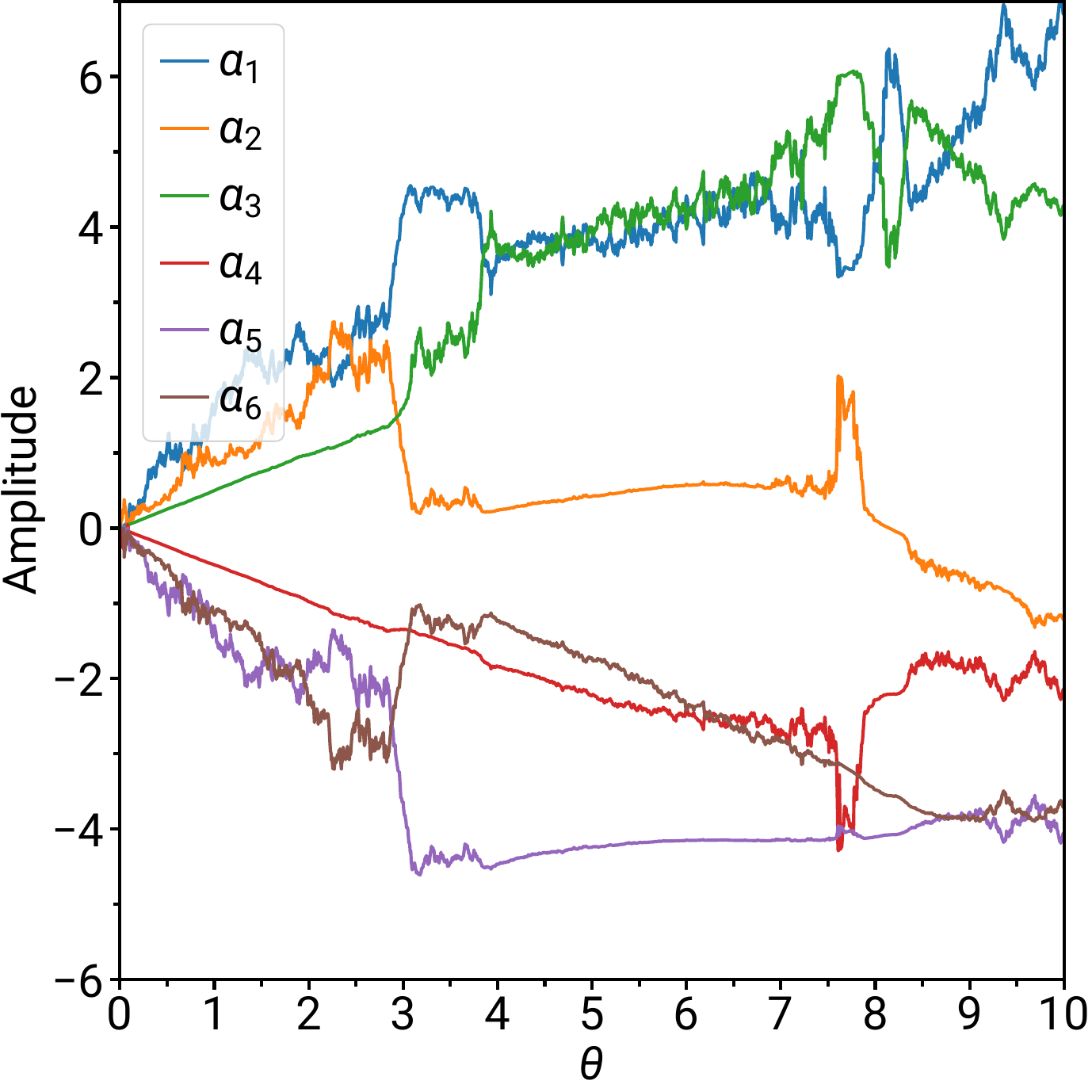}
	\caption{\label{sfig:ababab}
		Numerical values of the $\theta$-dependent parameters defining the Trotter--Suzuki-like decomposition of \cref{seq:trot_like}.}
\end{figure}

As mentioned in the main text, since the spin orbital operators that define spin-adapted unitaries generate finite Lie algebras, it might be possible to construct exact, finite product formulas reminiscent of finite-order Trotter--Suzuki decompositions.
We tested this numerically in the case of the simplest, yet nontrivial, spin-adapted unitary, $\exp[\theta (A_{\Pu\Pd}^{\Qu\Rd} - A_{\Pu\Pd}^{\Qd\Ru})/\sqrt{2}]$.
For the sake of brevity, we set $A\equiv A_{\Pu\Pd}^{\Qu\Rd}$ and $B\equiv A_{\Pu\Pd}^{\Qd\Ru}$.
Based on our numerical explorations, the minimum number of exponentials required is six, leading to the product formula
\begin{equation}\label{seq:trot_like}
	e^{\frac{\theta}{\sqrt{2}} (A-B)} = e^{\alpha_1(\theta) A} e^{\alpha_2(\theta) B} e^{\alpha_3(\theta) A} e^{\alpha_4(\theta) B} e^{\alpha_5(\theta) A} e^{\alpha_6(\theta) B},
\end{equation}
with the optimum values of the $\theta$-dependent parameters given in \cref{sfig:ababab}.

In contrast to the Wei--Norman decomposition presented in the main text, the parameters entering \cref{seq:trot_like} are not uniquely defined.
Nevertheless, the algebraic structure does impose constraints on the parameters.
For example, we find that
\begin{equation}
	\alpha_1(\theta) + \alpha_3(\theta) + \alpha_5(\theta) = \frac{\theta}{\sqrt{2}}
\end{equation}
and
\begin{equation}
	\alpha_2(\theta) + \alpha_4(\theta) + \alpha_6(\theta) = -\frac{\theta}{\sqrt{2}}.
\end{equation}
A major issue in the practical utility of such decompositions is that the parameters are highly oscillatory functions of $\theta$, rendering them more sensitive to device noise and complicating the computation of derivatives.

\section{Symmetry-Adapted Circuits via Linear Combination of Unitaries}

As mentioned in the main text, spin-adapted unitaries can be exactly represented as a finite series \cite{Magoulas.2025.10.1080/00268976.2025.2534672,Kjellgren.2025.10.1063/5.0278717}, enabling their circuit implementation as a linear combination of unitaries (LCU) \cite{Childs.2012.10.26421/QIC12.11-12,Berry.2015.10.1103/PhysRevLett.114.090502}.
For example, the unitary generated by $\intT$ is exactly expressed as
\begin{equation}\label{eq:sa_unitary_closed_sorb}
	\resizebox{\textwidth}{!}{$
		\begin{split}
			&e^{\theta \intT} = I\\
			&+\left[ -\frac{16 \sin \left(\frac{\sqrt{3} \theta}{2}\right)}{25 \sqrt{3}}-\frac{54}{25} \sqrt{3} \sin \left(\frac{\theta}{\sqrt{3}}\right)+\frac{8}{5} \sqrt{2} \sin \left(\frac{\theta}{\sqrt{2}}\right)+\frac{432}{115} \sqrt{3} \sin \left(\frac{\theta}{2 \sqrt{3}}\right)+\frac{\sin \left(\sqrt{2} \theta\right)}{575 \sqrt{2}} \right] \tensor*[^{[1]}]{A}{_{PQ} ^{RS}}\\
			&+\left[ \frac{32}{75} \cos \left(\frac{\sqrt{3} \theta}{2}\right)-\frac{16}{5} \cos \left(\frac{\theta}{\sqrt{2}}\right)+\frac{162}{25} \cos \left(\frac{\theta}{\sqrt{3}}\right)-\frac{2592}{115} \cos \left(\frac{\theta}{2 \sqrt{3}}\right)-\frac{\cos \left(\sqrt{2} \theta\right)}{1150}+\frac{113}{6}\right] \left(\tensor*[^{[1]}]{A}{_{PQ} ^{RS}}\right)^2\\
			&+\left[-\frac{56 \sin \left(\frac{\sqrt{3} \theta}{2}\right)}{5 \sqrt{3}}-\frac{171}{5} \sqrt{3} \sin \left(\frac{\theta}{\sqrt{3}}\right)+\frac{404}{15} \sqrt{2} \sin \left(\frac{\theta}{\sqrt{2}}\right)+\frac{2952}{115} \sqrt{3} \sin \left(\frac{\theta}{2 \sqrt{3}}\right)+\frac{11 \sin \left(\sqrt{2} \theta\right)}{345 \sqrt{2}} \right] \left(\tensor*[^{[1]}]{A}{_{PQ} ^{RS}}\right)^3\\
			&+\left[ \frac{112}{15} \cos \left(\frac{\sqrt{3} \theta}{2}\right)-\frac{11}{690} \cos \left(\sqrt{2} \theta\right)-\frac{808}{15} \cos \left(\frac{\theta}{\sqrt{2}}\right)+\frac{513}{5} \cos \left(\frac{\theta}{\sqrt{3}}\right)-\frac{17712}{115} \cos \left(\frac{\theta}{2 \sqrt{3}}\right)+\frac{587}{6} \right] \left(\tensor*[^{[1]}]{A}{_{PQ} ^{RS}}\right)^4\\
			&+\left[ -\frac{1192 \sin \left(\frac{\sqrt{3} \theta}{2}\right)}{25 \sqrt{3}}-\frac{2718}{25} \sqrt{3} \sin \left(\frac{\theta}{\sqrt{3}}\right)+\frac{133 \sqrt{2} \sin \left(\sqrt{2} \theta\right)}{1725}+\frac{308}{3} \sqrt{2} \sin \left(\frac{\theta}{\sqrt{2}}\right)+\frac{1368}{23} \sqrt{3} \sin \left(\frac{\theta}{2 \sqrt{3}}\right)\right] \left(\tensor*[^{[1]}]{A}{_{PQ} ^{RS}}\right)^5\\
			&+\left[ \frac{2384}{75} \cos \left(\frac{\sqrt{3} \theta}{2}\right)-\frac{616}{3} \cos \left(\frac{\theta}{\sqrt{2}}\right)+\frac{8154}{25} \cos \left(\frac{\theta}{\sqrt{3}}\right)-\frac{8208}{23} \cos \left(\frac{\theta}{2 \sqrt{3}}\right)-\frac{133 \cos \left(\sqrt{2} \theta\right)}{1725}+\frac{613}{3} \right] \left(\tensor*[^{[1]}]{A}{_{PQ} ^{RS}}\right)^6\\
			&+\left[ \frac{16}{115} \sqrt{2} \sin \left(\sqrt{2} \theta\right)+\frac{608}{5} \sqrt{2} \sin \left(\frac{\theta}{\sqrt{2}}\right)-\frac{112}{5} \sqrt{3} \sin \left(\frac{\sqrt{3} \theta}{2}\right)-\frac{576}{5} \sqrt{3} \sin \left(\frac{\theta}{\sqrt{3}}\right)+\frac{6192}{115} \sqrt{3} \sin \left(\frac{\theta}{2 \sqrt{3}}\right)\right] \left(\tensor*[^{[1]}]{A}{_{PQ} ^{RS}}\right)^7\\
			&+\left[ \frac{224}{5} \cos \left(\frac{\sqrt{3} \theta}{2}\right)-\frac{16}{115} \cos \left(\sqrt{2} \theta\right)-\frac{1216}{5} \cos \left(\frac{\theta}{\sqrt{2}}\right)+\frac{1728}{5} \cos \left(\frac{\theta}{\sqrt{3}}\right)-\frac{37152}{115} \cos \left(\frac{\theta}{2 \sqrt{3}}\right)+176 \right] \left(\tensor*[^{[1]}]{A}{_{PQ} ^{RS}}\right)^8\\
			&+\left[ \frac{48}{575} \sqrt{2} \sin \left(\sqrt{2} \theta\right)+\frac{192}{5} \sqrt{2} \sin \left(\frac{\theta}{\sqrt{2}}\right)-\frac{192}{25} \sqrt{3} \sin \left(\frac{\sqrt{3} \theta}{2}\right)-\frac{864}{25} \sqrt{3} \sin \left(\frac{\theta}{\sqrt{3}}\right)+\frac{1728}{115} \sqrt{3} \sin \left(\frac{\theta}{2 \sqrt{3}}\right)\right] \left(\tensor*[^{[1]}]{A}{_{PQ} ^{RS}}\right)^9\\
			&+\left[ \frac{384}{25} \cos \left(\frac{\sqrt{3} \theta}{2}\right)-\frac{48}{575} \cos \left(\sqrt{2} \theta\right)-\frac{384}{5} \cos \left(\frac{\theta}{\sqrt{2}}\right)+\frac{2592}{25} \cos \left(\frac{\theta}{\sqrt{3}}\right)-\frac{10368}{115} \cos \left(\frac{\theta}{2 \sqrt{3}}\right)+48 \right] \left(\tensor*[^{[1]}]{A}{_{PQ} ^{RS}}\right)^{10}.
		\end{split}
		$}
\end{equation}
with $\theta$ being a real parameter.
By translating \cref{eq:sa_unitary_closed_sorb} to the qubit space using the fermionic encoding of choice, the spin-adapted unitary is expressed as a linear combination of Pauli unitaries with $\theta$-dependent coefficients,
\begin{equation}
	\label{eq:sa_unitary_closed_LCU}
	e^{\theta \intT} = \sum_{i=0}^{M-1} \lambda_i(\theta) U_i,
\end{equation}
where $\lambda_i(\theta) \ge 0$, and any signs or phases have been absorbed into the unitary operators $U_i$.
Since each $U_i$ is a Pauli unitary, \foreign{i.e.}, a Pauli string possibly multiplied by a phase factor, \cref{eq:sa_unitary_closed_LCU} is an LCU.

\begin{figure*}[h!]
	\centering
	\includegraphics[width=\linewidth]{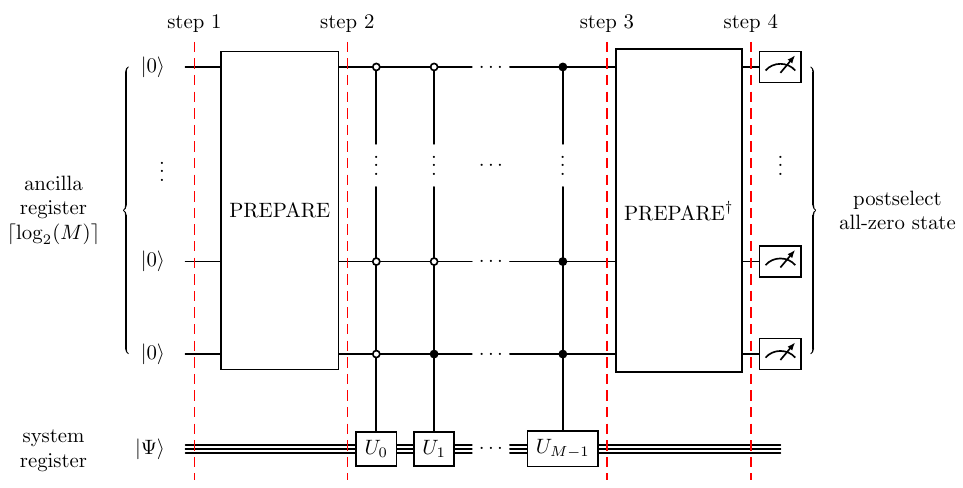}
	\caption{
		Quantum circuit implementation of unitary generated by a singlet spin-adapted double excitation operator.
		The system register is initialized in the state on which the spin-adapted unitary will act.
		Open and full circles denote controls based on the $\ket{0}$ and $\ket{1}$ states, respectively.
		The gates $U_0$ through $U_{M-1}$ represent the Pauli unitaries appearing in the LCU decomposition of the spin-adapted unitary.
	}
	\label{fig:LCU}
\end{figure*}

The LCU circuit implementation of spin-adapted unitaries is shown in \cref{fig:LCU}.
In addition to the number of qubits needed to realize the trial state $\ket{\Psi}$ on which the spin-adapted unitary will act, the circuit requires $\left\lceil\log_2 (M)\right\rceil$ ancillas, with $M$ being the number of terms in the LCU [\cref{eq:sa_unitary_closed_LCU}].
Besides the PREPARE gate and its adjoint acting on the ancillas, the circuit applies the unitaries $U_i$ to $\ket{\Psi}$, controlled by the ancilla qubits. 
To understand the mechanism by which the LCU circuit realizes $\exp(\theta A) \ket{\Psi}$, with $A$ being a spin-adapted operator, we examine the state of the quantum computer $\ket{\text{QC}}$ at the time slices shown in \cref{fig:LCU}:
\begin{equation}
	\begin{split}
		\ket{\mathrm{QC}} &\xrightarrow{\mathrm{step\ 1}} \ket{0}^{\otimes \left\lceil\log_2(M)\right\rceil} \otimes \ket{\Psi}\\
		&\xrightarrow{\mathrm{step\ 2}} \sum_{i=0}^{M-1} \sqrt{\frac{\lambda_i(\theta)}{\norm{\boldsymbol{\lambda}(\theta)}_1}} \ket{i} \otimes \ket{\Psi}\\
		&\xrightarrow{\mathrm{step\ 3}} \sum_{i=0}^{M-1} \sqrt{\frac{\lambda_i(\theta)}{\norm{\boldsymbol{\lambda}(\theta)}_1}} \ket{i} \otimes U_i \ket{\Psi}\\
		&\xrightarrow{\mathrm{step\ 4}} \frac{1}{\norm{\boldsymbol{\lambda}(\theta)}_1}\ket{0}^{\otimes \left\lceil\log_2(M)\right\rceil} \otimes \sum_{i=0}^{M-1} \lambda_i(\theta) U_i \ket{\Psi} + \cdots\\
		&= \frac{1}{\norm{\boldsymbol{\lambda}(\theta)}_1}\ket{0}^{\otimes \left\lceil\log_2(M)\right\rceil} \otimes e^{\theta A} \ket{\Psi} + \cdots,
	\end{split}
\end{equation}
with
\begin{equation}
	\norm{\boldsymbol{\lambda}(\theta)}_1 \equiv \sum_{i=0}^{M-1} \lambda_i(\theta).
\end{equation}
Therefore, if the measurement outcome on the ancilla register is the all-zero state, the quantum circuit in \cref{fig:LCU} has applied the spin-adapted unitary to the trial state $\ket{\Psi}$.
The probability of obtaining this outcome is the inverse square of the 1-norm of the LCU coefficients.

The advantage of the quantum circuit of \cref{fig:LCU} is that it implements an exact representation of the spin-adapted unitary, rigorously enforcing all desired symmetries.
Furthermore, the number of terms $M$ in the LCU is independent of the identity of the excitation indices, depending only on the type of generator ($A_{PP}^{QR}$ \foreign{vs.} $\intS$ \foreign{vs.} $\intT$) and on the chosen fermionic encoding.

Unfortunately, the LCU approach has its own complications.
To begin with, the number of terms $M$ is extremely large for the generators considered here.
For example, under the Jordan--Wigner \cite{Jordan.1928.10.1007/BF01331938} mapping, we obtain that the number of unique Pauli strings for the LCU decomposition of $\exp(\theta \intT)$ is 2,985.
This, in turn, implies that the corresponding quantum circuit requires 12 ancilla qubits and 2,985 12-qubit-controlled Pauli gates.
The hardware implementation of the PREPARE gate will introduce additional multiqubit-controlled gates, although their number and nature depend on the form of the $\lambda_i(\theta)$ coefficients.
Furthermore, as described above, the success probability of postselecting the ancillas in the all-zero state is $1/\norm{\boldsymbol{\lambda}(\theta)}_1^2$.
Oblivious amplitude amplification can improve the success rate to near-deterministic \cite{Berry.2014.10.1145/2591796.2591854,Berry.2015.10.1109/FOCS.2015.54,Berry.2015.10.1103/PhysRevLett.114.090502}, while postselection can be completely bypassed by viewing the LCU as a block-encoding \cite{Gilyen.2019.10.1145/3313276.3316366} of the spin-adapted unitary and taking advantage of the qubitization algorithm \cite{Low.2017.10.1103/PhysRevLett.118.010501,Low.2019.10.22331/q-2019-07-12-163}.
However, even with these approaches, the tremendous burden on the quantum resources remains.
To make matters worse, the circuit of \cref{fig:LCU} implements a single spin-adapted unitary, while typical ans\"{a}tze are expressed as a product of many such operators.
Although formally exact, these considerations render the LCU approach unattractive for implementing spin-adapted unitaries, even in a fault-tolerant setting.

\section{Classification of the Lie Algebra $\boldsymbol{\Lie\left(A_{\Pu\Pd}^{\Qu\Rd}, A_{\Pu\Pd}^{\Qd\Ru}\right)}$}\label{ssec:ppqr_Lie}

In this section, we show that the 5-dimensional dynamical Lie algebra $\Lie\left(A_{\Pu\Pd}^{\Qu\Rd}, A_{\Pu\Pd}^{\Qd\Ru}\right)$ is isomorphic to the Lie algebra direct sum $\mathbb{R}^2 \oplus \mathfrak{so}(3)$ or, equivalently, $\mathbb{R}^2 \oplus \mathfrak{su}(2)$.

As shown in the main text, a convenient basis is $\Lie\left(A_{\Pu\Pd}^{\Qu\Rd}, A_{\Pu\Pd}^{\Qd\Ru}\right) = \Span(\{E_i\}_{i=1}^5)$, with $E_1 \equiv A_{\Pu\Pd}^{\Qu\Rd}$, $E_2 \equiv A_{\Pu\Pd}^{\Qd\Ru}$, $E_3 \equiv A_{\Pu\Pd}^{\Qu\Rd} \left( h_{\Qd\Ru} + n_{\Qd\Ru} \right)$, $E_4 \equiv A_{\Pu\Pd}^{\Qd\Ru} \left( h_{\Qu\Rd} + n_{\Qu\Rd} \right)$, and $E_5 \equiv A_{\Qu\Rd}^{\Qd\Ru} \left( h_{\Pu\Pd} - n_{\Pu\Pd} \right)$.
Note that, for the purposes of this derivation, the basis elements have been reordered when compared to the main text.
The commutation matrix $\mathbf{C}$, whose elements are defined as $C_{ij} = [E_i, E_j]$, is
\begin{equation}\label{seq:com_matrix}
	\mathbf{C} = 
	\left(\begin{tabular}{ccccc}
		0      & $E_5$  & 0    & $E_5$  & $-E_4$\\
		$-E_5$ & 0      & $-E_5$ & 0    &  $E_3$\\
		0      & $E_5$  & \cellcolor{blue!10} 0      & \cellcolor{blue!10} $E_5$  & \cellcolor{blue!10} $-E_4$\\
		$-E_5$ & 0      & \cellcolor{blue!10} $-E_5$ & \cellcolor{blue!10} 0      & \cellcolor{blue!10}  $E_3$\\
		$E_4$ & $-E_3$ & \cellcolor{blue!10}  $E_4$ & \cellcolor{blue!10} $-E_3$ & \cellcolor{blue!10}  0
	\end{tabular}\right).
\end{equation}
A quick inspection of \cref{seq:com_matrix} reveals that $E_3$, $E_4$, and $E_5$ satisfy commutation relations of the form $[E_i, E_j] = \varepsilon_{ij}^k E_k$, with $\varepsilon_{ij}^k$ denoting the Levi-Civita symbol. Thus, the set $\{A_{\Pu\Pd}^{\Qu\Rd} \left( h_{\Qd\Ru} + n_{\Qd\Ru} \right),A_{\Pu\Pd}^{\Qd\Ru} \left( h_{\Qu\Rd} + n_{\Qu\Rd} \right), A_{\Qu\Rd}^{\Qd\Ru} \left( h_{\Pu\Pd} - n_{\Pu\Pd} \right)\}$ generates a Lie subalgebra isomorphic to $\mathfrak{so}(3)\cong\mathfrak{su}(2)$.

Subsequently, we change the basis by replacing the elements $E_1$ and $E_2$ by $\tilde{E}_1 = E_3 - E_1$ and $\tilde{E}_2 = E_4 - E_2$, respectively.
In doing so, we obtain
\begin{equation}
	\begin{split}
		[\tilde{E}_1, \tilde{E}_2] &= [E_3 - E_1, E_4 - E_2]\\
		&= [E_3,E_4] - [E_3,E_2] - [E_1,E_4] + [E_1,E_2]\\
		&= 0,
	\end{split}
\end{equation}
where we used \cref{seq:com_matrix}.
Thus, the set $\{A_{\Pu\Pd}^{\Qu\Rd} \left( h_{\Qd\Ru} + n_{\Qd\Ru} \right) - A_{\Pu\Pd}^{\Qu\Rd}, A_{\Pu\Pd}^{\Qd\Ru} \left( h_{\Qu\Rd} + n_{\Qu\Rd} \right) - A_{\Pu\Pd}^{\Qd\Ru}\}$ spans an abelian 2-dimensional Lie algebra and, therefore, it is isomorphic to $\mathbb{R}^2$.
Furthermore, from \cref{seq:com_matrix} we immediately observe that $[\tilde{E}_i, E_j] = 0$ for $i = 1$, 2 and $j = 3$, 4, 5. 
As a result, we obtain $\Lie(A_{\Pu\Pd}^{\Qu\Rd}, A_{\Pu\Pd}^{\Qd\Ru}) \cong \mathbb{R}^2 \oplus \mathfrak{so}(3) \cong \mathbb{R}^2 \oplus \mathfrak{su}(2)$.

\section{Closed-Form Unitary Transformation for Fermionic Generators with Particle--Hole-Conjugate Number-Operator Pairs}

To derive the closed-form expression of the unitary transformation shown in \cref{eq:wn_fst} in the main text, we work as follows.
First, we note that a generic element $E_j$ of the dynamical Lie algebras obtained in this work has the form
\begin{equation}
	E_j = A_j N_j,
\end{equation}
where $A_j$ is an anti-Hermitian two-body fermionic excitation operator,
\begin{equation}
	A_j = F_j - F_j^\dagger,
\end{equation}
with $F_j$ being a two-body fermionic string, and $N_j$ being a linear combination of two particle--hole-conjugate number-operator strings, satisfying $N_j^2 = N_j$.
Note that $A_j$ and $N_j$ share no indices by construction and, thus, they commute, $[A_j, N_j] = 0$.
Following our previous work on deriving closed-form fermionic unitary transformations (Ref.\ \cite{Evangelista.2025.10.1103/PhysRevA.111.042825} in the main text), we examine the various powers of $E_j$, obtaining
\begin{equation}
	\begin{split}
		E_j^2 &= A_j N_j A_j N_j\\
		&= A_j^2 N_j^2\\
		&= - (F_j F_j^\dagger + F_j^\dagger F_j) N_j
	\end{split}
\end{equation}
and
\begin{equation}
	\begin{split}
		E_j^3 &= A_j^3 N_j^3\\
		&= - A_j N_j\\
		&= -E_j,
	\end{split}
\end{equation}
where we used the fact that the fermionic string $F_j$ is nilpotent, $F_j^2 = 0$.
As shown in Ref.\ \cite{Evangelista.2025.10.1103/PhysRevA.111.042825} in the main text, the above closure relation directly leads to
\begin{equation}
	[[[E_i, E_j], E_j], E_j] = -[E_i, E_j] -3E_j[E_i, E_j]E_j.
\end{equation}

To prove that the unitary transformation $\exp(\alpha_j E_j) E_i \exp(-\alpha_j E_j)$ proceeds via \cref{eq:wn_fst} in the main text, it suffices to show that $E_j [E_i, E_j] E_j = 0$ for every pair of basis elements $E_i$ and $E_j$ (see Ref.\ \cite{Evangelista.2025.10.1103/PhysRevA.111.042825} for the details).
The above equality holds in the trivial case $[E_i, E_j] = 0$, \textit{i.e.}, when $A_i$ and $A_j$ have no common indices or when $A_i = A_j$.
When $A_i$ and $A_j$ share only a subset of their indices, we use the fact that $\{E_i\}$ is a basis of the pertinent dynamical Lie algebra, allowing us to write
\begin{equation}\label{seq:strct_const}
	[E_i, E_j] = c_{ij}^k E_k,
\end{equation}
where we use the Einstein summation convention of repeated upper and lower indices.
Using \cref{seq:strct_const}, we arrive at
\begin{equation}\label{seq:discriminant}
	E_j[E_i, E_j]E_j = c_{ij}^k E_j E_k E_j. 
\end{equation}
Since each $A_\mu$ is an anti-Hermitian linear combination of two fermionic strings and $N_\mu$ is a linear combination of two particle--hole conjugate number operator strings, each $E_j E_k E_j$ term in \cref{seq:discriminant} gives rise to a linear combination of, at most, 64 fermionic strings.
Subsequently, we use the fact that all indices of $E_j$ appear in each $E_k$.
Indeed, the number operator component $N_k$ of $E_k$ contains the indices shared between $E_i$ and $E_j$, while the remaining indices of $E_i$ and $E_j$ form the excitation part of $E_k$, namely $A_k$.
Because of this structure, equation \cref{seq:discriminant} equals 0 due to the nilpotency of elementary annihilation and creation operators, as in each of the fermionic strings appearing in it there will be at least one pair of annihilation/creation operators with the same index.
Consequently, we obtain \cref{eq:wn_fst} in the main text.	

\section{Effect of Permutations in the Wei--Norman Decompositions of $\boldsymbol{\exp(A_{PP}^{QR})}$}

As an illustration of the effect of permuting the exponentials appearing in the Wei--Norman decomposition of spin-adapted unitaries, we computed all 120 permutations of the five exponentials defining the Wei--Norman decomposition of $\exp(A_{PP}^{QR})$.
Note that the permutations are given relative to the ordering of basis elements used in \cref{ssec:ppqr_Lie}, \foreign{i.e.}, the ordering used in the main text corresponds to the (1,3,2,4,5) permutation.

\FloatBarrier
\begin{figure}[h!]
	\centering
	\includegraphics[height=0.9\textheight]{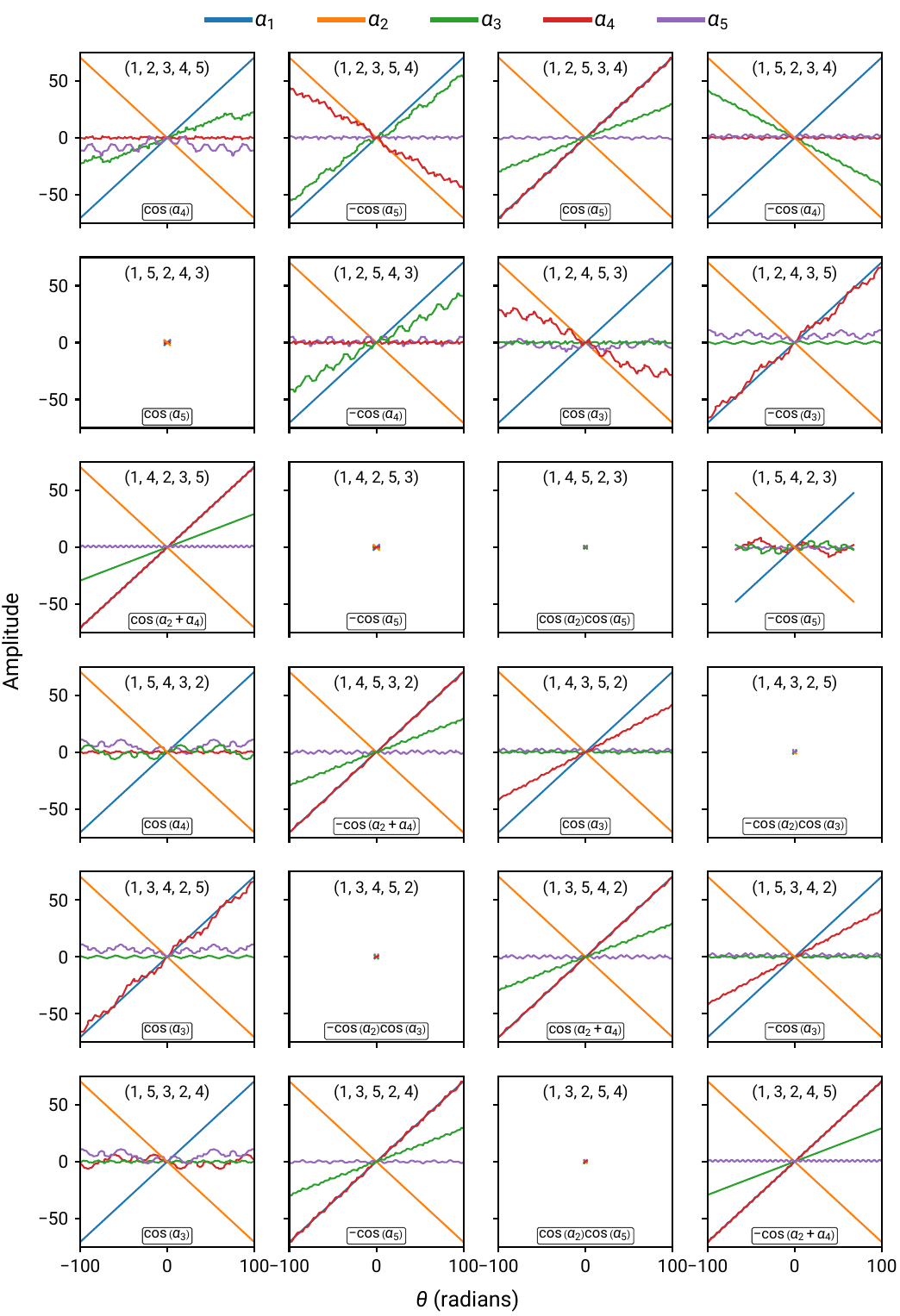}
	\caption{
		Parameters defining the Wei--Norman decompositions of $\exp\left( \theta A_{PP}^{QR}\right)$ [\cref{eq:wn_decomposition}], with $\exp(\alpha_1 E_1)$ in the first position. In each panel, the permutation is given in parentheses, and the determinant of the Wei--Norman coefficient matrix is shown in the rounded box.
	}
	\label{sfig:wn_ppqr_perms_full_range_1}
\end{figure}

\begin{figure}[h!]
	\centering
	\includegraphics[height=0.95\textheight]{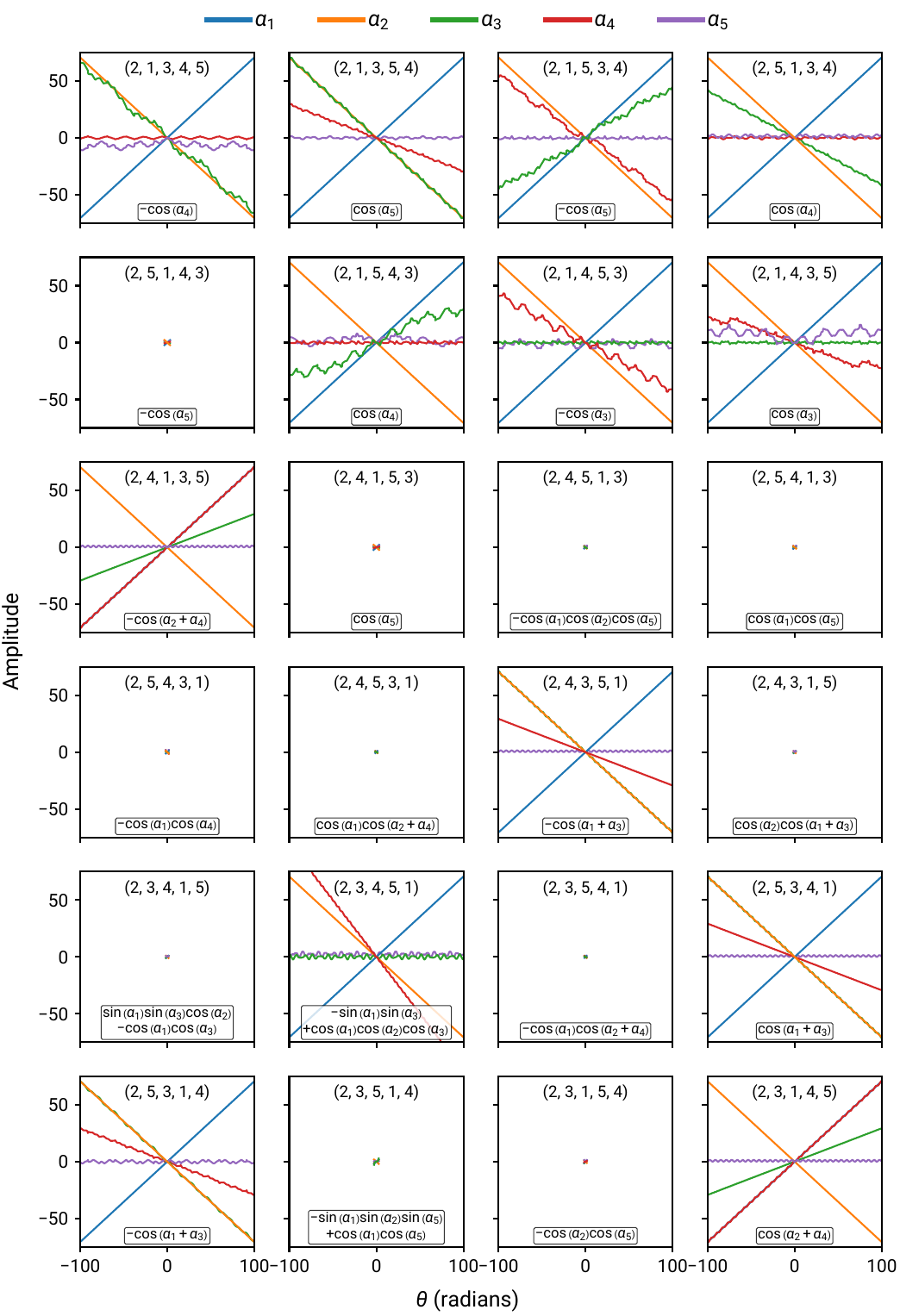}
	\caption{
		Same as \cref{sfig:wn_ppqr_perms_full_range_1}, but with $\exp(\alpha_2 E_2)$ in the first position.
	}
	\label{sfig:wn_ppqr_perms_full_range_2}
\end{figure}

\begin{figure}[h!]
	\centering
	\includegraphics[height=0.95\textheight]{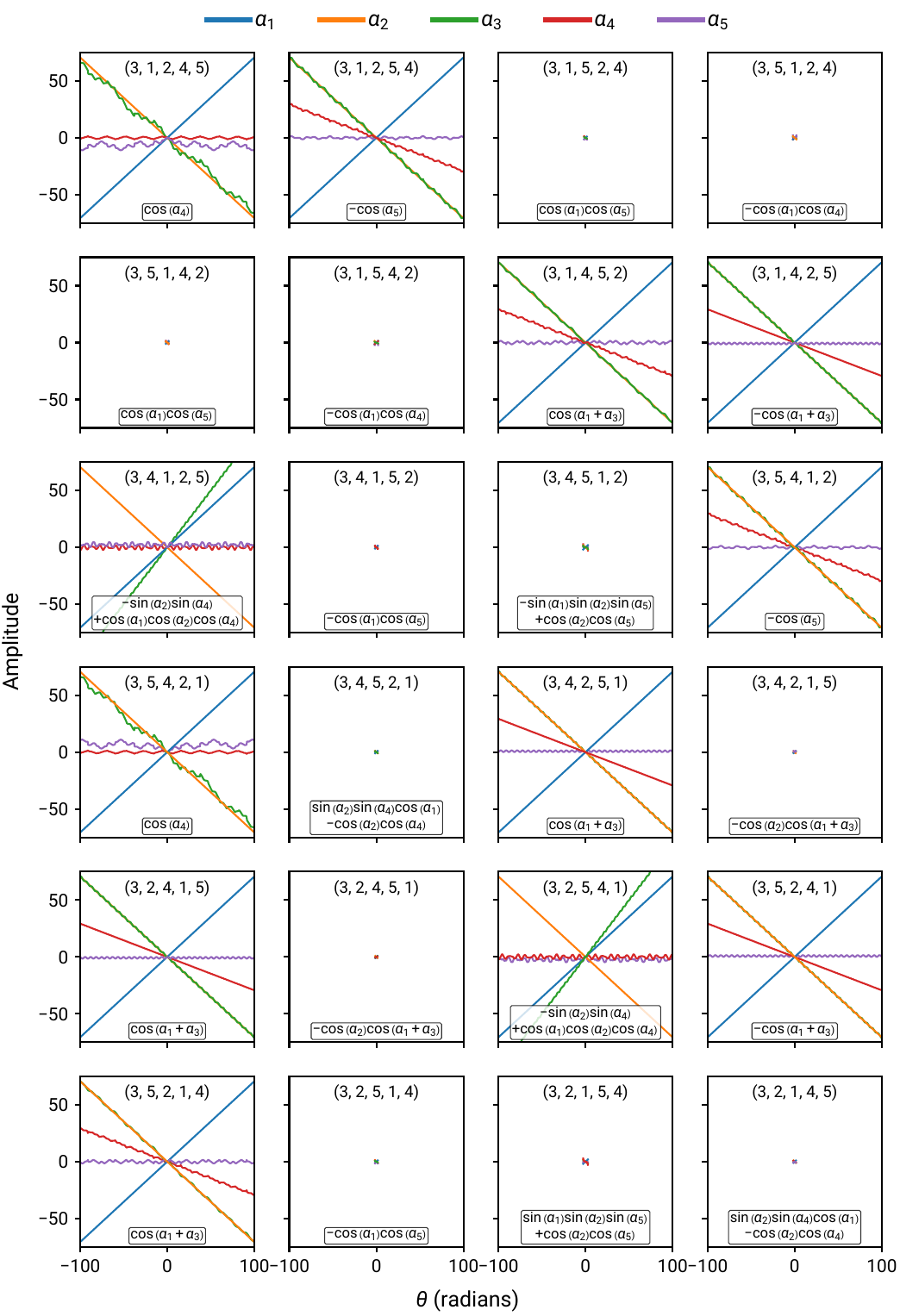}
	\caption{
		Same as \cref{sfig:wn_ppqr_perms_full_range_1}, but with $\exp(\alpha_3 E_3)$ in the first position.
	}
	\label{sfig:wn_ppqr_perms_full_range_3}
\end{figure}

\begin{figure}[h!]
	\centering
	\includegraphics[height=0.95\textheight]{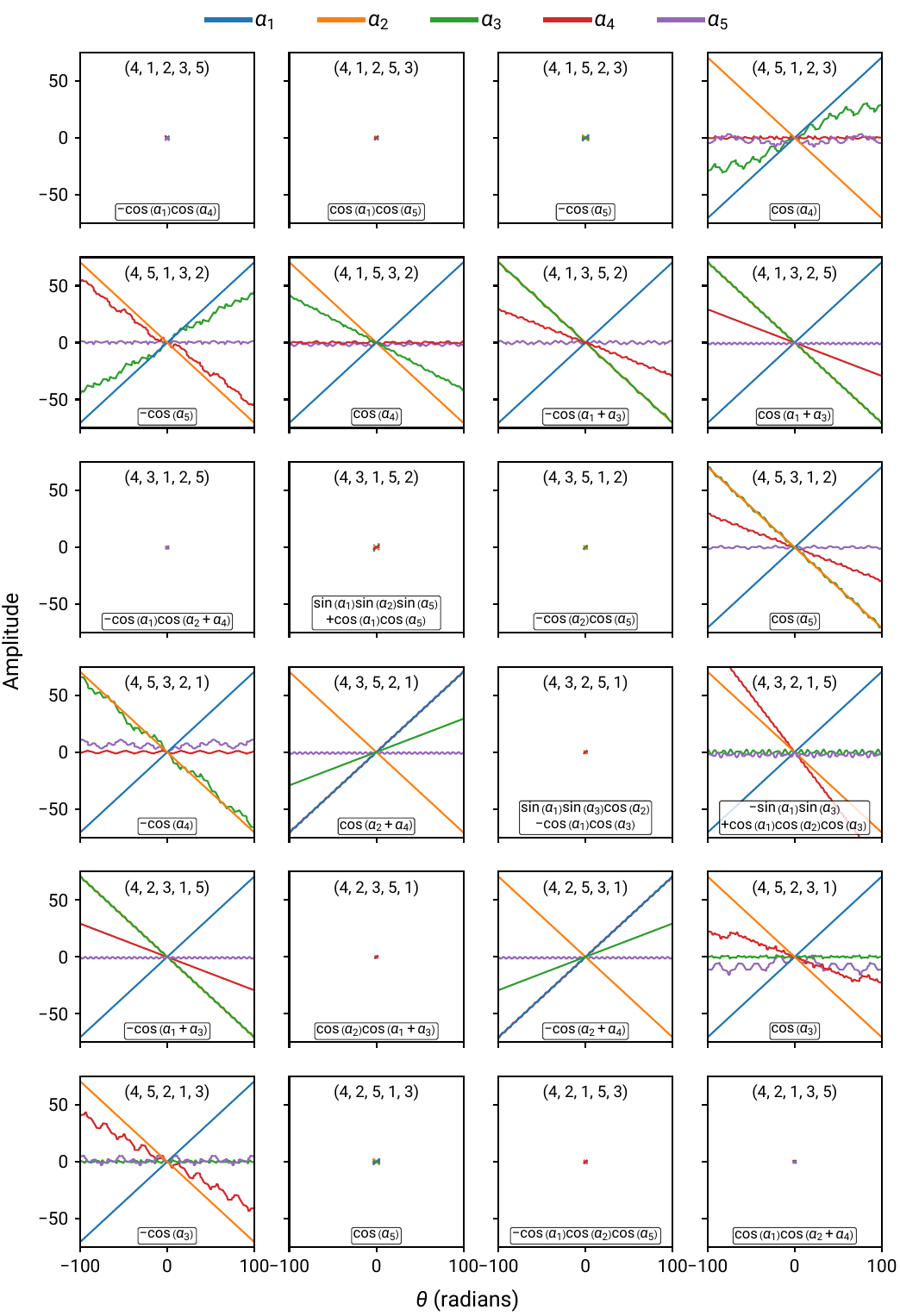}
	\caption{
		Same as \cref{sfig:wn_ppqr_perms_full_range_1}, but with $\exp(\alpha_4 E_4)$ in the first position.
	}
	\label{sfig:wn_ppqr_perms_full_range_4}
\end{figure}

\begin{figure}[h!]
	\centering
	\includegraphics[height=0.95\textheight]{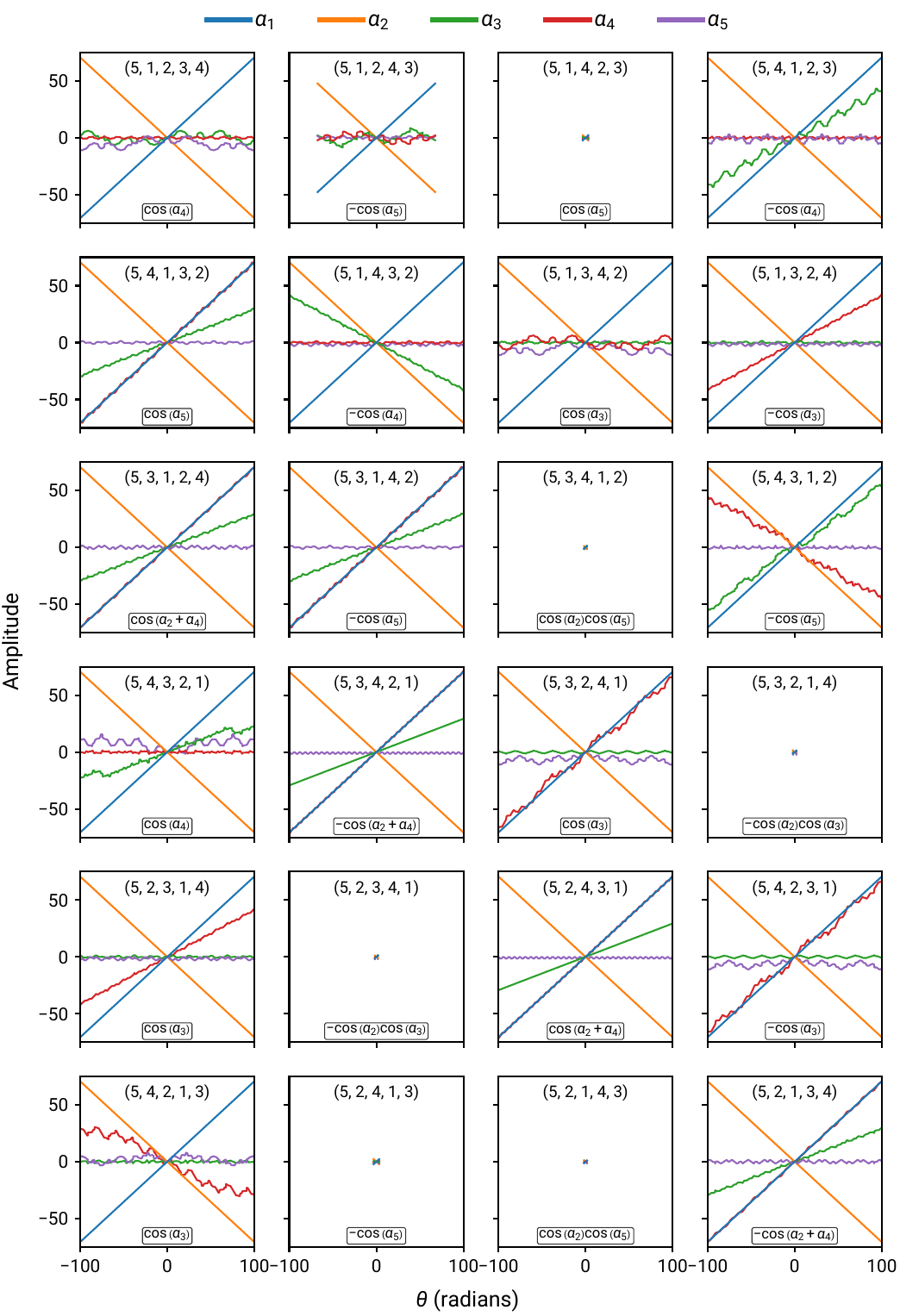}
	\caption{
		Same as \cref{sfig:wn_ppqr_perms_full_range_1}, but with $\exp(\alpha_5 E_5)$ in the first position.
	}
	\label{sfig:wn_ppqr_perms_full_range_5}
\end{figure}
\clearpage

\section{Dynamical Lie Algebras of $\boldsymbol{^{[0]}A_{PQ}^{RS}}$ and $\boldsymbol{^{[1]}A_{PQ}^{RS}}$}

\begin{longtable}{LC}
	\caption{\label{tab:basis_int0_pqrs}
		Basis elements spanning the dynamical Lie algebra of $^{[0]}A_{PQ}^{RS}$, $\mathfrak{g}_0 \equiv \Lie\{A_{\Pu\Qd}^{\Ru\Sd}, A_{\Pd\Qu}^{\Rd\Su}, A_{\Pu\Qd}^{\Rd\Su}, A_{\Pd\Qu}^{\Ru\Sd}\}$.\textsuperscript{\textit{a}}}\\
	\toprule
	\multirow{2}{*}{\text{Basis element}} & \text{Wei--Norman}\\
	& \text{coefficient}\\
	\midrule
	\endfirsthead
	
	\multicolumn{2}{c}{{\tablename\ \thetable{}: \textit{Continued.}}} \\
	\toprule
	\multirow{2}{*}{\text{Basis element}} & \text{Wei--Norman}\\
	& \text{coefficient}\\
	\midrule
	\endhead
	
	\midrule
	\multicolumn{2}{r}{\small\textit{Continued on next page.}}\\
	\endfoot
	
	\bottomrule
	\multicolumn{2}{@{}l@{}}{%
		\parbox{0.7\linewidth}{\footnotesize
			\textsuperscript{\textit{a}}Thin partial horizontal rules separate
			basis elements that differ in the anti-Hermitian excitation part.
			Thick horizontal rules separate mutually commuting sets.}}\\
	\endlastfoot
	
	A_{\Pu\Qd}^{\Ru\Sd} & \frac{\theta}{2} \\
	A_{\Pu\Qd}^{\Ru\Sd} \left(h_{\Pd\Qu}+n_{\Pd\Qu}\right) & b_{1}(\theta) \\
	A_{\Pu\Qd}^{\Ru\Sd} \left(h_{\Rd\Su}+n_{\Rd\Su}\right) & b_{1}(\theta) \\
	A_{\Pu\Qd}^{\Ru\Sd} \left(h_{\Pd\Qu\Rd\Su}+n_{\Pd\Qu\Rd\Su}\right) & -2b_{1}(\theta) \\
	A_{\Pu\Qd}^{\Ru\Sd} \left(n_{\Rd\Su}h_{\Pd\Qu}+n_{\Pd\Qu}h_{\Rd\Su}\right) & \frac{1}{2}b_{1}(2\theta) - 2b_1(\theta) \\
	\AntiHBoundary
	A_{\Pd\Qu}^{\Rd\Su} & \frac{\theta}{2} \\
	A_{\Pd\Qu}^{\Rd\Su} \left(h_{\Pu\Qd}+n_{\Pu\Qd}\right) & b_{1}(\theta) \\
	A_{\Pd\Qu}^{\Rd\Su} \left(h_{\Ru\Sd}+n_{\Ru\Sd}\right) & b_{1}(\theta) \\
	A_{\Pd\Qu}^{\Rd\Su} \left(h_{\Pu\Qd\Ru\Sd}+n_{\Pu\Qd\Ru\Sd}\right) & -2b_{1}(\theta) \\
	A_{\Pd\Qu}^{\Rd\Su} \left(n_{\Ru\Sd}h_{\Pu\Qd}+n_{\Pu\Qd}h_{\Ru\Sd}\right) & \frac{1}{2}b_{1}(2\theta) - 2b_1(\theta) \\
	\CommuteBoundary
	A_{\Pu\Qd}^{\Rd\Su} & -\frac{\theta}{2} \\
	A_{\Pu\Qd}^{\Rd\Su} \left(h_{\Pd\Qu}+n_{\Pd\Qu}\right) & b_{3}(\theta) \\
	A_{\Pu\Qd}^{\Rd\Su} \left(h_{\Ru\Sd}+n_{\Ru\Sd}\right) & b_{3}(\theta) \\
	A_{\Pu\Qd}^{\Rd\Su} \left(h_{\Pd\Qu\Ru\Sd}+n_{\Pd\Qu\Ru\Sd}\right) & -2b_{3}(\theta) \\
	A_{\Pu\Qd}^{\Rd\Su} \left(n_{\Ru\Sd}h_{\Pd\Qu}+n_{\Pd\Qu}h_{\Ru\Sd}\right) & \frac{1}{2}b_{3}(2\theta) - 2b_3(\theta) \\
	\AntiHBoundary
	A_{\Pd\Qu}^{\Ru\Sd} & -\frac{\theta}{2} \\
	A_{\Pd\Qu}^{\Ru\Sd} \left(h_{\Pu\Qd}+n_{\Pu\Qd}\right) & b_{3}(\theta) \\
	A_{\Pd\Qu}^{\Ru\Sd} \left(h_{\Rd\Su}+n_{\Rd\Su}\right) & b_{3}(\theta) \\
	A_{\Pd\Qu}^{\Ru\Sd} \left(h_{\Pu\Qd\Rd\Su}+n_{\Pu\Qd\Rd\Su}\right) & -2b_{3}(\theta) \\
	A_{\Pd\Qu}^{\Ru\Sd} \left(n_{\Rd\Su}h_{\Pu\Qd}+n_{\Pu\Qd}h_{\Rd\Su}\right) & \frac{1}{2}b_{3}(2\theta) - 2b_3(\theta) \\
	\CommuteBoundary
	A_{\Pu\Qd}^{\Pd\Qu} \left(h_{\Ru\Sd}-n_{\Ru\Sd}\right) & b_{5}(\theta) \\
	A_{\Pu\Qd}^{\Pd\Qu} \left(h_{\Rd\Su}-n_{\Rd\Su}\right) & -b_{5}(\theta) \\
	A_{\Pu\Qd}^{\Pd\Qu} \left(h_{\Ru\Rd\Su\Sd}-n_{\Ru\Rd\Su\Sd}\right) & \boldsymbol{0} \\
	A_{\Pu\Qd}^{\Pd\Qu} \left(n_{\Rd\Su}h_{\Ru\Sd}-n_{\Ru\Sd}h_{\Rd\Su}\right) & \frac{1}{2}b_{5}(2\theta) - 2b_5(\theta) \\
	\AntiHBoundary
	A_{\Ru\Sd}^{\Rd\Su} \left(h_{\Pu\Qd}-n_{\Pu\Qd}\right) & b_{5}(\theta) \\
	A_{\Ru\Sd}^{\Rd\Su} \left(h_{\Pd\Qu}-n_{\Pd\Qu}\right) & -b_{5}(\theta) \\
	A_{\Ru\Sd}^{\Rd\Su} \left(h_{\Pu\Pd\Qu\Qd}-n_{\Pu\Pd\Qu\Qd}\right) & \boldsymbol{0} \\
	A_{\Ru\Sd}^{\Rd\Su} \left(n_{\Pd\Qu}h_{\Pu\Qd}-n_{\Pu\Qd}h_{\Pd\Qu}\right) & \frac{1}{2}b_{5}(2\theta) - 2b_5(\theta) \\
\end{longtable}

\begin{longtable}{LC}
	\caption{Basis elements spanning the dynamical Lie algebra of $^{[1]}A_{PQ}^{RS}$, $\mathfrak{g}_1 \equiv \Lie\{A_{\Pu\Qu}^{\Ru\Su}, A_{\Pd\Qd}^{\Rd\Sd}, A_{\Pu\Qd}^{\Ru\Sd}, A_{\Pd\Qu}^{\Rd\Su}, A_{\Pu\Qd}^{\Rd\Su}, A_{\Pd\Qu}^{\Ru\Sd}\}$.\textsuperscript{\textit{a}}}
	\label{tab:basis_int1_pqrs}\\
	\toprule
	\multirow{2}{*}{\text{Basis element}} & \text{Wei--Norman}\\
	& \text{coefficient}\\
	\midrule
	\endfirsthead
	
	\multicolumn{2}{c}{{\tablename\ \thetable{}: Continued from previous page}} \\
	\toprule
	\multirow{2}{*}{\text{Basis element}} & \text{Wei--Norman}\\
	& \text{coefficient}\\
	\midrule
	\endhead
	
	\midrule
	\multicolumn{2}{r}{\small\textit{Continued on next page.}}\\
	\endfoot
	
	\bottomrule
	\multicolumn{2}{@{}l@{}}{%
		\parbox{0.7\linewidth}{\footnotesize
			\textsuperscript{\textit{a}}Thin partial horizontal rules separate
			basis elements that differ in the anti-Hermitian excitation part.
			Thick horizontal rules separate mutually commuting sets.}}\\
	\endlastfoot
	
	A_{\Pu\Qu}^{\Ru\Su} & \frac{\theta}{\sqrt{3}} \\
	A_{\Pu\Qu}^{\Ru\Su} \left(n_{\Sd}h_{\Qd}+n_{\Qd}h_{\Sd}\right) & c_{1}(\theta) \\
	A_{\Pu\Qu}^{\Ru\Su} \left(n_{\Rd}h_{\Pd}+n_{\Pd}h_{\Rd}\right) & -c_{1}(\theta) \\
	A_{\Pu\Qu}^{\Ru\Su} \left(n_{\Rd}h_{\Qd}+n_{\Qd}h_{\Rd}\right) & \boldsymbol{0} \\
	A_{\Pu\Qu}^{\Ru\Su} \left(n_{\Sd}h_{\Pd}+n_{\Pd}h_{\Sd}\right) & \boldsymbol{0} \\
	A_{\Pu\Qu}^{\Ru\Su} \left(n_{\Rd\Sd}h_{\Qd}+n_{\Qd}h_{\Rd\Sd}\right) & -2c_{1}(\theta) \\
	A_{\Pu\Qu}^{\Ru\Su} \left(n_{\Sd}h_{\Pd\Qd}+n_{\Pd\Qd}h_{\Sd}\right) & -2c_{1}(\theta) \\
	A_{\Pu\Qu}^{\Ru\Su} \left(n_{\Rd\Sd}h_{\Pd\Qd}+n_{\Pd\Qd}h_{\Rd\Sd}\right) & 4c_1(\theta) - \frac{1}{2} c_1(2\theta)\\
	\AntiHBoundary
	A_{\Pd\Qd}^{\Rd\Sd} & \frac{\theta}{\sqrt{3}} \\
	A_{\Pd\Qd}^{\Rd\Sd} \left(n_{\Ru}h_{\Pu}+n_{\Pu}h_{\Ru}\right) & c_{1}(\theta) \\
	A_{\Pd\Qd}^{\Rd\Sd} \left(n_{\Su}h_{\Qu}+n_{\Qu}h_{\Su}\right) & -c_{1}(\theta) \\
	A_{\Pd\Qd}^{\Rd\Sd} \left(n_{\Su}h_{\Pu}+n_{\Pu}h_{\Su}\right) & \boldsymbol{0} \\
	A_{\Pd\Qd}^{\Rd\Sd} \left(n_{\Ru}h_{\Qu}+n_{\Qu}h_{\Ru}\right) & \boldsymbol{0} \\
	A_{\Pd\Qd}^{\Rd\Sd} \left(n_{\Ru\Su}h_{\Pu}+n_{\Pu}h_{\Ru\Su}\right) & -2c_{1}(\theta) \\
	A_{\Pd\Qd}^{\Rd\Sd} \left(n_{\Ru}h_{\Pu\Qu}+n_{\Pu\Qu}h_{\Ru}\right) & -2c_{1}(\theta) \\
	A_{\Pd\Qd}^{\Rd\Sd} \left(n_{\Ru\Su}h_{\Pu\Qu}+n_{\Pu\Qu}h_{\Ru\Su}\right) & 4c_1(\theta) - \frac{1}{2} c_1(2\theta) \\
	\CommuteBoundary
	A_{\Pu\Qd}^{\Ru\Sd} & \frac{\theta}{2\sqrt{3}} \\
	A_{\Pu\Qd}^{\Ru\Sd} \left(n_{\Su}h_{\Qu}+n_{\Qu}h_{\Su}\right) & c_{3}(\theta) \\
	A_{\Pu\Qd}^{\Ru\Sd} \left(n_{\Rd}h_{\Pd}+n_{\Pd}h_{\Rd}\right) & c_{4}(\theta) \\
	A_{\Pu\Qd}^{\Ru\Sd} \left(h_{\Rd\Su}+n_{\Rd\Su}\right) & \boldsymbol{0} \\
	A_{\Pu\Qd}^{\Ru\Sd} \left(h_{\Pd\Qu}+n_{\Pd\Qu}\right) & \boldsymbol{0} \\
	A_{\Pu\Qd}^{\Ru\Sd} \left(n_{\Rd\Su}h_{\Qu}+n_{\Qu}h_{\Rd\Su}\right) & c_{4}(\theta) - c_3(\theta) \\
	A_{\Pu\Qd}^{\Ru\Sd} \left(n_{\Su}h_{\Pd\Qu}+n_{\Pd\Qu}h_{\Su}\right) & c_{4}(\theta) - c_3(\theta) \\
	A_{\Pu\Qd}^{\Ru\Sd} \left(n_{\Qu\Rd}h_{\Pd\Su}+n_{\Pd\Su}h_{\Qu\Rd}\right) & -c_{3}(\theta) - c_4(\theta) \\
	\AntiHBoundary
	A_{\Pd\Qu}^{\Rd\Su} & \frac{\theta}{2\sqrt{3}} \\
	A_{\Pd\Qu}^{\Rd\Su} \left(n_{\Ru}h_{\Pu}+n_{\Pu}h_{\Ru}\right) & c_{3}(\theta) \\
	A_{\Pd\Qu}^{\Rd\Su} \left(n_{\Sd}h_{\Qd}+n_{\Qd}h_{\Sd}\right) & c_{4}(\theta) \\
	A_{\Pd\Qu}^{\Rd\Su} \left(h_{\Pu\Qd}+n_{\Pu\Qd}\right) & \boldsymbol{0} \\
	A_{\Pd\Qu}^{\Rd\Su} \left(h_{\Ru\Sd}+n_{\Ru\Sd}\right) & \boldsymbol{0} \\
	A_{\Pd\Qu}^{\Rd\Su} \left(n_{\Ru}h_{\Pu\Qd}+n_{\Pu\Qd}h_{\Ru}\right) & c_{4}(\theta) - c_3(\theta) \\
	A_{\Pd\Qu}^{\Rd\Su} \left(n_{\Ru\Sd}h_{\Pu}+n_{\Pu}h_{\Ru\Sd}\right) & c_{4}(\theta) - c_3(\theta) \\
	A_{\Pd\Qu}^{\Rd\Su} \left(n_{\Qd\Ru}h_{\Pu\Sd}+n_{\Pu\Sd}h_{\Qd\Ru}\right) & -c_{3}(\theta) - c_4(\theta) \\
	\CommuteBoundary
	A_{\Pu\Qd}^{\Rd\Su} & \frac{\theta}{2\sqrt{3}} \\
	A_{\Pu\Qd}^{\Rd\Su} \left(n_{\Ru}h_{\Qu}+n_{\Qu}h_{\Ru}\right) & c_{7}(\theta) \\
	A_{\Pu\Qd}^{\Rd\Su} \left(n_{\Sd}h_{\Pd}+n_{\Pd}h_{\Sd}\right) & c_{8}(\theta) \\
	A_{\Pu\Qd}^{\Rd\Su} \left(h_{\Ru\Sd}+n_{\Ru\Sd}\right) & \boldsymbol{0} \\
	A_{\Pu\Qd}^{\Rd\Su} \left(h_{\Pd\Qu}+n_{\Pd\Qu}\right) & \boldsymbol{0} \\
	A_{\Pu\Qd}^{\Rd\Su} \left(n_{\Ru\Sd}h_{\Qu}+n_{\Qu}h_{\Ru\Sd}\right) & c_{8}(\theta) - c_7(\theta) \\
	A_{\Pu\Qd}^{\Rd\Su} \left(n_{\Ru}h_{\Pd\Qu}+n_{\Pd\Qu}h_{\Ru}\right) & c_{8}(\theta) - c_7(\theta) \\
	A_{\Pu\Qd}^{\Rd\Su} \left(n_{\Qu\Sd}h_{\Pd\Ru}+n_{\Pd\Ru}h_{\Qu\Sd}\right) & -c_{7}(\theta) - c_8(\theta) \\
	\AntiHBoundary
	A_{\Pd\Qu}^{\Ru\Sd} & \frac{\theta}{2\sqrt{3}} \\
	A_{\Pd\Qu}^{\Ru\Sd} \left(n_{\Su}h_{\Pu}+n_{\Pu}h_{\Su}\right) & c_{7}(\theta) \\
	A_{\Pd\Qu}^{\Ru\Sd} \left(n_{\Rd}h_{\Qd}+n_{\Qd}h_{\Rd}\right) & c_{8}(\theta) \\
	A_{\Pd\Qu}^{\Ru\Sd} \left(h_{\Pu\Qd}+n_{\Pu\Qd}\right) & \boldsymbol{0} \\
	A_{\Pd\Qu}^{\Ru\Sd} \left(h_{\Rd\Su}+n_{\Rd\Su}\right) & \boldsymbol{0} \\
	A_{\Pd\Qu}^{\Ru\Sd} \left(n_{\Su}h_{\Pu\Qd}+n_{\Pu\Qd}h_{\Su}\right) & c_{8}(\theta) - c_7(\theta) \\
	A_{\Pd\Qu}^{\Ru\Sd} \left(n_{\Rd\Su}h_{\Pu}+n_{\Pu}h_{\Rd\Su}\right) & c_{8}(\theta) - c_7(\theta) \\
	A_{\Pd\Qu}^{\Ru\Sd} \left(n_{\Qd\Su}h_{\Pu\Rd}+n_{\Pu\Rd}h_{\Qd\Su}\right) & -c_{7}(\theta) - c_8(\theta) \\
	\CommuteBoundary
	A_{\Qu\Sd}^{\Qd\Su} \left(n_{\Ru}h_{\Pu}-n_{\Pu}h_{\Ru}\right) & c_{11}(\theta) \\
	A_{\Qu\Sd}^{\Qd\Su} \left(n_{\Rd}h_{\Pd}-n_{\Pd}h_{\Rd}\right) & -c_{11}(\theta) \\
	A_{\Qu\Sd}^{\Qd\Su} \left(n_{\Ru}h_{\Pu\Rd}-n_{\Pu\Rd}h_{\Ru}\right) & c_{12}(\theta) \\
	A_{\Qu\Sd}^{\Qd\Su} \left(n_{\Pd\Ru}h_{\Pu}-n_{\Pu}h_{\Pd\Ru}\right) & c_{12}(\theta) \\
	A_{\Qu\Sd}^{\Qd\Su} \left(n_{\Pd}h_{\Pu\Rd}-n_{\Pu\Rd}h_{\Pd}\right) & c_{12}(\theta) \\
	A_{\Qu\Sd}^{\Qd\Su} \left(n_{\Rd}h_{\Pd\Ru}-n_{\Pd\Ru}h_{\Rd}\right) & -c_{12}(\theta) \\
	\AntiHBoundary
	A_{\Pu\Rd}^{\Pd\Ru} \left(n_{\Su}h_{\Qu}-n_{\Qu}h_{\Su}\right) & c_{11}(\theta) \\
	A_{\Pu\Rd}^{\Pd\Ru} \left(n_{\Sd}h_{\Qd}-n_{\Qd}h_{\Sd}\right) & -c_{11}(\theta) \\
	A_{\Pu\Rd}^{\Pd\Ru} \left(n_{\Qd\Su}h_{\Qu}-n_{\Qu}h_{\Qd\Su}\right) & c_{12}(\theta) \\
	A_{\Pu\Rd}^{\Pd\Ru} \left(n_{\Su}h_{\Qu\Sd}-n_{\Qu\Sd}h_{\Su}\right) & c_{12}(\theta) \\
	A_{\Pu\Rd}^{\Pd\Ru} \left(n_{\Sd}h_{\Qd\Su}-n_{\Qd\Su}h_{\Sd}\right) & -c_{12}(\theta) \\
	A_{\Pu\Rd}^{\Pd\Ru} \left(n_{\Qd}h_{\Qu\Sd}-n_{\Qu\Sd}h_{\Qd}\right) & c_{12}(\theta) \\
	\CommuteBoundary
	A_{\Qu\Rd}^{\Qd\Ru} \left(n_{\Su}h_{\Pu}-n_{\Pu}h_{\Su}\right) & c_{13}(\theta) \\
	A_{\Qu\Rd}^{\Qd\Ru} \left(n_{\Sd}h_{\Pd}-n_{\Pd}h_{\Sd}\right) & -c_{13}(\theta) \\
	A_{\Qu\Rd}^{\Qd\Ru} \left(n_{\Su}h_{\Pu\Sd}-n_{\Pu\Sd}h_{\Su}\right) & c_{14}(\theta) \\
	A_{\Qu\Rd}^{\Qd\Ru} \left(n_{\Pd\Su}h_{\Pu}-n_{\Pu}h_{\Pd\Su}\right) & c_{14}(\theta) \\
	A_{\Qu\Rd}^{\Qd\Ru} \left(n_{\Pd}h_{\Pu\Sd}-n_{\Pu\Sd}h_{\Pd}\right) & c_{14}(\theta) \\
	A_{\Qu\Rd}^{\Qd\Ru} \left(n_{\Sd}h_{\Pd\Su}-n_{\Pd\Su}h_{\Sd}\right) & -c_{14}(\theta) \\
	\AntiHBoundary
	A_{\Pu\Sd}^{\Pd\Su} \left(n_{\Ru}h_{\Qu}-n_{\Qu}h_{\Ru}\right) & c_{13}(\theta) \\
	A_{\Pu\Sd}^{\Pd\Su} \left(n_{\Rd}h_{\Qd}-n_{\Qd}h_{\Rd}\right) & -c_{13}(\theta) \\
	A_{\Pu\Sd}^{\Pd\Su} \left(n_{\Qd\Ru}h_{\Qu}-n_{\Qu}h_{\Qd\Ru}\right) & c_{14}(\theta) \\
	A_{\Pu\Sd}^{\Pd\Su} \left(n_{\Ru}h_{\Qu\Rd}-n_{\Qu\Rd}h_{\Ru}\right) & c_{14}(\theta) \\
	A_{\Pu\Sd}^{\Pd\Su} \left(n_{\Rd}h_{\Qd\Ru}-n_{\Qd\Ru}h_{\Rd}\right) & -c_{14}(\theta) \\
	A_{\Pu\Sd}^{\Pd\Su} \left(n_{\Qd}h_{\Qu\Rd}-n_{\Qu\Rd}h_{\Qd}\right) & c_{14}(\theta) \\
	\CommuteBoundary
	A_{\Ru\Sd}^{\Rd\Su} \left(h_{\Pu\Qd}-n_{\Pu\Qd}\right) & c_{15}(\theta) \\
	A_{\Ru\Sd}^{\Rd\Su} \left(h_{\Pd\Qu}-n_{\Pd\Qu}\right) & -c_{15}(\theta) \\
	A_{\Ru\Sd}^{\Rd\Su} \left(n_{\Qu}h_{\Pu\Qd}-n_{\Pu\Qd}h_{\Qu}\right) & c_{16}(\theta) \\
	A_{\Ru\Sd}^{\Rd\Su} \left(n_{\Pd\Qu}h_{\Pu}-n_{\Pu}h_{\Pd\Qu}\right) & c_{16}(\theta) \\
	A_{\Ru\Sd}^{\Rd\Su} \left(n_{\Pd}h_{\Pu\Qd}-n_{\Pu\Qd}h_{\Pd}\right) & c_{16}(\theta) \\
	A_{\Ru\Sd}^{\Rd\Su} \left(n_{\Qd}h_{\Pd\Qu}-n_{\Pd\Qu}h_{\Qd}\right) & -c_{16}(\theta) \\
	\AntiHBoundary
	A_{\Pu\Qd}^{\Pd\Qu} \left(h_{\Ru\Sd}-n_{\Ru\Sd}\right) & c_{15}(\theta) \\
	A_{\Pu\Qd}^{\Pd\Qu} \left(h_{\Rd\Su}-n_{\Rd\Su}\right) & -c_{15}(\theta) \\
	A_{\Pu\Qd}^{\Pd\Qu} \left(n_{\Su}h_{\Ru\Sd}-n_{\Ru\Sd}h_{\Su}\right) & c_{16}(\theta) \\
	A_{\Pu\Qd}^{\Pd\Qu} \left(n_{\Rd\Su}h_{\Ru}-n_{\Ru}h_{\Rd\Su}\right) & c_{16}(\theta) \\
	A_{\Pu\Qd}^{\Pd\Qu} \left(n_{\Rd}h_{\Ru\Sd}-n_{\Ru\Sd}h_{\Rd}\right) & c_{16}(\theta) \\
	A_{\Pu\Qd}^{\Pd\Qu} \left(n_{\Sd}h_{\Rd\Su}-n_{\Rd\Su}h_{\Sd}\right) & -c_{16}(\theta) \\
\end{longtable}

\pagebreak

\section{Quantum Circuit of Spin-Adapted Unitary $\boldsymbol{\exp\left(\theta \intT\right)}$}

\begin{figure}[h!]
	\centering
	
	\subfloat[]{
		\includegraphics[width=0.95\linewidth]{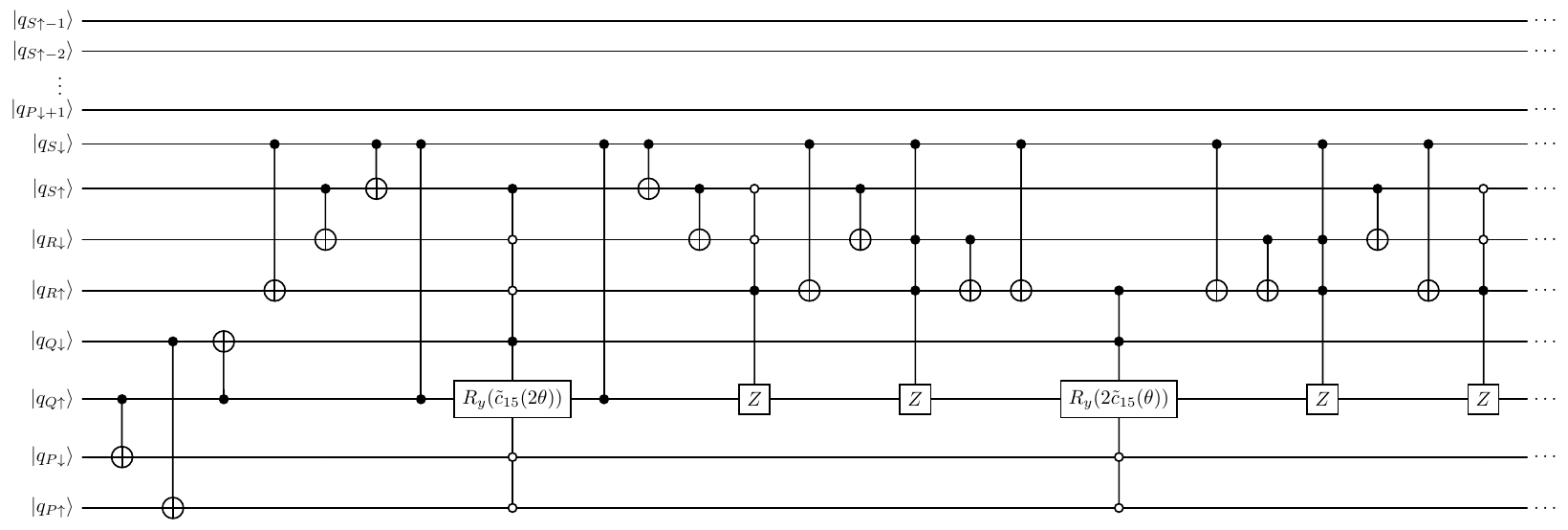}
	}\\
	\subfloat[]{
		\includegraphics[width=0.95\linewidth]{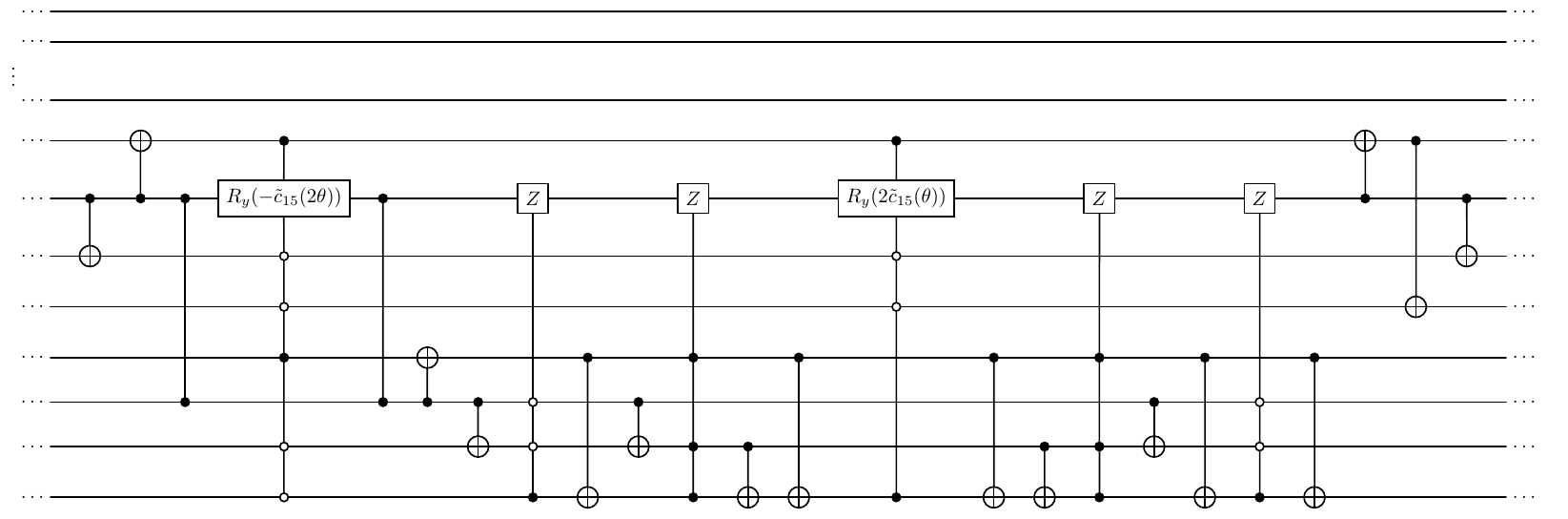}
	}\\
	\subfloat[]{
		\includegraphics[width=0.95\linewidth]{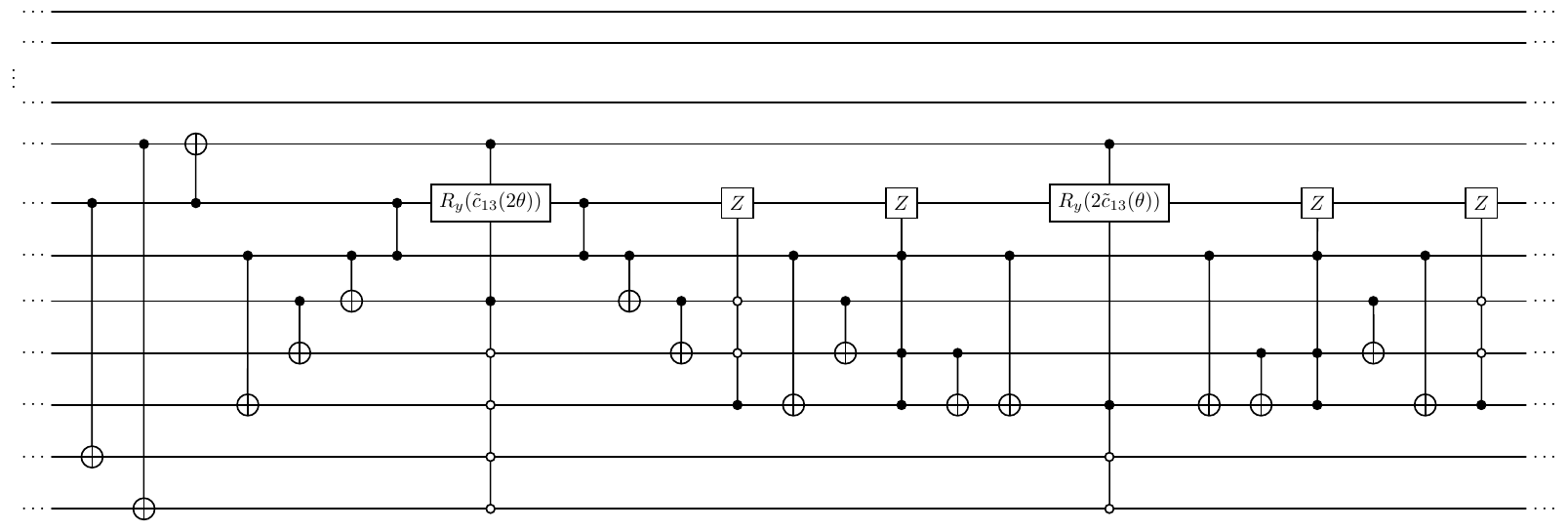}
	}
	\caption{
		First part of the full quantum circuit implementing the spin-adapted unitary $\exp(\theta \intT)$.
		Panels are read top to bottom and show consecutive segments of the same circuit. Continuation ellipses indicate omitted columns at the panel boundaries.
	}
	\label{fig:int1_circ_a}
\end{figure}
\FloatBarrier

\begin{figure}[t]
	\centering
	
	\subfloat[]{
		\includegraphics[width=\linewidth]{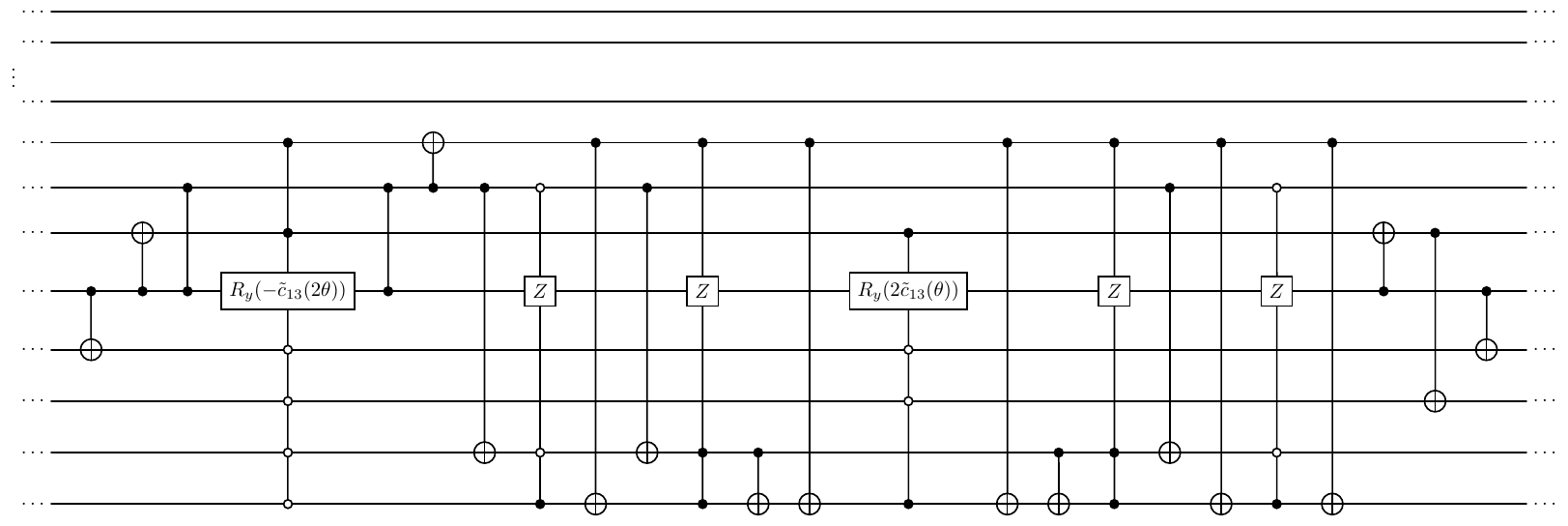}
	}\\
	\subfloat[]{
		\includegraphics[width=\linewidth]{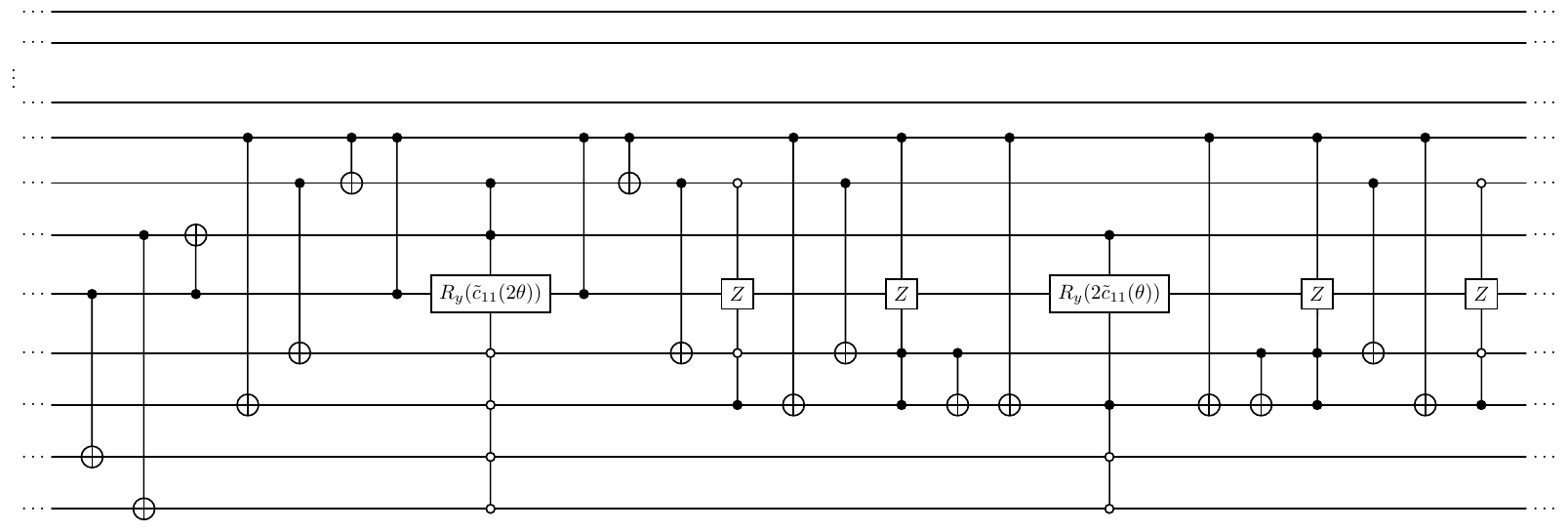}
	}\\
	\subfloat[]{
		\includegraphics[width=\linewidth]{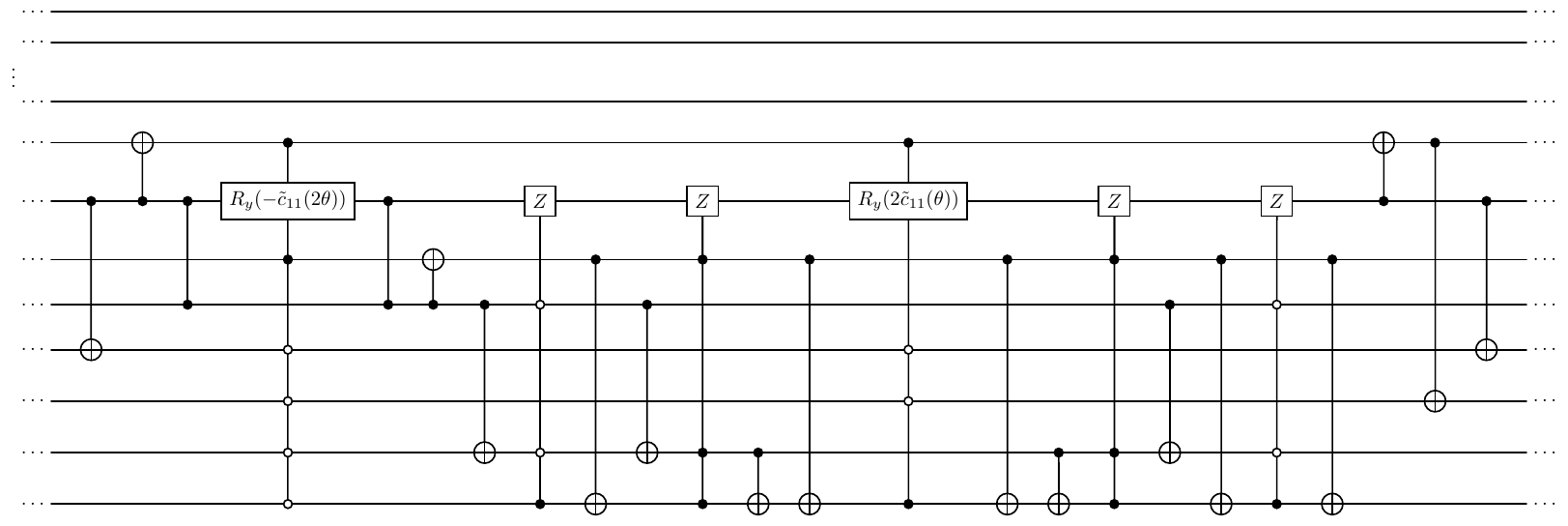}
	}
	\caption{
		Second part of the full quantum circuit implementing the spin-adapted unitary $\exp(\theta \intT)$. Panels are read top to bottom and show consecutive segments of the same circuit. Continuation ellipses indicate omitted columns at the panel boundaries.
	}
	\label{fig:int1_circ_b}
\end{figure}

\begin{figure}[t]
	\centering
	
	\subfloat[]{
		\includegraphics[width=0.69\linewidth]{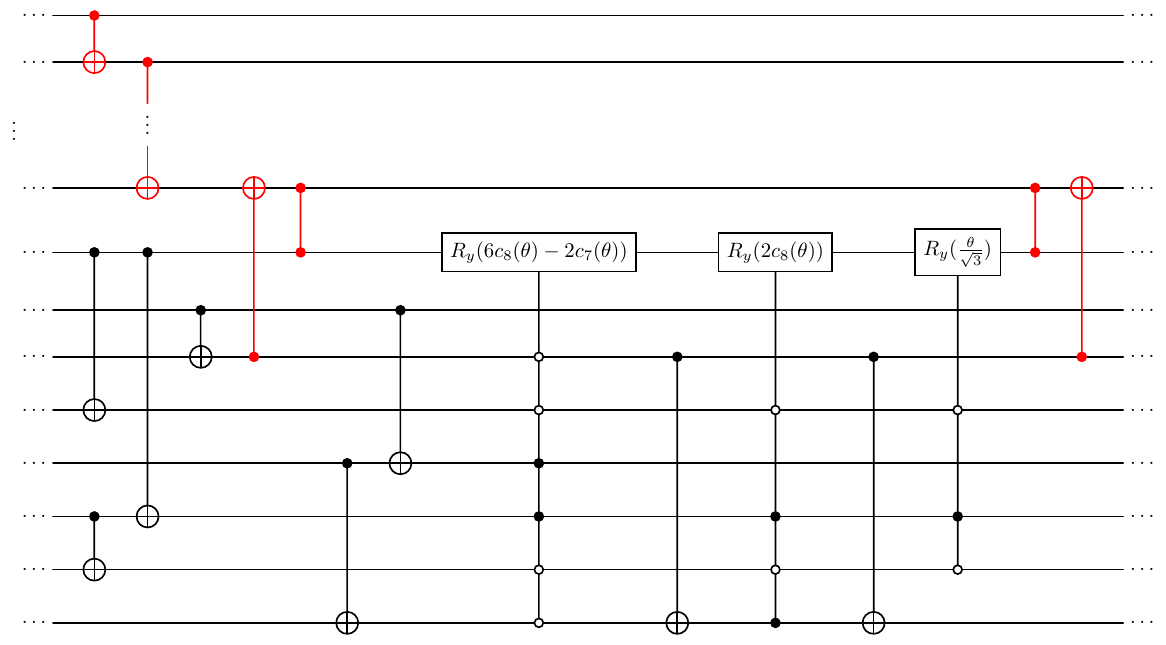}
	}\\
	\subfloat[]{
		\includegraphics[width=0.69\linewidth]{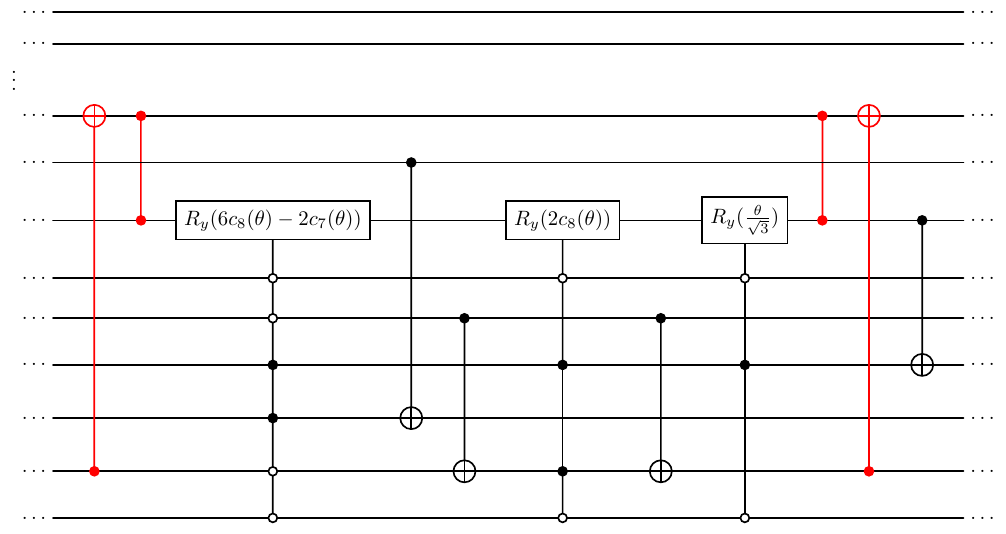}
	}\\
	\subfloat[]{
		\includegraphics[width=0.69\linewidth]{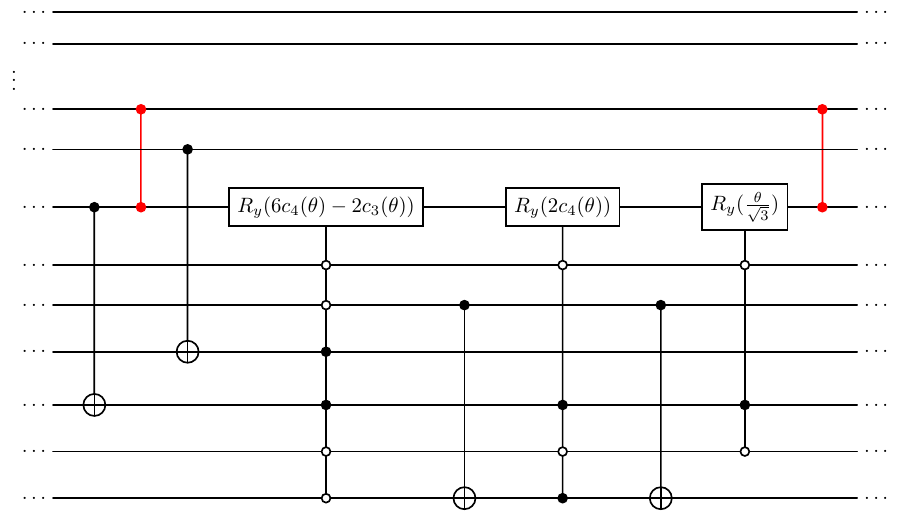}
	}
	\caption{
		Third part of the full quantum circuit implementing the spin-adapted unitary $\exp(\theta \intT)$. Panels are read top to bottom and show consecutive segments of the same circuit. Continuation ellipses indicate omitted columns at the panel boundaries.
	}
	\label{fig:int1_circ_c}
\end{figure}

\begin{figure}[t]
	\centering
	
	\subfloat[]{
		\includegraphics[width=0.75\linewidth]{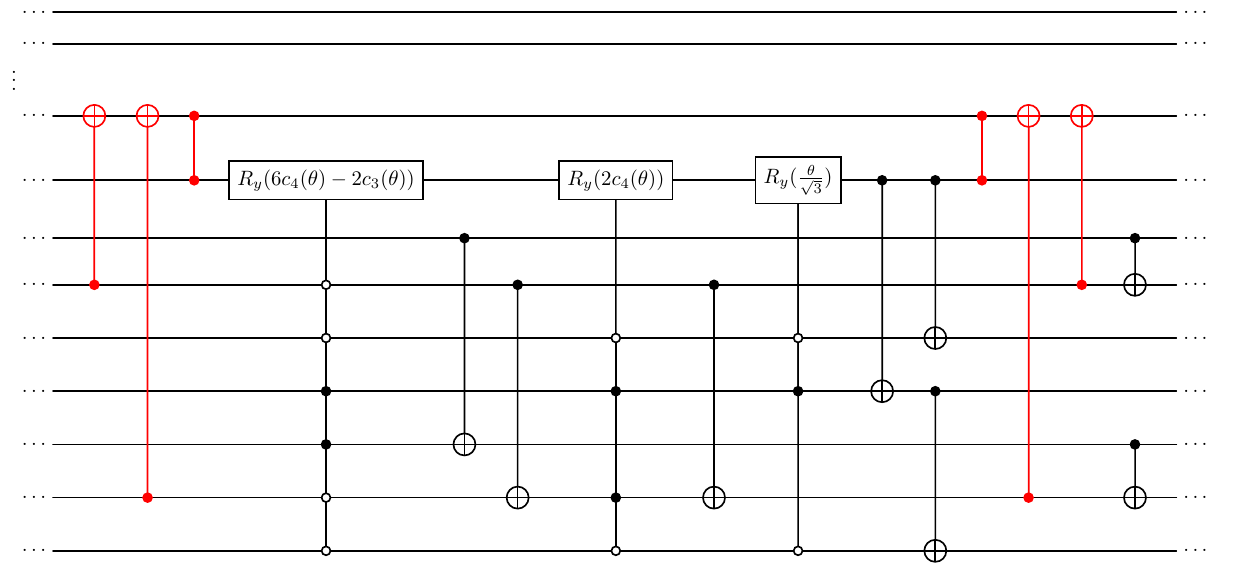}
	}\\
	\subfloat[]{
		\includegraphics[width=0.75\linewidth]{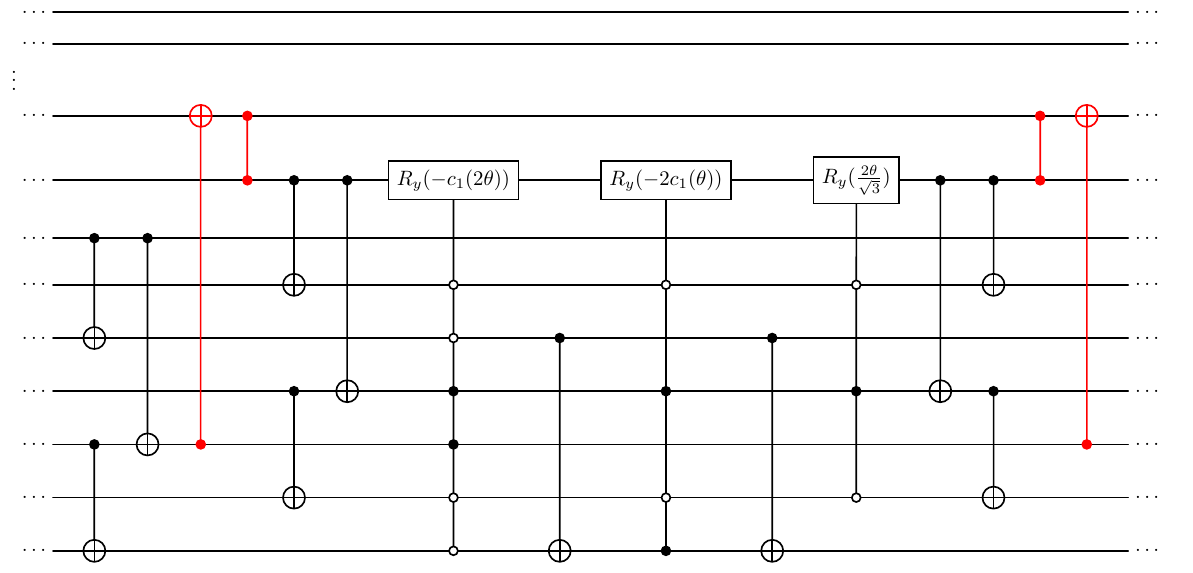}
	}\\
	\subfloat[]{
		\includegraphics[width=0.75\linewidth]{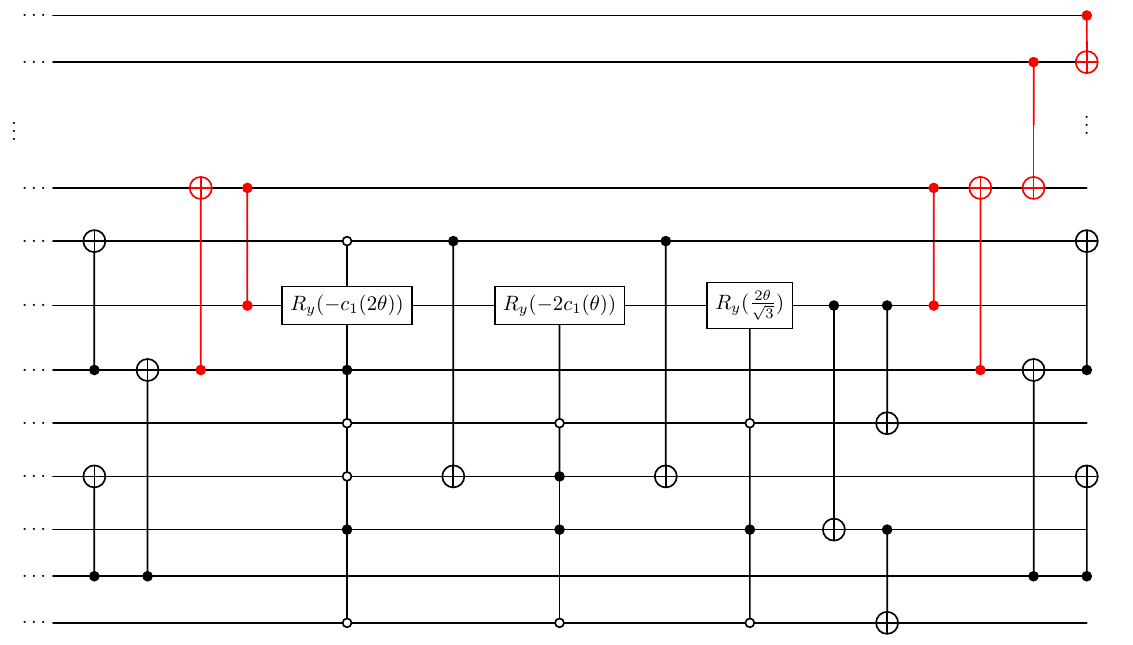}
	}
	\caption{
		Final part of the full quantum circuit implementing the spin-adapted unitary $\exp(\theta \intT)$. Panels are read top to bottom and show consecutive segments of the same circuit. Continuation ellipses indicate omitted columns at the panel boundaries.
	}
	\label{fig:int1_circ_d}
\end{figure}

\FloatBarrier
\section{Nuclear Coordinates of the $\text{H}_6$ system}

\begin{table}[h]
	\centering
	\caption{Nuclear coordinates (in \AA) defining the $D_{2h}$-symmetric distorted hexagonal $\text{H}_6$ structure used in the numerical simulations reported in the main text.}
	\begin{tabular}{cccc}
		\toprule
		atom & $x$  & $y$         & $z$\\
		\midrule
		H1   &    3 &    0        & 0\\
		H2   & $-$3 &    0        & 0\\
		H3   &    2 &  $\sqrt{3}$ & 0\\
		H4   &    2 & $-\sqrt{3}$ & 0\\
		H5   & $-$2 &  $\sqrt{3}$ & 0\\
		H6   & $-$2 & $-\sqrt{3}$ & 0\\
		\bottomrule
	\end{tabular}
\end{table}
\pagebreak

\section{Exact Potential-Energy Curves for the $\boldsymbol{C_s}$-Symmetric Bending Motion of $\boldsymbol{\text{H}_2\text{O}}$/STO-6G}

\begin{figure}[h]
	\includegraphics[width=0.65\linewidth]{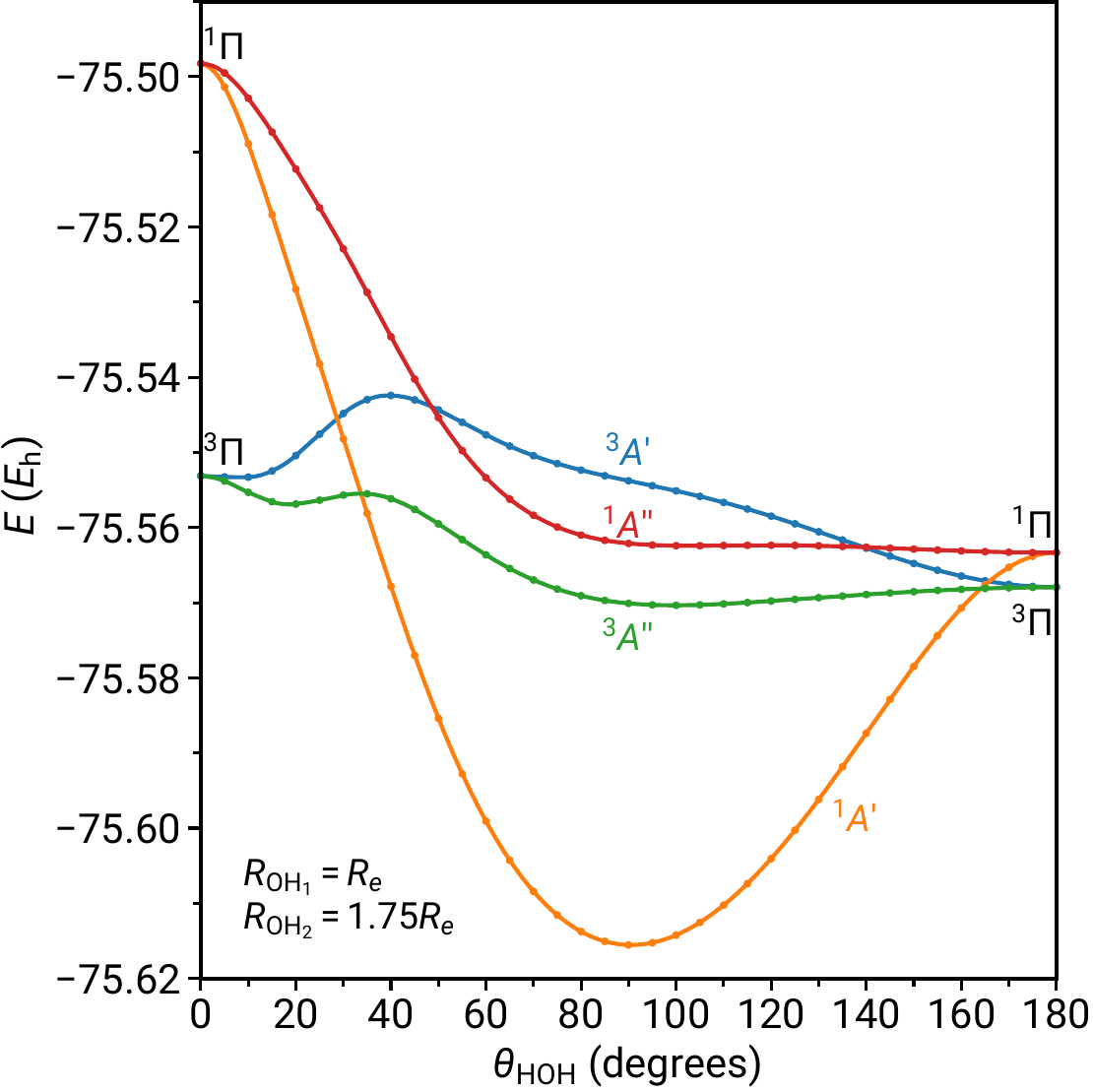}
	\caption{
		Exact potential-energy curves of the lowest-energy $^1A^\prime$, $^3A^\prime$, $^1A^{\prime\prime}$, and $^3A^{\prime\prime}$ states along the $C_s$-symmetric bending coordinate of $\text{H}_2\text{O}$ in the STO-6G basis.
		The O--H equilibrium distance, $R_e = 1.0264$ {\AA}, was determined using exact diagonalization.
	}
	\label{fig:fci_h20}
\end{figure}

\end{document}